\documentclass[usenatbib]{mnras}
\usepackage{times}
\usepackage{flushend}
\usepackage{lscape}
\usepackage{graphicx}
\usepackage{amssymb}
\usepackage{amsmath}
\usepackage{longtable}
\usepackage{url}
\usepackage{bm}
\usepackage{aas_macros}
\usepackage{pdflscape}

\newcommand{\masyr}{mas\,yr$^{-1}$}
\newcommand{\kms}{\textrm{km\,s$^{-1}$}}

\usepackage{multirow}

\title[32 Orionis group]{A stellar census of the nearby, young 32 Orionis group}

\author[C.~P.~M.~Bell et al.]{Cameron~P.~M.~Bell\thanks{E-mail:
  cbell@aip.de (CPMB)},$^{1,2}$ Simon~J.~Murphy$^{3}$ and Eric~E.~Mamajek$^{4,5}$\\
$^{1}$Institute for Astronomy, ETH Z{\"u}rich, Wolfgang-Pauli-Strasse 27, 8093 Z{\"u}rich, Switzerland\\
$^{2}$Leibniz Institute for Astrophysics Potsdam (AIP), An der Sternwarte 16, 14482 Potsdam, Germany\\
$^{3}$School of Physical, Environmental and Mathematical Sciences, University of New South Wales, Northcott Drive, Canberra, ACT 2600, Australia\\
$^{4}$Department of Physics \& Astronomy, University of Rochester, Rochester, NY 14627, USA\\
$^{5}$Jet Propulsion Laboratory, California Institute of Technology, 4800 Oak Grove Drive, Pasadena, CA 91109, USA}

\begin{document}

\date{Accepted ?, Received ?; in original form ?}

\pubyear{2017}

\maketitle

\label{firstpage}

\begin{abstract}
The 32~Orionis group was discovered almost a decade ago and despite
the fact that it represents the first northern, young (age
$\sim$\,25\,Myr) stellar aggregate within 100\,pc of the Sun ($d
\simeq$\,93\,pc), a comprehensive survey for members and detailed
characterisation of the group has yet to be performed. We present the
first large-scale spectroscopic survey for new (predominantly M-type)
members of the group after combining kinematic and photometric data to
select candidates with Galactic space motion and positions in
colour-magnitude space consistent with membership. We identify 30 new
members, increasing the number of known 32~Ori group members by a
factor of three and bringing the total number of identified members to
46, spanning spectral types B5 to L1. We also identify the lithium
depletion boundary (LDB) of the group, i.e. the luminosity at which
lithium remains unburnt in a coeval population. We estimate the age of
the 32~Ori group independently using both isochronal fitting and LDB
analyses and find it is essentially coeval with the $\beta$ Pictoris
moving group, with an age of $24\pm4\,\rm{Myr}$. Finally, we have also
searched for circumstellar disc hosts utilising the All\emph{WISE}
catalogue. Although we find no evidence for warm, dusty discs, we
identify several stars with excess emission in the \emph{WISE}
$W4$-band at $22\,\rm{\mu m}$. Based on the limited number of $W4$
detections we estimate a debris disc fraction of $32^{+12}_{-8}$ per
cent for the 32~Ori group.
\end{abstract}


\section{Introduction}
\label{introduction}

The region surrounding the Sun out to a distance of $\sim$\,100\,pc is
often referred to as the `Local Bubble' on account of the relatively
low density of the interstellar medium and the accompanying lack of
active star-forming regions. Hence the discovery just over three
decades ago of young, seemingly isolated T-Tauri stars in close
proximity to the Sun was a watershed moment and precipitated a massive
observational effort to characterise the young population of the Local
Bubble (see
e.g. \citealp{Rucinski83,delaReza89,Gregorio-Hetem92,Webb99}).  Our
understanding of this young population has increased dramatically
since these early discoveries with the advent of all-sky astrometric
and X-ray/UV surveys.  Not only have hundreds of additional young
stars been identified and spectroscopically characterised within this
region, but more importantly, it has been demonstrated that many of
these stars comprise kinematically distinct, yet spatially dispersed 
groups within which the members share a common motion through space
(see e.g. \citealp{Zuckerman00,Torres00}). To date approximately one
dozen such groups have been identified within 100\,pc with ages
ranging from $\sim$\,10 to 200\,Myr (see reviews by
\citealp{Zuckerman04a}, \citealp{Torres08} and \citealp{Mamajek16}).

The study of nearby young moving groups plays an important role in
constraining theories of star and planet formation. Given their youth
and proximity to Earth, these moving groups provide the best available
samples to investigate the early evolution of low- to
intermediate-mass stars (see e.g. \citealp{Mamajek14}). In addition to
their obvious importance regarding stellar astrophysics, and in
particular the pre-main-sequence (pre-MS) phase, these groups also
represent the most readily accessible targets for direct imaging (and
other measurements/studies) of dusty circumstellar discs; especially
during the epoch of terrestrial planet formation (see
e.g. \citealp{Canup04}). Finally, constituent members of these groups
are ideal candidates for the discovery and characterisation of young,
substellar objects and, of course, extrasolar planets (see
e.g. \citealp{Lagrange10}).

\subsection{The 32 Orionis Group}

\begin{figure}
\centering
\includegraphics[width=\columnwidth]{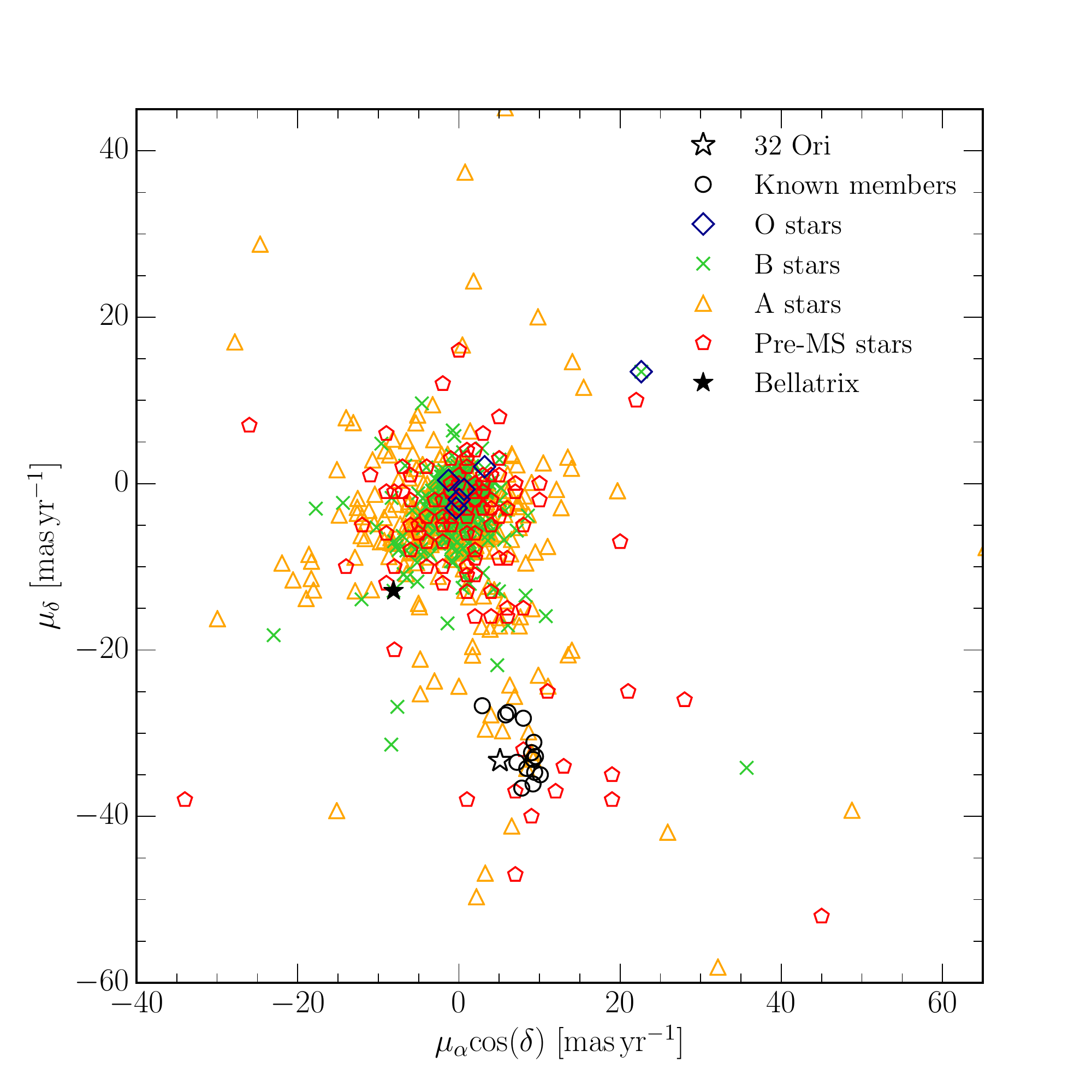}
\caption[]{Proper motions of stars within $10\degr$ of 32~Ori. Proper
  motions for the O-, B- and A-type stars (including 32~Ori and
  Bellatrix; see Introduction) are taken from the revised
  \emph{Hipparcos} reduction, whereas the pre-MS stars are from the
  catalogue of \cite{Ducourant05}. Those for the known members are
  compiled from a combination of the revised \emph{Hipparcos}
  reduction, PPMXL and UCAC4 (see
  Table~\ref{tab:bonafide_members}). The majority of stars have small
  proper motions and are associated with the more distant Orion~OB1
  and $\lambda$~Orionis associations at $\sim$\,400\,pc, however there is
  an appreciable concentration of stars with proper motions similar to
  that of 32~Ori.}
\label{fig:proper_motion}
\end{figure}

Unlike other nearby young moving groups and associations, the 32~Ori
group has received little attention in the literature despite its
discovery almost a decade ago, and as such its stellar population
remains poorly characterised. \cite{Mamajek07} was the first to
present evidence of a young stellar aggregate (age $\sim$25\,Myr;
designated Mamajek~3) associated with the B5IV+B7V binary 32~Ori based
on a concentration of co-moving stars in a proper motion
diagram. Fig.~\ref{fig:proper_motion} shows the proper motions of
stars within $10\degr$ of 32~Ori, for which the O-, B- and A-type
stars (including Bellatrix and 32~Ori itself) are taken from the
revised \emph{Hipparcos} reduction of \cite{vanLeeuwen07}. Also
plotted are pre-MS stars from the catalogue of \cite{Ducourant05} and
the 15 known stellar members of the 32~Ori group, compiled from a
combination of \emph{Hipparcos}, PPMXL \citep*{Roeser10} and the
Fourth U.S. Naval Observatory CCD Astrograph Catalog (UCAC4;
\citealp{Zacharias13}).

While the majority of stars in Fig.~\ref{fig:proper_motion} have small
proper motions ($\lesssim$\,10~\masyr) and are associated with the
more distant ($\sim$\,400\,pc) Orion~OB1 and $\lambda$~Orionis
associations, there is a clear concentration of stars in proper motion
space in the vicinity of 32~Ori. Within this concentration there are
three A-type stars with \emph{Hipparcos} entries (HR~1807 [HIP~25453],
HD~36823 [HIP~26161] and HD~35714 [HIP~25483]) which are all
approximately co-distant with 32~Ori, and which together provide a
weighted mean group distance of $92.9\pm2.3\,\rm{pc}$. Also appearing
to be co-moving with 32~Ori are several X-ray bright late-type stars
discovered by the \emph{ROSAT} All-Sky Survey
\citep{Voges99}. Spectroscopic follow-up of a subset of these was
performed by \cite{Alcala96,Alcala00} who measured strong lithium (Li)
absorption in addition to deriving radial velocities which are
consistent with that of 32~Ori ($18.6\pm1.2\,\rm{km\,s^{-1}}$;
\citealp{Barbier-Brossat00}).

\begin{table*}
\caption[]{Properties of the 15 known stellar members of the 32~Ori group.}
\begin{tabular}{l c c c c c c c c c}
\hline
Name                    &   $\alpha$ (J2000.0)   &   $\delta$ (J2000.0)   &   SpT   &   Ref.   &   $\mu_{\alpha} \mathrm{cos} (\delta)$   &   $\mu_{\delta}$                       &   Ref.   &   RV                                        &   Ref.\\
                               &   (hh mm ss.ss)            &   (hh mm ss.s)             &              &             &   ($\mathrm{mas\,yr^{-1}}$)                       &   ($\mathrm{mas\,yr^{-1}}$)   &              &   ($\mathrm{km\,s^{-1}}$)   &   \\
\hline
32 Ori                     &   05 30 47.05               &   +05 56 53.3   &   B5IV+B7V    &   1   &   $5.10\pm0.67$                                           &   $-33.30\pm0.35$                 &   2        &   $18.6\pm1.2$                    &   3    \\
HR 1807                    &   05 26 38.83               &   +06 52 07.2   &   B9.5V           &   4   &   $9.22\pm0.56$                                          &   $-33.15\pm0.31$                  &   2        &   $13.1\pm2.5$                    &   5    \\
HD 35714                   &   05 26 59.99               &   +07 10 13.0   &   A3                 &   6   &   $9.34\pm0.04$                                          &   $-32.97\pm0.03$                  &   7        &   $35.4\pm1.0$                     &   8   \\
HD 36823                   &   05 34 38.42               &   +06 07 36.7   &   A7.5               &   9   &   $9.07\pm0.05$                                          &   $-31.67\pm0.02$                  &   7        &   --                                          &   --   \\
HD 35499                   &   05 25 14.56               &   +04 11 48.2   &   F4                 &   6   &   $8.29\pm0.54$                                          &   $-29.02\pm0.56$                  &   7        &   --                                            &   --           \\
HD 36338                   &   05 31 15.70               &   +05 39 46.4   &   F4.5              &   6   &   $9.48\pm0.48$                                         &   $-32.76\pm0.54$                   &   7        &   --                                            &   --           \\
HD 35695                   &   05 26 52.03               &   +06 28 22.7   &   F9                 &   6   &   $9.20\pm0.80$                                          &   $-36.10\pm1.20$                   &   10        &           --                                       &   --     \\
HD 245567                  &   05 37 18.43               &   +13 34 52.5   &   G5                &   6   &   $7.20\pm0.90$                                          &   $-33.50\pm0.80$                   &   10        &   $14.9\pm0.8$                      &   11    \\
HD 245059                  &   05 34 34.91               &   +10 07 06.4   &   G7                &   6   &   $10.10\pm1.30$                                        &   $-35.00\pm1.20$                   &   10        &   $19.8\pm1.0$                      &   12      \\
TYC 112-1486-1            &   05 20 31.82               &   +06 16 11.6   &   K3                &   6   &   $9.50\pm1.80$                                            &   $-32.80\pm2.10$                  &   10        &   $18.5\pm0.2$                      &   13      \\
TYC 112-917-1             &   05 20 00.29               &   +06 13 03.7   &   K4                 &   6   &   $9.40\pm1.90$                                            &   $-34.70\pm2.10$                 &   10        &   $18.8\pm0.1$                      &   13      \\
2MASS J05234246+0651581   &   05 23 42.46              &   +06 51 58.2   &   K6.5              &   6   &   $7.80\pm2.30$                                            &   $-36.60\pm2.80$                 &   10        &   $18.4\pm1.0$                      &   12      \\
V1874 Ori$^{\dagger}$            &   05 29 19.00             &   +12 09 29.6   &   K6.5               &   6   &   $2.90\pm2.10$                                            &   $-26.70\pm3.00$                 &   10        &   $18.4\pm0.3$                      &   14    \\
2MASS J05253253+0625336   &   05 25 32.54             &   +06 25 33.7   &   M3                  &   6   &   $8.00\pm5.80$                                            &   $-28.20\pm6.10$                 &   10        &   --                                            &   --            \\
2MASS J05194398+0535021   &   05 19 43.98             &   +05 35 02.2   &   M3                  &   6   &   $5.80\pm4.00$                                            &   $-27.80\pm4.00$                 &   15      &   --                                            &   --           \\
\hline
\end{tabular}

\vspace{1pt}
\begin{flushleft}
$^{\dagger}$Double-lined spectroscopic binary. The quoted radial
velocity is the centre-of-mass velocity.\\
%
References for spectral types, proper motions and radial
  velocities: (1) \protect\cite{Edwards76}, (2)
\protect\cite{vanLeeuwen07}, (3) \protect\cite{Barbier-Brossat00}, (4)
\protect\cite{Abt95}, (5) \protect\cite*{Bobylev06}, (6) \protect\cite{Shvonski16},
(7) \protect\cite{Gaia16}, (8) \protect\cite{Gontcharov06},
(9) Mark Pecaut (priv. comm.), (10)
\protect\cite{Zacharias13}, (11) \protect\cite*{White07}, (12)
\protect\cite{Alcala00}, (13) \protect\cite{Elliott14}, (14)
\protect\cite{Mace09}, (15) \protect\cite{Roeser10}.
\end{flushleft}
\label{tab:bonafide_members}
\end{table*}

\begin{table*}
\caption[]{Mean properties of the 32~Ori group (Mamajek~3).}
\begin{tabular}{c c c c c c c c c}
\hline
$\alpha$ (J2000.0)   &   $\delta$ (J2000.0)   &   $\pi$$^{\dagger}$           &   $\mu_{\alpha} \mathrm{cos} (\delta)$   &   $\mu_{\delta}$                       &   RV$^{\ddag}$                            &   $U$                                     &   $V$                                      &   $W$   \\
(hh mm ss.ss)            &   (hh mm ss.s)             &   (mas)                      &   ($\mathrm{mas\,yr^{-1}}$)                       &   ($\mathrm{mas\,yr^{-1}}$)   &   ($\mathrm{km\,s^{-1}}$)   &   ($\mathrm{km\,s^{-1}}$)   &   ($\mathrm{km\,s^{-1}}$)   &   ($\mathrm{km\,s^{-1}}$)   \\
\hline
05 27 16.32               &   +06 40 37.2              &   $10.49\pm0.22$   &   $8.6\pm0.5$                                              &   $-32.6\pm0.5$                       &   $18.6\pm0.2$                     &   $-11.8\pm0.3$                   &   $-18.9\pm0.3$                  &   $-8.9\pm0.3$\\
\hline
\end{tabular}
\vspace{1pt}
\begin{flushleft}
$^{\dagger}$Calculated using the seven stars with
trigonometric parallax measurements from the revised \emph{Hipparcos}
reduction and \emph{Gaia} DR1 (32~Ori, HR~1807, HD~35714, HD~36823,
HD~35499, HD~36338, and TYC 112-1486-1).\\
$^{\ddag}$The mean group radial velocity does not include the velocities
of HR~1807 and HD~35714. Unlike 32~Ori, for which the value in
Table~\ref{tab:bonafide_members} represents the mean of 39
observations, single-epoch measurements of rapidly-rotating early-type
stars are notoriously unreliable.
\end{flushleft}
\label{tab:group_mean}
\end{table*}

\begin{figure*}
\centering
\includegraphics[width=\textwidth]{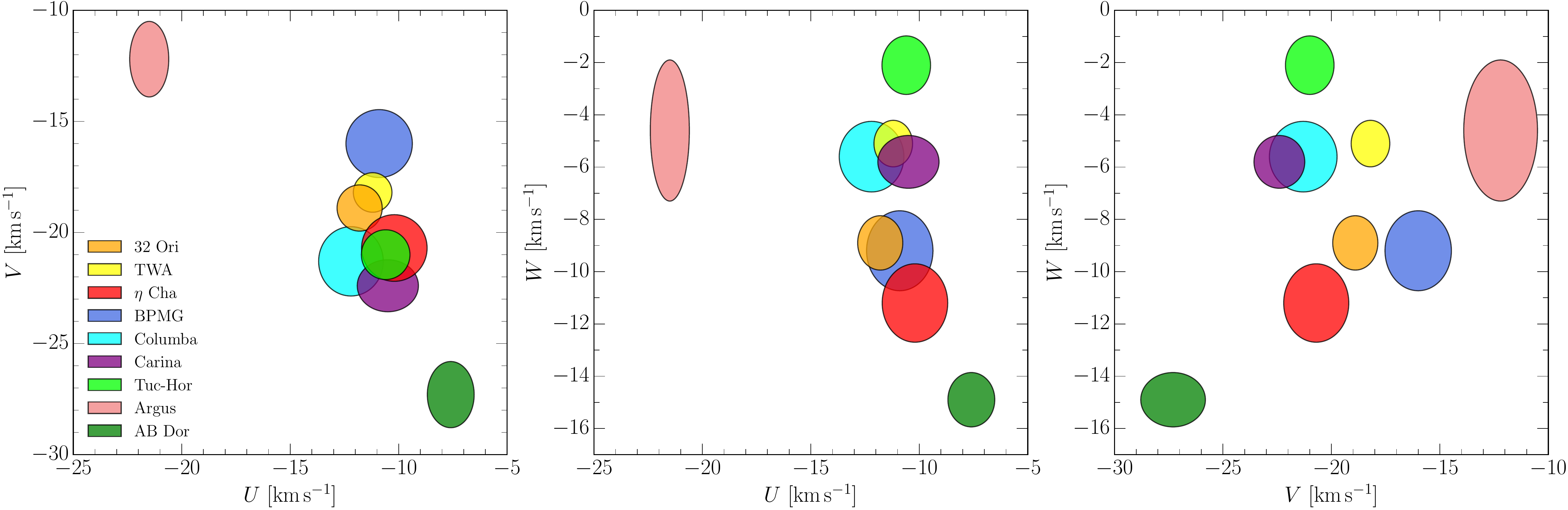}
\caption[]{Galactic $UVW$ velocity of the 32~Ori group relative to
  other young groups and associations within 100\,pc.  Ellipses
  represent the observed velocity dispersion of each group, i.e. the
  uncertainties in the mean velocities ($\sigma_{U}, \sigma_{V},
  \sigma_{W}$) added in quadrature with the intrinsic one-dimensional
  velocity dispersion (typically $\lesssim 1.5\,\rm{km\,s^{-1}}$). In
  all three planes the 32 Ori group lies close in velocity space to
  the $\beta$ Pictoris moving group (BPMG).}
\label{fig:uvw}
\end{figure*}

To date, a total of 15 stars have been identified as members of the
32~Ori group, in addition to a potential planetary-mass object
(\citealp{Burgasser16}; see Section~\ref{sec:burgasser}). The stellar
members have been compiled primarily by Mamajek and collaborators
(\citealp{Mamajek07, Shvonski16}) -- but see also additions by
\cite{Mace09} and \cite{Franciosini11} -- over the last decade, with
new members being assigned on the basis of various diagnostics
including common proper motion, radial velocities, H$\alpha$ emission
and Li absorption. Note that the recent study of \cite*{Bell15} only
included 14 stellar members of the group, having neglected
HD~36823. This star has been shown to be a wide, common proper motion
companion to HD~35714 \citep{Shaya11}, and so we include it here in
our list of known group members. Table~\ref{tab:bonafide_members}
provides positions, spectral types, proper motions and radial
velocities for the 15 known stellar members of the 32~Ori group, while
Table~\ref{tab:group_mean} lists the mean group position, parallax,
proper motion, radial velocity and Galactic $UVW$ velocity. The
spectral types listed in Table~\ref{tab:bonafide_members} for all
stars other than 32~Ori and HR~1807 have been determined
through visual comparison of flux-calibrated spectra acquired over the
past decade against a dense grid of MK spectral standards (see
\citealp{Shvonski16}).
Fig.~\ref{fig:uvw} depicts the $UVW$ velocity of the 32~Ori group
relative to other young groups and associations within
100\,pc. Velocities for the other young groups were taken from the
recent compilation of \cite{Mamajek16}, except for Argus which is from
\cite{Gagne14}. Interestingly, not only does the 32~Ori group appear very
close in velocity to the $\beta$ Pictoris moving group (BPMG) in all three
$UVW$ planes, but the two groups also have very similar ages (see~\citealp{Bell15}
and the discussion in Section~\ref{properties_32_ori_group}).

Two of the 15 stars listed in Table~\ref{tab:bonafide_members}
(TYC~112-1486-1 and TYC~112-917-1) have previously been classified as
members of the BPMG by \cite{Elliott14}. Using the Bayesian Analysis
for Nearby Young AssociatioNs (BANYAN) II web tool
\citep{Gagne14,Malo13}, we calculate that both stars have BPMG
membership probabilities of $\lesssim$\,2 per cent, whereas the
probabilities that they are associated with the young field are much
higher ($\simeq$\,90 per cent). Furthermore, their kinematic distances
(see Section~\ref{initial_input_catalogue_and_search_criteria} for
details) are $\sim$\,80\,pc for BPMG (adopting the $UVW$ velocity from
\citealp{Mamajek16}). Hence, if both stars are genuine BPMG members
they would be two of the most distant members of the group, at
approximately twice the $\sim$\,40\,pc median distance of the `classic'
membership list of \citet[see also
  \citealp{Mamajek14}]{Zuckerman04a}. Based on this evidence we retain
both TYC~112-1486-1 and TYC~112-917-1 as 32~Ori group members for this
study and include them in our determination of the mean group
properties listed in Table~\ref{tab:group_mean}.

\cite{Shvonski16} describes a 
\emph{Spitzer} IRAC and MIPS survey of the 32~Ori group which was
performed during 2007/08, covering all bandpasses from 3.6 to
$70\,\rm{\mu m}$ (see also \citealp{Shvonski10}). In this study the
authors combined the \emph{Spitzer} photometry with optical and near-infrared
(IR) data to quantify excess emission arising from circumstellar
material. \citeauthor{Shvonski16} report that 4/14 members exhibit excess
$24\,\rm{\mu m}$ emission; HR~1807 ($f_{24}$ = 88.45\,$\pm$\,0.36\,mJy),
HD~35499 ($f_{24}$ = 8.45\,$\pm$\,0.10\,mJy), HD~36338 ($f_{24}$ =
14.79\,$\pm$\,0.15\,mJy) and TYC~112-1486-1 ($f_{24}$ =
3.87\,$\pm$\,0.09\,mJy). HR~1807 also exhibits a $70\,\rm{\mu m}$
excess ($f_{70}$ = 91.0\,$\pm$\,4.2\,mJy). 
Note that the quoted fluxes represent total fluxes
and correspond to excess emission more than 4$\sigma$ above typical
photospheric levels. Modelling the excess
emission, \citeauthor{Shvonski16} determined that the dust
temperatures associated with these debris disc candidates were $\lesssim$\,200\,K.

Recently, \cite{Bouy15} argued that the 32~Ori group should in fact be
termed the `Bellatrix cluster' on the basis that the sky position,
distance ($77^{+4}_{-3}\,\rm{pc}$; \citealp{vanLeeuwen07}) and age
($20^{+2}_{-4}\,\rm{Myr}$; \citealp{Janson11}) of the B2V star
Bellatrix ($\gamma$~Ori) are similar to those of the 32~Ori group.
There are three additional remarkable coincidences among its stellar
observables:
\begin{enumerate}
\item Bellatrix's radial velocity
  ($18.2\pm0.8\,\rm{km\,s^{-1}}$; \citealp{Barbier-Brossat00}) is
  consistent with the mean group radial velocity of the 32 Ori group
  ($18.6\pm0.2\,\rm{km\,s^{-1}}$).
\item Bellatrix's reddening [$E(B-V)=0.02$\,mag,
    \citealp{Friedemann92,Zorec09,Bhatt15}] is a good match for the
  mean of the 32 Ori group [($E(B-V)=0.03\pm0.01$\,mag; see
    Table~\ref{tab:red}]\footnote{Using the revised Johnson Q-method
    calibration of \citet{Pecaut13}, and the $U-B$ and $B-V$ colours
    from \citet{Mermilliod06}, we independently estimate the reddening
    of Bellatrix to be $E(B-V)=0.017$\,mag.}.
\item Bellatrix is the only star in
  the northern hemisphere within 100\,pc of Earth of spectral type B2 or earlier.
\end{enumerate}

\noindent Despite these coincidences, however, we find that Bellatrix's deviant
proper motion (see Fig.~\ref{fig:proper_motion}) translates to a 3D
velocity of ($U, V, W) = (-14.7\pm0.7, -7.1\pm0.4,
-9.8\pm0.3)\,\rm{km\,s^{-1}}$ which is inconsistent ($>$10$\sigma$
in the $V$ component) with the mean velocity of the 32 Ori group shown
in Table~\ref{tab:group_mean}. In light of this kinematic
inconsistency we do not believe there is sufficient evidence at present
to include Bellatrix in the group, and do not recommend use of
the name Bellatrix cluster at this time. Despite this inconsistency, however,
we believe further
astrometric, spectroscopic, and high-contrast imaging of Bellatrix is warranted to see whether it could harbour a dark companion which
could be responsible for its deviant velocity\footnote{Bellatrix has a mass of $\sim$ $8.7\,\rm{M_{\odot}}$
  \citep*{Hohle10}, and its projected tangential velocity is peculiar
  compared to the 32 Ori group velocity by $\Delta \mu_{\alpha},
  \Delta \mu_{\delta} \simeq -16, +20 $\,\masyr ($\sim$\,12\,\kms\, if
  the star is actually at $d \simeq$\,93\,pc), with negligible
  difference in radial velocity. A normal dwarf, white dwarf, or neutron star companion can not
reconcile this velocity offset and the consistency of Bellatrix's
proper motion over the past decades. The star is too bright for \emph{Gaia} DR1, but the revised \emph{Hipparcos} catalogue
\citep{vanLeeuwen07} reports a 7-parameter solution for Bellatrix with a
2.5$\sigma$ significance acceleration in $\mu_{\alpha}$
(4.85\,$\pm$\,1.94 mas\,yr$^{-2}$) and negligible acceleration in
$\mu_{\delta}$. Bellatrix's velocity and properties can be reconciled
with 32 Ori group membership if it is in a face-on orbit perturbed by
a distant (a $\sim$\,10$^2$\,au, $P$ $\sim$ centuries) black hole
companion. We are currently investigating Bellatrix further to test
this idea.}.

We hereby
present the first large-scale spectroscopic survey for new low-mass
members of the 32 Ori group to better understand its stellar
population and properties. In Section~\ref{candidate_selection} we
discuss our candidate selection
process. Section~\ref{spectroscopic_observations} details the
medium-resolution optical spectroscopic observations in addition to
describing how the spectroscopic properties were determined. In
Section~\ref{32_ori_low_mass_population} we combine various indicators
of stellar youth and group membership to identify new bona fide
members, and place these in context by comparing them against the
findings of previous surveys. Finally,
Section~\ref{properties_32_ori_group} synthesises our stellar census
results to discuss the global properties of the 32 Ori group,
including its age, circumstellar disc frequency and spatial structure.

\section{Candidate selection}
\label{candidate_selection}

Members of a moving group typically cover both a large range of
distances and a large area on the sky; unlike compact clusters
whose members essentially share the same proper motion, the proper
motions and tangential velocities of individual group members can vary
systematically and significantly. To identify potential kinematic
members of a group one must therefore project its fixed Galactic $UVW$
velocity onto the sky over a range of distances and search for objects
sharing a common motion. Furthermore, given the young age of the
32~Ori group, any genuine low-mass members will not only share a
common space motion, but will also be over-luminous with respect to older
 main sequence stars of the same spectral type.  In the absence
of trigonometric parallaxes (soon to be rectified by \emph{Gaia}), the
combination of proper motions and photometry provides a
robust method with which to identify potential group members, whilst
also efficiently removing a substantial number of field interlopers
which are naturally included in large area searches (see \citealp{Kraus14,Murphy15}).

\subsection{Input catalogues and search criteria}
\label{initial_input_catalogue_and_search_criteria}

We adopt UCAC4 as our primary input catalogue, which provides
positions, absolute proper motions and instrumental magnitudes (in a
single, non-standard bandpass similar to $R$, hereafter termed
$R_{\rm{UCAC}}$) complete to $\simeq$\,16~mag across the entire
sky. In addition to these instrumental magnitudes, the catalogue also
includes APASS \citep{Henden12} DR6 $BVgri$ and 2MASS Point Source
Catalog \citep{Cutri03} $JHK_{\rm{s}}$ photometry. Note that as a
consequence of the APASS limiting magnitude ($V \simeq 16\,\rm{mag}$),
standard $BVgri$ photometry is only available for approximately 50 per
cent of the sources in UCAC4.

\begin{figure}
\centering
\includegraphics[width=\columnwidth]{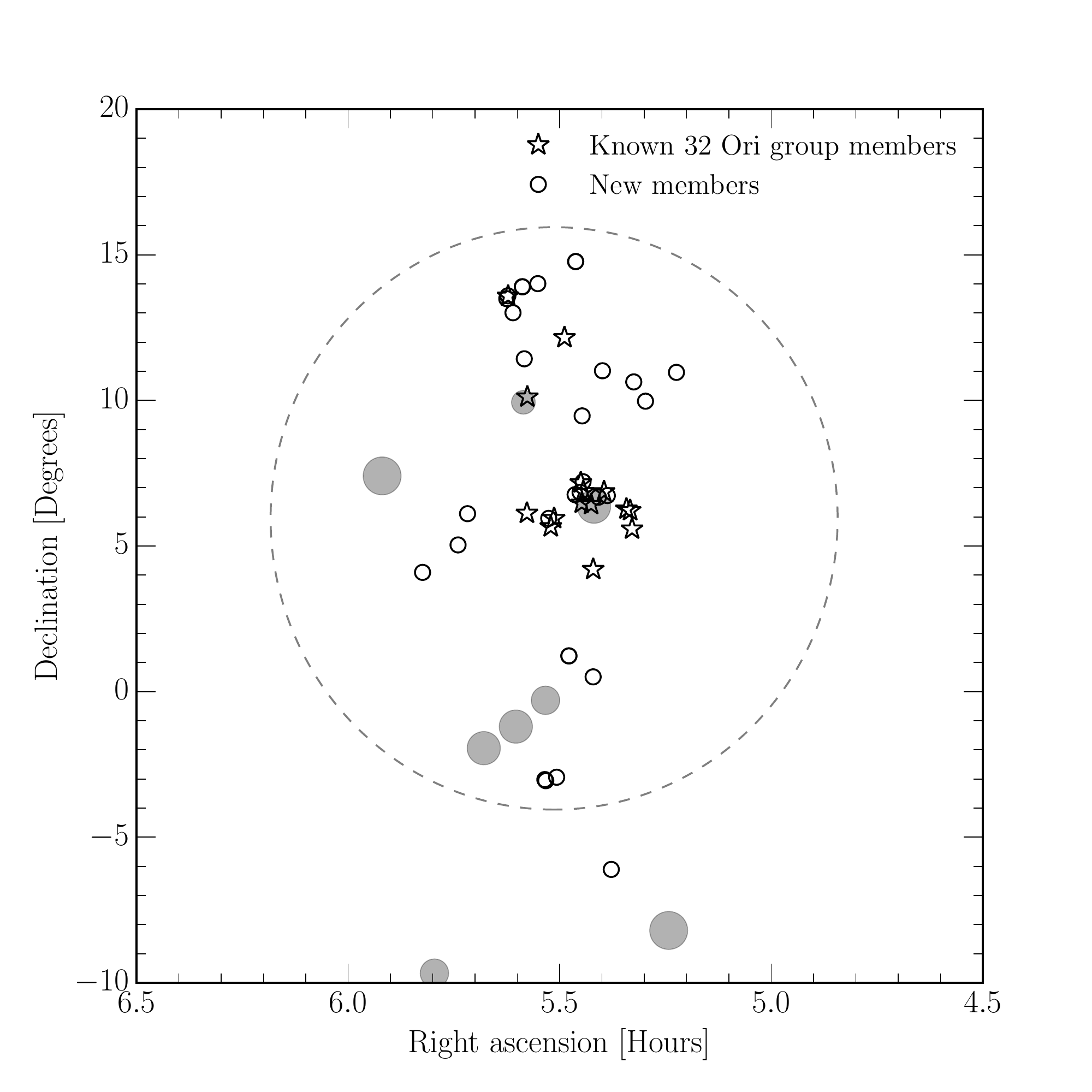}
\caption[]{Spatial distribution of the 15 known stellar members of the 32~Ori
  group listed in Table~\ref{tab:bonafide_members}, in addition to the
  30 new members we have identified in this study. The dashed circle
  denotes our 10$\degr$ search radius and the large grey circles
  represent the bright stars outlining the constellation Orion. The new member
  outside of our search radius is discussed in Section~\ref{sec:burgasser}
  and represents confirmation of a previously identified potential member.}
\label{fig:search_radius}
\end{figure}

Given the sky positions of the known group members (see
Fig.~\ref{fig:search_radius}), a 10$\degr$ search radius around 32~Ori
itself represents a reasonable compromise between survey area and the
telescope time required for spectroscopic follow-up. Within this
region, UCAC4 returned a total of $\sim$\,$7.1 \times 10^{5}$
sources. To identify potential kinematic members from this sample we
follow a formalism similar to that described in
\cite{Murphy15}. Adopting the mean $UVW$ velocity for the 32~Ori group\footnote{Our selection of candidate members
preceded \emph{Gaia} DR1 and so the actual $UVW$ velocity we adopted
was slightly different to that stated in Table~\ref{tab:group_mean}, which we present
as the current best estimate for the mean group velocity. The omission of
the \emph{Gaia} DR1 astrometry modifies the mean group parallax by $+0.27\,\rm{mas}$ and
the mean group proper motion by $-0.4$ and $0.0\,\rm{mas\,yr^{-1}}$ in
$\mu_{\alpha} \mathrm{cos} (\delta)$ and $\mu_{\delta}$, respectively, which results in
a mean space motion of $(U, V, W) = (11.9\pm0.3, -18.6\pm0.4, -9.0\pm0.3)\,\rm{km\,s^{-1}}$.
We note, however, that this subtle difference in the adopted $UVW$ velocity does not
have a significant impact on the calculated $\Delta_{\rm{PM}}$ and kinematic distances,
on average affecting these at the $< 1\,\rm{mas\,yr^{-1}}$ and $<$\,2\,pc level, respectively.},
we project this over a
range of distances ($50 \leq d \leq 150\,\rm{pc}$, in 1\,pc
increments) for each object to calculate the expected proper motion
and radial velocity, retaining only those objects whose proper motions
and kinematic distances satisfied both of the following criteria:

\begin{enumerate}
\item The lowest total difference between the expected and observed
  proper motion,
\begin{equation}
\Delta_{\mathrm{PM}} = \left[ (\mu_{\alpha} \mathrm{cos}\, \delta_{\mathrm{\,obs}} - \mu_{\alpha} \mathrm{cos}\, \delta_{\mathrm{\,expt}})^2 + (\mu_{\delta\, \mathrm{obs}} - \mu_{\delta\, \mathrm{expt}})^2\right]^{1/2},
\end{equation}

\noindent must be $\leq$10\,\masyr.\\

\item The `best' kinematic distance corresponding to this proper
  motion must be $70 \leq d_{\rm kin} \leq 110\,\rm{pc}$.
\end{enumerate}

These thresholds returned 5349 potential kinematic members of the
32~Ori group, each of which had an associated $\Delta_{\rm PM}$,
$d_{\rm kin}$ and expected radial velocity.  The $d_{\rm kin}$ and
$\Delta_{\mathrm{PM}}$ limits are somewhat arbitrary, but are
motivated by the bona fide members as found in UCAC4 (14/15 with
$\Delta_{\rm PM}< 10$\,\masyr\ and 13/15 with $70<d_{\rm
  kin}<110$\,pc). During the spectroscopic follow-up we also observed
stars whose distance and $\Delta_{\mathrm{PM}}$ values fell outside
these limits but otherwise resembled strong candidates.

\subsection{Colour-magnitude selection}
\label{colour_magnitude_selection}

\begin{figure}
\centering
\includegraphics[width=\columnwidth]{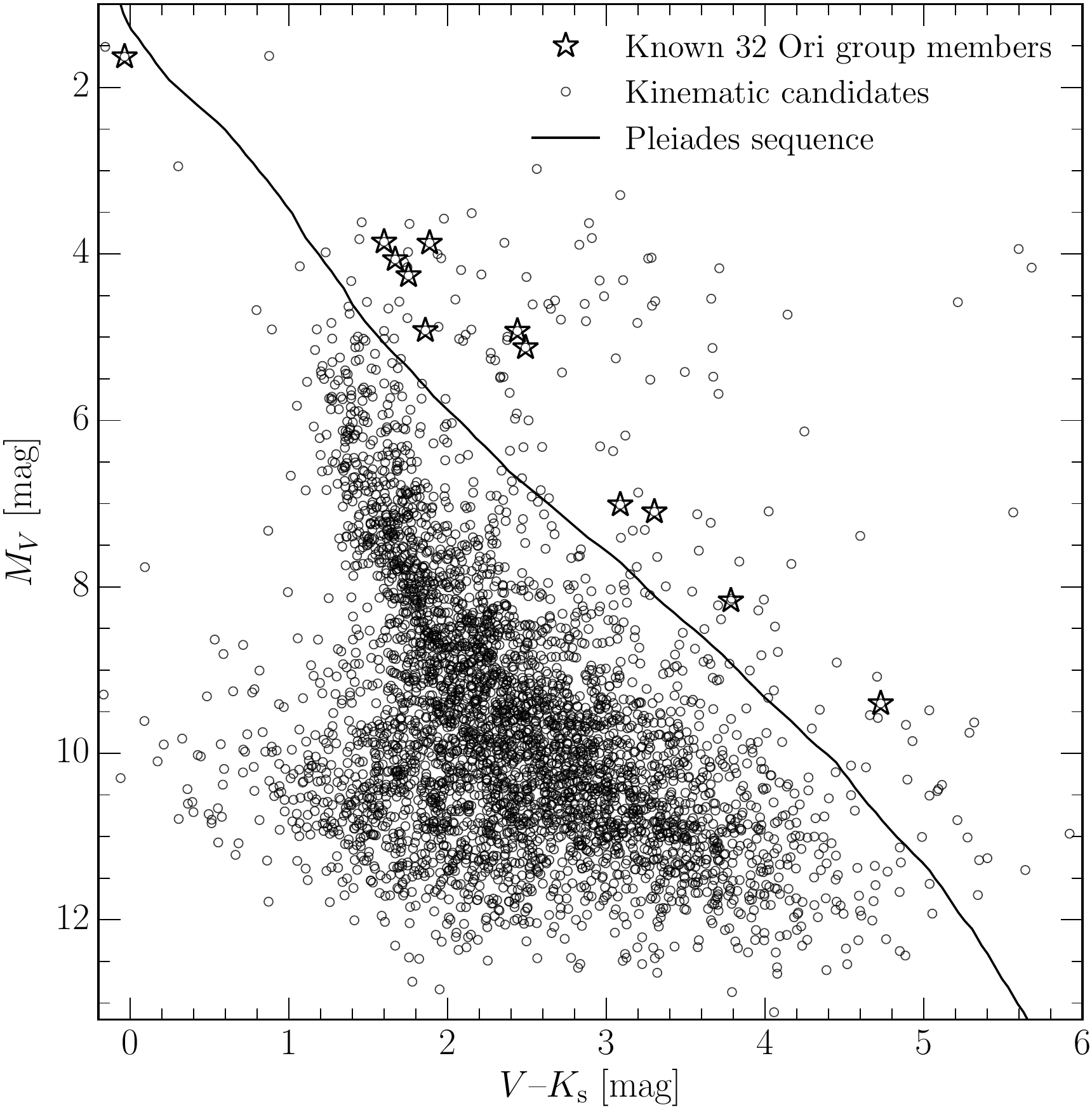}
\caption[]{UCAC4/APASS DR6 $M_{V}, V-K_{\rm{s}}$ colour-magnitude
  diagram. The circles denote potential kinematic members and the
  stars are known 32~Ori group members in UCAC4 which satisfy the same
  kinematic selection criteria (see
  Section~\ref{initial_input_catalogue_and_search_criteria}). The
  solid line is the empirical fit to the Pleiades single-star sequence
  by \cite{Stauffer07} which has been corrected for the effects of
  both distance and interstellar extinction.}
\label{fig:cand_cmd}
\end{figure}

The kinematic distances determined in
Section~\ref{initial_input_catalogue_and_search_criteria} must
also be consistent with
the colour-magnitude diagram (CMD) positions of young, low-mass stars
above the main-sequence. Our primary photometric selection was
performed using the UCAC4/APASS DR6 $M_{V}, V-K_{\rm{s}}$
CMD. Fig.~\ref{fig:cand_cmd} shows the 3730 (out of 5349) potential
kinematic members with APASS $V$- and 2MASS $K_{\rm{s}}$-band photometry, in
addition to known members of the 32~Ori group which satisfy the same
kinematic selection criteria. Also plotted is the empirical fit to the
$\sim$\,130\,Myr-old Pleiades single-star sequence by
\cite{Stauffer07}, corrected for the effects of both distance and
extinction ($d=136$\,pc and $A_{V}=0.12$~mag from \citealp{Melis14} and
\citealp{Stauffer98a}, respectively).  As expected, the vast majority
of sources lie below the Pleiades sequence, indicating that they are field
stars with coincidental proper motions and not photometric members of
the 32~Ori group. However, there are a significant number of candidates
above the sequence whose CMD positions are consistent with known
members, i.e. their kinematic distances agree with photometric distances
for a putative $\sim$\,25\,Myr population. These stars are 
potential new 32~Ori group members and we selected $\sim$100 of them for
spectroscopic follow-up, with a particular emphasis on late-type stars
($V-K_{\rm{s}} \geq 3$\,mag; spectral type $\gtrsim$\,K4).

Of the 15 known stellar members listed in
Table~\ref{tab:bonafide_members}, three do not have counterparts in
Fig.\,\ref{fig:cand_cmd}. While 32~Ori itself satisfies both criteria
in Section~\ref{initial_input_catalogue_and_search_criteria}, it is
saturated in APASS $V$. In contrast, the best-fit distances for both
V1874~Ori and 2MASS~J05194398+0535021 fall outside of the allowed
range (131 and 116\,pc, respectively). The latter member also has
$\Delta_{\rm PM}=12.5$~\masyr, but with large ($>$10~\masyr) errors on
its UCAC4 proper motion\footnote{If we instead adopt proper motions
  from URAT1 for V1874~Ori and PPMXL for 2MASS~J05194398+0535021 (as
  found in Table~\ref{tab:bonafide_members} and which has a smaller
  uncertainty than UCAC4), we calculate distances of 88 and 110\,pc,
  respectively. Both stars would then satisfy the selection
  criteria.}.

Given the incompleteness of APASS $BVgri$ photometry in UCAC4, we also
searched for additional low-mass members in the $M_{R_{\rm{UCAC}}},
R_{\rm{UCAC}}-K_{\rm{s}}$ CMD using the stars selected in
Fig.\,\ref{fig:cand_cmd} as a reference, and identified $\sim$30
additional candidates with consistent CMD positions not present in the
$M_{V}, V-K_{\rm{s}}$ diagram. Furthermore, we also looked for
potential members in the First U.S. Naval Observatory Robotic
Astrometric Telescope Catalog (URAT1; \citealp{Zacharias15}), which
extends approximately 2\,mag fainter than UCAC4 in an almost identical
non-standard pseudo-$R$ bandpass. We used the same method as above,
but allowed a $\Delta_{\mathrm{PM}}<15$~\masyr\ limit due to the
two-epoch astrometry having larger proper motion uncertainties. From
a similar kinematic and CMD analysis we identified another $\sim$20
candidates only present in URAT1 down to $R\approx16\,\rm{mag}$,
giving a total of $\sim$150 potential members from the combined
searches.

The fainter $R_{\rm{UCAC}}$ and URAT1 samples are crucial for better
estimating the age of the group. The limiting magnitude of APASS is $V
\simeq 16~\rm{mag}$, which at a distance of 90\,pc, corresponds to the
expected location of the lithium depletion boundary (LDB) in a
$\sim$25 Myr population. Only by confirming fainter Li-rich members
can we can identify the precise position of the 32 Ori group LDB and
calculate a semi-fundamental \citep{Soderblom14} age for the group
(see Section \ref{ldb}).

\section{Spectroscopic observations and data analysis}
\label{spectroscopic_observations}

To unambiguously differentiate between genuine 32 Ori group members
and field interlopers or other contaminants, additional spectroscopic
diagnostics are required. These typically include measuring radial
velocities to ascertain whether candidates share similar
three-dimensional space motions, as well as identifying spectroscopic
features associated with stellar youth (e.g. Li\,\textsc{i}
$6708\,\rm{\AA}$ absorption and H$\alpha$ emission).

We observed 124 candidate members (plus 11 of the known members listed in
\citealp{Bell15}) in four runs between 2015 September and 2016
February using the Wide Field Spectrograph (WiFeS; \citealp{Dopita07})
on the ANU 2.3-m telescope at Siding Spring Observatory (SSO). Poorer than
expected conditions meant we were unable to observe all of the
faintest ($V>15\,\rm{mag}$) candidates and so we prioritised
objects with small $\Delta_{\rm PM}$ values, CMD or sky positions
similar to known members and candidates coincident with \emph{ROSAT}
X-ray sources \citep{Voges99}. We later took observations 
in 2016 October and 2017 January to revisit several possible spectroscopic binaries
identified in earlier runs.

WiFeS is an image-slicing integral field spectrograph with a nominal
25$\times$38\,arcsec field-of-view and 0.5\,arcsec sampling along
twenty-five 38$\times$1\,arcsec slitlets. Observations were made in
half-field (12$\times$38\,arcsec) `stellar' mode with 2$\times$
spatial binning (1\,arcsec spaxels; well-matched to typical
1.5--2.5\,arcsec SSO seeing). The $R$7000 grating and $RT$480 dichroic
gave a resolution of $\lambda/\Delta\lambda\approx7000$ and wavelength
coverage from 5300--7000\,\AA. The field-of-view was aligned to the
parallactic angle prior to each exposure, with exposure times up to
3$\times$1200\,sec. In addition to science targets, we also obtained
spectra for 4--10 FGKM radial velocity standards each night from the
list of \citet{Nidever02} and white dwarf flux calibrators following
\citet{Bessell99}.

\begin{figure*}
\centering
\includegraphics[width=\linewidth]{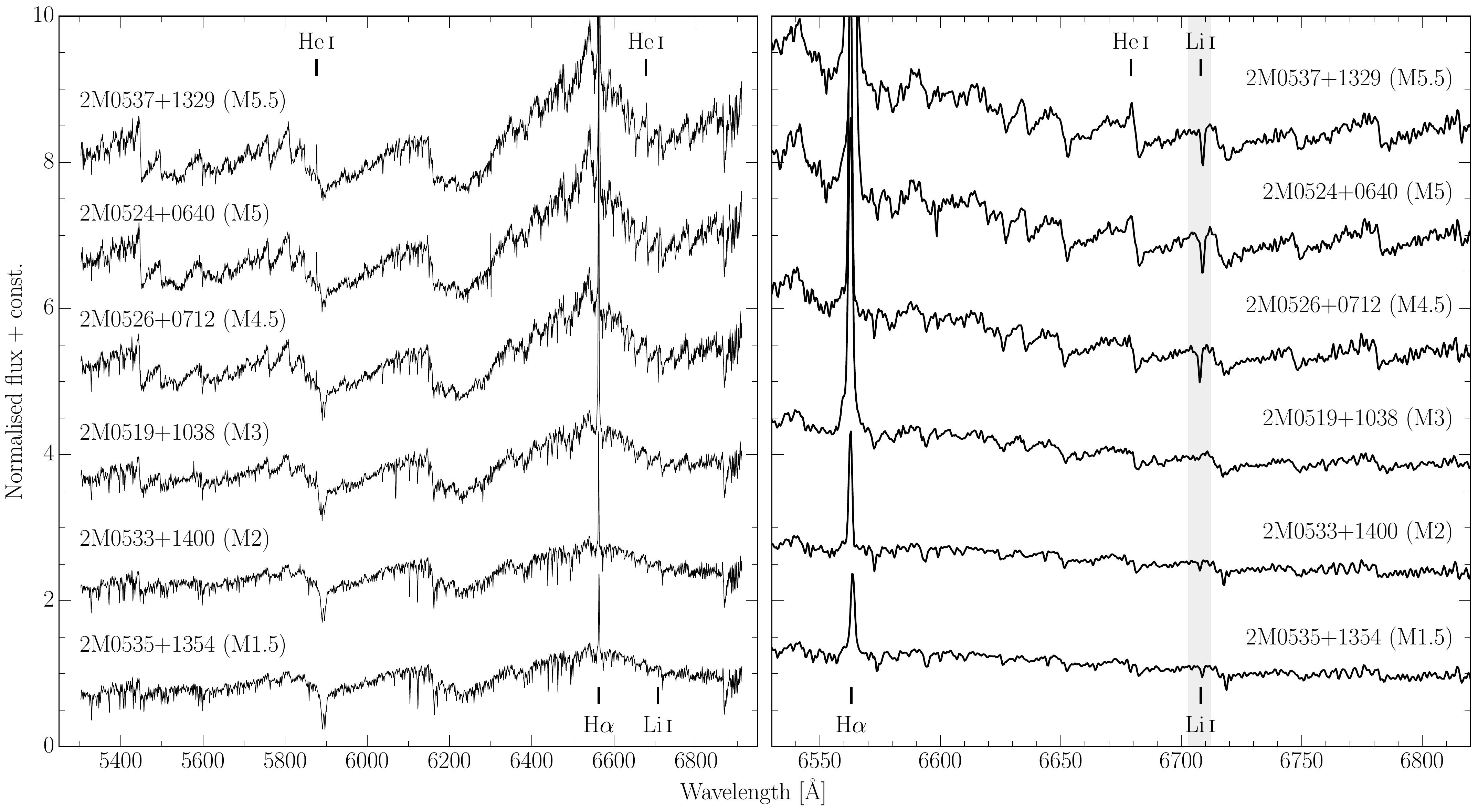}
\caption[]{WiFeS/$R7000$ spectra for a selection of new M dwarf 32~Ori
  group members identified in this study. Fluxes are normalised around
  6100\,\AA\ and key youth indicators are labelled. Li\,\textsc{i}
  $6708\,\rm{\AA}$ absorption decreases in strength with decreasing
  effective temperature through the early-M spectral types, before
  sharply returning to undepleted levels below the $\sim$\,25\,Myr
  lithium depletion boundary (LDB) at around M4.5 (see also
  Fig.~\ref{fig:ew_li}).}
\label{fig:spectra}
\end{figure*}

After basic image processing, we used \textsc{iraf}, \textsc{figaro}
and \textsc{python} routines to rectify, extract, wavelength-calibrate
and combine the 3--6 image slices (depending on seeing) that contained
the majority of the stellar flux. The slices were treated like
long-slit spectra and individually extracted and wavelength calibrated
against NeAr arc frames taken following each exposure. The rms of the
final wavelength solution in all cases was better than
0.02\,\AA. Typical signal-to-noise ratios were 50--100 around
H$\alpha$ and Li\,\textsc{i}, decreasing to $\sim$\,20 for the
faintest candidates. Table~\ref{tab:candidate_spectra} (the full
version of which is available as Supporting Information with the
online version of the paper) details our
spectroscopic measurements for the observed candidates and 
known members listed in \cite{Bell15}. Typical WiFeS/$R7000$ spectra
of several new M dwarf 32~Ori group members confirmed in this work are
shown in Fig.~\ref{fig:spectra}.

\subsection{Spectral types}
\label{spectral_types}

Spectral types for each candidate were first estimated by visual
comparison against the \citet{Pickles98} stellar spectral flux library
and the Sloan Digital Sky Survey (SDSS) average M dwarf templates of
\citet{Bochanski07}, with the WiFeS spectra Gaussian-smoothed to the
approximate resolution of the templates prior to comparison. This
analysis showed a substantial fraction of the candidates were clearly
reddened, motivating us to fit their spectra to the 
templates with interstellar reddening as a free parameter. For this
we adopted the \cite*{Cardelli89} reddening law and a total-to-selective
extinction ratio of $R_{V}=3.1$ and calculated
the best-fit reddening [in the range $0<E(B-V)<3$\,mag, in 0.05\,mag
  increments], after resampling the template onto the same wavelength
array as the star. The unreduced $\chi^{2}$ statistic was used as the
goodness of fit.

The resulting spectral types for all candidates are listed in
Table~\ref{tab:candidate_spectra}, with the estimated reddenings
provided in the electronic version of the table (see also
Table~\ref{tab:red} for the early type members). Given the modest
wavelength range of the WiFeS data, spectral type coverage of the
Pickles library and degeneracies with reddening, we estimate these
types are accurate to $\pm$1--2 subtypes for A-, F-, G- and K-type
candidates. The homogeneity and high quality of the SDSS M dwarf
templates permit spectral types for these candidates to approximately half a
subtype.

\subsection{Radial velocities}

Radial velocities for each candidate were measured by
cross-correlation over 5500--6500\,\AA\ against standards of similar
spectral type observed that run using a \textsc{python} implementation
of the \textsc{fxcor} algorithm \citep{Tonry79}. Following the
procedure described in \citet{Murphy15}, each spectrum was first
resampled onto a log-linear wavelength scale and normalised by
subtracting a boxcar-smoothed copy and dividing by the standard
deviation. The waveforms were then cross-correlated and a Gaussian
fitted to the cross-correlation function (CCF) peak, before
transforming to a heliocentric frame. The velocities reported in
Table~\ref{tab:candidate_spectra} are the mean and standard deviation
against standards observed that run. Repeat and standard star
observations demonstrate that the instrument and this technique
provide an external velocity precision of $\lesssim$\,1\,\kms\ for
bright stars. For the faintest candidates in
Table~\ref{tab:candidate_spectra} this falls to 2--3\,\kms\ per
exposure.

\subsubsection{Spectroscopic binaries and fast rotators}
\label{sec:binaries}

\begin{figure}
\centering
\includegraphics[width=\columnwidth]{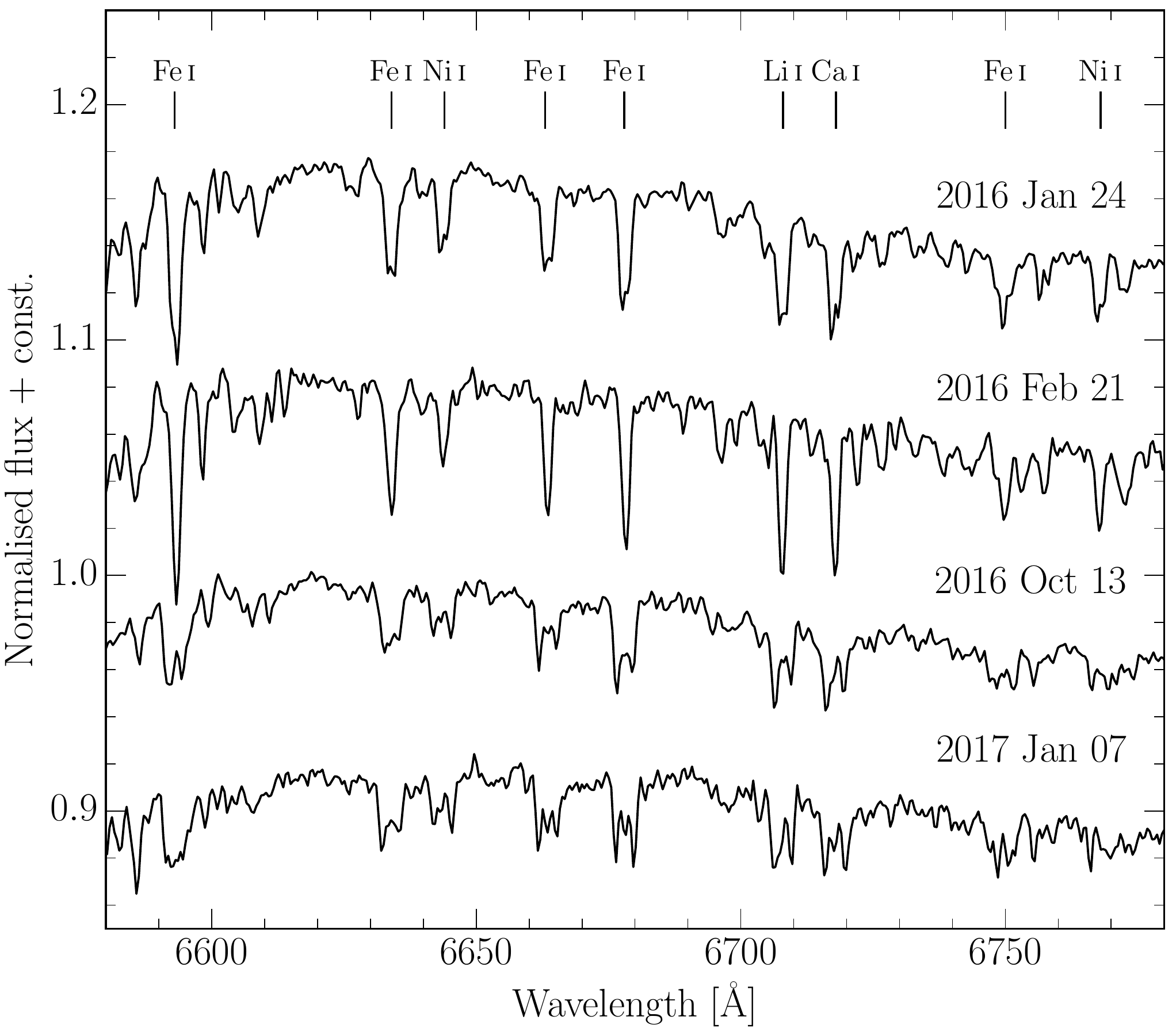}
\caption[]{Li\,\textsc{i}-region spectra of the new SB3 system and possible member 
  HD\,37825 (spectral type F5), showing clear evolution of its
  neutral metal lines with time. The 2016 February spectrum is
  single-lined ($\textrm{RV}\approx28$~\kms), while in 2016 October and 2017 January  it is resolved into three components (only two of which are visible in the CCF). The
  two main components are marginally resolved in 2016 January.}
\label{fig:sb3}
\end{figure}

Six candidates (HD\,37825, BD+08\,900B, 2MASS J05442447+0502114,
2MASS J05363692+1300369, 2MASS J05320596$-$0301159, 2MASS J05561307+0803034) and the
previously known 32~Ori group members HD\,35499 and HD\,35695 showed
average CCF widths significantly larger than other stars
\citep[$\textrm{FWHM}\gtrsim 2.5$\,px; see][]{Murphy15}, with
BD+08\,900A and the known member 2MASS\,J05234246+0651581 borderline cases.  This broadening can be attributed to
either fast rotation or unresolved spectroscopic binarity at the
modest ($c\Delta\lambda/\lambda\approx45$~\kms) velocity resolution of
WiFeS \citep[see discussion in][]{Murphy15}. Three of these stars --
HD\,37825, 2M0536+1300 and 2M0556+0803 -- showed double line cores and
broad, asymmetric (though unresolved) CCFs indicative of double-lined spectroscopic
binary systems (SB2). Notably, the 2016 October and 2017 January CCFs of HD\,37825 were
clearly resolved into two near-equal amplitude peaks (the latter spectrum showing the system to
be an SB3 with a weak tertiary component; see Fig.\,\ref{fig:sb3}), whilst the 2016
February 21 spectrum had much narrower, single profiles, ostensibly close
to the systemic velocity ($\textrm{RV}\approx28$\,\kms).

In the absence of resolved double lines, the origin of the broadening
seen in the other stars is less clear. Given their smoother
spectral features and symmetric CCFs, it is likely HD~35499, HD~35695,
2M0532$-$0301 and 2M0544+0502 are fast rotators ($v\sin i \gtrsim
45$~\kms), while BD+08\,900B may be a binary (with BD+08\,900A more
likely a fast rotator). Higher resolution observations are necessary
to confirm these predictions.

Of the 21 candidates or previously known members observed more than
once with WiFeS (see Tables~\ref{tab:rvs} and \ref{tab:sb2}), eight non-SB2s have maximum radial
velocity differences larger than 5\,\kms\ and therefore may be single-lined (SB1) 
spectroscopic binaries. One of these stars (2MASS\,J05525572$-$0044266) is
a known $\sim$0.86\,d period eclipsing binary \citep[][see
  Section~\ref{2massj0552-0044}]{Drake14}, while three more have broad CCFs
(and thus less precise velocities), yielding differences marginally
greater than 5\,\kms\ (HD\,35499, 2M0532$-$0301 and 2M0544+0502). The
Li-rich candidate 2MASS\,J05350092+1125423 (RV$_\textrm{expt}=
17.4$\,\kms) was observed twice in 2016 February (19.6 and 10.9\,\kms,
$\Delta t=3$\,d), with further observations in 2016 October
(21.0\,\kms) and 2017 January (19.5\,\kms). Given the agreement of the three other epochs, with
no other indications of binarity it is likely that the second February velocity was erroneous.
The known K6.5 member 2M0523+0651 (RV$_\textrm{expt}= 18.5$\,\kms) has two
WiFeS velocities separated by 442 days which agree and a third which is discrepant by
$\sim$9\,\kms, well outside the expected errors. \citet{Alcala00} reported a $18.4\pm1.0$\,\kms\
velocity from a high resolution ($R\approx30\,000$) spectrum which agrees with the expected
velocity. Interestingly, that study noted 2M0523+0651 may be a spectroscopic binary based on
an asymmetric CCF and we too find a somewhat broadened CCF in our observations. Finally, the Li-rich, early M-type candidates 2MASS J05351761+1354180 and 2MASS J05330574+1400365 both have velocity differences of $\sim$7\,\kms\ over three epochs. They do not show broadened CCFs and their mean velocities are close to those expected of bona fide members.

\subsection{H$\alpha$ and Li equivalent widths}
\label{ha_li_equivalent_widths}

Young stars are predominantly active and one of the most common
diagnostics of activity is the Balmer H$\alpha$ line at 6563~\AA. For
late-type stars (spectral type $\gtrsim$ M0) H$\alpha$ is typically
seen in emission to ages of several Gyr \citep*[e.g.][]{Gizis02},
whereas in early and solar-type stars it can be seen in various levels
of emission or absorption depending on activity. We measured the
H$\alpha$ equivalent width (EW) of our candidates by fitting a single
Gaussian or Voigt profile with linear continuum to the
emission/absorption feature. For stars where the line could not be
well fitted analytically, EW[H$\alpha$] was calculated by direct
integration of the line profile. Due to uncertainties in the placement
of the continuum and integration limits, we estimate uncertainties of
up to 1\,\AA\ in the measured EW[H$\alpha$] values. Several of the
M-type stars in our sample also showed strong He\,{\footnotesize{I}}
emission at 5876 and $6678\,\rm{\AA}$. This is chromospheric in nature
and is observed in active M-dwarfs with ages of up to several Gyr
\citep{Gizis02}.

Another key indicator of stellar youth is the resonant Li\,\textsc{i}
$6708\,\rm{\AA}$ absorption feature. As a coeval group of young stars
contracts toward the zero-age main-sequence their core temperatures
increase until at $\sim$\,$3 \times 10^{6}$\,K Li fuses. Such
temperatures can be reached in either fully convective mid- to late-M
dwarfs or at the base of the convective zone in late-K/early-M
dwarfs. For stars between these luminosities, rapid Li depletion
ensues and the Li\,\textsc{i} feature is no longer visible. At an age
of $\sim$\,25\,Myr we expect Li to be fully depleted in stars with
spectral types between $\sim$\,M1.5 and M4.5, however it should still
be present in both earlier and later spectral types (see
e.g. \citealp{Mentuch08,Jeffries13}). As in the case of H$\alpha$, we
determined EW[Li] by Gaussian profile fitting if the line was present
and estimated an upper limit from the local pseudo-continuum if it was
not. No attempt was made to deblend Li\,\textsc{i} from the nearby
Fe\,\textsc{i} feature at $6707.4\,\rm{\AA}$, which is unresolved at
our resolution and is typically much weaker (EW[Fe]\,$\lesssim
30\,\rm{m\AA}$; \citealp{Soderblom93}) compared to the expected EW[Li]
of 32 Ori group members.  WiFeS EW[H$\alpha$] and EW[Li] values for
all observed candidate and known members of the 32 Ori group are listed in 
Table~\ref{tab:candidate_spectra}.

\begin{landscape}
\begin{table}
\scriptsize
\caption[]{Spectroscopic parameters and membership assignments for the
  candidate members and 11 known members of the 32~Ori group
  observed with WiFeS. Four additional early-type stars and a brown dwarf
  confirmed as members but not observed in this work are also included. HD\,36002 and SCR 0522$-$0606
  were not in the original candidate list but were added and confirmed during observations. The full table, which also includes
  possible and non-members as well as derived $E(B-V)$ values and object-specific comments (omitted in the print
  version due to space restrictions), is available 
  as Supporting Information with the online version of
  the paper.}
\begin{tabular}{r l c c c c c c c c c c c c c c c c c}
\hline
THOR & 2MASS J                             &   Other name                         &   $V$     &   Ref.     &   $K_{\rm{s}}$   &   $\mu_{\alpha}\cos(\delta)$  &   $\mu_{\delta}$          &   Ref.   &   $d_{\rm{kin}}$   &   $d_{\rm{trig}}$   &   Ref.   &   RV                                        &   $\Delta_{\mathrm{RV}}$   &   SpT   &   EW[H$\alpha$]       &   EW[Li]                          &   \multicolumn{2}{c}{Membership}\\ \cline{18-19}
\#        & designation                          &                                               &   (mag)  &               &   (mag)              &           (mas\,yr$^{-1}$)          &    (mas\,yr$^{-1}$)       &             &            (pc)          &    (pc)                   &            &($\mathrm{km\,s^{-1}}$)          &   ($\mathrm{km\,s^{-1}}$)    &            &   ($\mathrm{\AA}$)   &   ($\mathrm{m\AA}$)      &   RV,H$\alpha$,Li   &   Final\\
\hline
1 & 05304706+0556536   &   32 Ori                                                    &  4.20     &     1         &   4.61                  &   $12.9\pm1.0$                     &   $-28.1\pm1.0$           &   2      &   102                  &   $93^{+6}_{-5}$         &   3              &   \dots                               &\dots                                       & B5+B7  &   \dots                        &   \dots                             &   \dots                    &   Y\\
2 & 05263883+0652071   &   HR 1807                                                &   6.41    &     4         &  6.41                    &   $9.2\pm1.0$                       &   $-33.2\pm1.0$          &   2       &   92                   &   $92\pm4$              &   3             &   \dots                               & \dots                                       & B9.5      &   \dots                       &   \dots                             &   \dots                    &   Y\\
3A & 05265999+0710131   &   HD 35714                                            &   7.03    &     5         &   6.75                  &   $8.4\pm1.0$                        &   $-34.2\pm1.0$         &   2       &   91                   &   $94^{+3}_{-4}$          &   6             &   \dots                               & \dots                                       & A3        &   \dots                        &   \dots                             &   \dots                    &   Y\\
3B & 05343842+0607367   &   HD 36823                                            &   7.73    &     5         &   7.14                  &   $9.0\pm0.6$                        &   $-31.1\pm0.5$          &   2       &   95                   &   $102\pm5$            &   6             &   \dots                               &   \dots                                     & A7.5     &   \dots                        &   \dots                             &   \dots                    &   Y\\
4A & 05284209+0113369 & HD 36002                                                &   7.46     &    5         &   6.71                  &   $6.9\pm1.0$                        &   $-25.6\pm1.0$          &   2       &   105                 &   $109^{+6}_{-5}$       &  6              & SB2                                 &   \dots                                     &  A7       &   14                         &   $\sim$70                             &   ?,?,?                    &   Y\\
4B & 05284050+0113333   &                                                               &  14.74    &    7         &   10.19                 &           $8.8\pm5.8$                &    $-26.6\pm5.8$         &    8      &    99                  &   \dots                     &   \dots      &           $18.4\pm1.0$         &  $-$1.5                                    &   M3.5  &   $-$4.8                       &   $<$50                           &    Y,?,?                      &    Y\\
5 & 05251457+0411482   &   HD 35499                                              &   8.66     &    5         &      7.50                &      $6.1\pm0.6$                     &   $-27.5\pm0.8$          &    2      &   106                 &   $105^{+8}_{-6}$      &   9             &       $20.6\pm2.6^{\star}$    &   $+$1.4                                  &   F4       &   8.0                             &   160                               &   Y,?,Y                       &   Y\\
6 & 05311570+0539461   &   HD 36338                                              &   8.52     &    5         &      7.40                &         $9.3\pm0.6$                  &   $-31.1\pm0.9$           &    2     &    95                  &   $96\pm5$            &   9             &       $21.3\pm0.2$              &   $+$2.4                                 &   F4.5     &   7.0                             &   110                               &   Y,?,Y                       &   Y\\
7 & 05265202+0628227   &   HD 35695                                              &   9.27     &    5         &      7.71                &         $9.2\pm0.8$                  &   $-36.1\pm1.2$           &    2     &    84                  & \dots                      &   \dots       &          $20.0\pm0.3^{\star}$ &   $+$1.4                                 &   F9        &   5.0                             &   280                               &   Y,?,Y                       &   Y\\
8A& 05371844+1334525   &   HD 245567                                           &   9.54     &    5         &      7.59                &         $7.2\pm0.9$                  &   $-33.5\pm0.8$           &    2     &   104                 & \dots                      &   \dots       &          $16.0\pm1.3$          &   $-$0.8                                  &   G5       &   $-$0.2$^{\S}$                       &   265                               &   Y,?,Y                       &   Y\\
8B& 05372061+1335310   &                                                                &  14.96    &    7         &   10.07                 &            $6.9\pm6.2$                &    $-30.8\pm5.7$          &    2     &    113               & \dots                      &   \dots       &            $15.7\pm0.8$         &   $-$1.1                                  &   M3     &   $-$4.8                       &   $<$30                           &   Y,?,?                       &   Y\\
9 & 05343491+1007062   &   HD 245059                                            &   9.96     &    5        &      7.41                 &         $10.1\pm1.3$                 &   $-35.0\pm1.2$           &     2     &   92                 & \dots                      &   \dots       &          $19.4\pm1.1$           &   $+$1.6                                 &   G7      &   $-$0.2                       &   285                               &   Y,?,Y                       &   Y\\
10 & 05203182+0616115   &   TYC 112-1486-1                                 &   11.87    &    7       &   8.57                    &          $9.5\pm1.8$                  &   $-32.8\pm2.1$           &    2     &   92                  &   $99^{+14}_{-11}$     &   9             &         $21.1\pm0.3$            &   $+$2.5                                 &   K3      &   $-$1.2                       &   500                              &   Y,?,Y                        &   Y\\
11 & 05200029+0613036   &   TYC 112-917-1                                   &   11.66    &    7       &   8.57                    &           $9.4\pm1.9$                  &   $-34.7\pm2.1$          &    2     &    86                 & \dots                      &   \dots       &         $21.0\pm0.3$           &   $+$2.4                                  &   K4      &   $-$0.3                       &   470                             &   Y,?,Y                        &   Y\\
12 & 05234246+0651581   &  1RXS J052342.7+065156                      &   12.81   &    7        &   9.03                    &           $7.8\pm2.3$                  &   $-36.6\pm2.8$          &    2     &    85                 & \dots                      &   \dots       &         $18.3\pm4.2^{\star}$  &   $-$0.2                                  &   K6.5   &   $-$5.0                       &   580                              &   Y,?,Y                        &   Y\\
13 & 05291899+1209295   &   V1874 Ori                                            &   13.22   &    7        &   9.19                    &           $2.9\pm2.1$                  &   $-26.7\pm3.0$           &    2     &    131              & \dots                      &   \dots        &         $22.4\pm0.3$           &   $+$5.3                                   &   K6.5   &   $-$1.7                       &   380                              &   Y,?,Y                       &   Y\\
14A & 05351761+1354180   &                                                             &   13.53   &    7        &   9.15                   &          $7.0\pm4.4$                   &    $-30.2\pm9.4$          &    2      &    116              & \dots                      &   \dots       &             $18.1\pm2.8^{\star}$       &   $+$1.4                                   &   M1.5   &   $-$1.7                       &   180                             &   Y,?,Y                        &   Y\\
14B & 05351625+1353594   &      1RXS J053516.6+135404                &   14.95   &    10       &   10.39                &           $7.6\pm4.3$                   &    $-31.8\pm4.6$          &    2     &    110              & \dots                      &   \dots        &            $19.2\pm1.0$       &   $+$2.5                                    &   M3.5   &   $-$6.1                      &   $<$40                         &    Y,?,?                       &   Y\\
15 & 05330574+1400365   &      1RXS J053306.7+140011                  &   13.80   &    7        &   9.35                   &           $9.4\pm2.4$                   &    $-37.4\pm2.2$          &    2     &    94                & \dots                      &   \dots        &        $19.2\pm2.9^{\star}$  &   $+$2.5                                   &   M2       &   $-$2.0                      &   160                              &   Y,?,Y                        &   Y\\
16 & 05430354+0606340   &      1RXS J054304.3+060646                  &   14.98   &    7         &   10.29                &            $11.5\pm5.4$                &    $-25.2\pm5.9$          &    2     &    114              & \dots                      &   \dots        &             $21.0\pm0.9$      &   $+$2.1                                    &   M2.5   &   $-$2.9                      &   $<$50                         &   Y,?,?                         &   Y\\
17A & 05274313+1446121   &      1RXS J052743.4+144609                &   14.13   &    7        &   9.10                   &           $8.6\pm3.8$                   &    $-41.1\pm3.8$          &    2     &    87                & \dots                      &   \dots        &     $15.8\pm1.1^{\star}$    &    $-$0.6                                   &   M3       &   $-$10.3$^{\S}$                   &   $<$100                       &   Y,?,?                          &   Y\\
17B & 05274404+1445584   &                                                              &   16.26   &    10      &    10.97                &           $6.5\pm6.0$                   &    $-44.4\pm6.7$          &    2     &    82                & \dots                      &   \dots        &           $15.4\pm0.8$       &   $-$1.0                                    &   M5       &   $-$5.9                     &   580                              &   Y,?,Y                         &    Y\\
18 & 05224069--0606238   &   SCR 0522--0606                                &   14.27   &    7        &   9.13                   &          $17.0\pm3.2$                  &   $-21.1\pm3.3$           &    2     &    88                & \dots                      &   \dots        &           $25.6\pm1.9^{\star}$        &   $+$4.3                                   &   M3       &   $-$7.0                      &   $<$20                       &   Y,?,?                          &   Y\\
19 & 05253253+0625336   &     1RXS J052532.3+062534                   &   14.51   &    7        &   9.78                   &           $8.0\pm5.8$                   &   $-28.2\pm6.1$           &    2     &    107              & \dots                      &   \dots        &         $19.6\pm0.4^{\star}$ &   $+$1.0                                   &   M3       &   $-$6.6                     &   $<$20                         &   Y,?,?                         &   Y  \\
20 & 05192941+1038081   &       1RXS J051930.4+103812                 &   14.66   &    7        &   9.77                   &          $8.5\pm7.1$                     &   $-32.9\pm5.1$           &    2    &    101              & \dots                      &   \dots        &          $17.7\pm0.8^{\star}$ &   $+$0.3                                   &   M3      &   $-$7.1                     &   $<$80                         &    Y,?,?                         &   Y\\
21 & 05251517+0030232   &     1RXS J052515.0+003027                   &   14.85   &    7        &   9.93                   &           $10.3\pm5.8$                  &    $-24.7\pm6.0$          &    2     &    103             & \dots                      &   \dots        &           $19.9\pm0.7$         &   $-$0.1                                    &   M3      &   $-$9.7$^{\S}$                     &   $<$20                        &   Y,?,?                          &   Y \\
22 & 05492632+0405379   &     1RXS J054926.3+040541                   &   14.96   &    7        &   10.07                 &          $11.2\pm5.5$                  &    $-33.3\pm5.9$            &    2    &    84               & \dots                      &   \dots        &             $20.9\pm1.0$       &   $+$1.4                                   &   M3       &   $-$2.5                    &   $<$50                         &   Y,?,?                          &    Y  \\
23 & 05442447+0502114   &      1RXS J054424.7+050153                  &   15.08   &    7        &   10.24                 &          $15.8\pm5.4$                  &    $-28.4\pm5.8$           &    2      &    98               & \dots                     &   \dots        &    $24.4\pm2.6^{\star}$      &   $+$5.2                                    &   M3       &   $-$4.4                     &   $<$30                        &   Y,?,?                         &    Y\\
24 & 05194398+0535021   &                                                               &   15.13   &    7        &   10.00                 &          $-4.2\pm16.4$                &   $-29.3\pm13.3$          &    2      &    116            & \dots                      &   \dots        &         $15.1\pm0.8$           &   $-$3.6                                    &   M3       &   $-$15.0                    &   $<$20                      &   Y,?,?                          &   Y  \\
25 & 05132631+1057439   &                                                               &   15.25   &    7        &   10.22                 &          $7.3\pm5.9$                    &   $-30.2\pm5.9$            &    8      &    113            & \dots                      &   \dots        &         $17.0\pm0.7$           &   $-$0.2                                    &   M3       &   $-$5.2                      &   $<$50                       &   Y,?,?                         &   Y\\
26 & 05302546--0256255   &                                                              &   15.65   &    7        &   10.66                &          $12.1\pm2.7$                   &    $-25.7\pm3.2$           &    2      &    88              & \dots                      &   \dots        &           $25.1\pm1.2$         &   $+$4.3                                   &   M3       &   $-$10.0$^{\S}$                   &   $<$90                         &   Y,?,?                         &   Y\\
27 & 05274855+0645459   &   1RXS J052748.7+064544                     &   14.75   &    7        &   9.47                  &           $10.9\pm5.9$                  &    $-29.1\pm6.3$           &    2      &    102            & \dots                      &   \dots        &           $18.8\pm0.8$        &   $+$0.2                                    &   M4       &   $-$6.6                     &   $<$50                         &   Y,?,?                          &   Y  \\
28 & 05264886+0928055   &                                                               &   15.64   &    11       &   10.47               &           $6.7\pm5.3$                    &    $-32.7\pm6.1$           &    2      &    100             & \dots                      &   \dots       &           $22.4\pm0.8$        &   $+$4.6                                     &   M4      &   $-$3.9                      &   $<$70                        &   Y,?,?                          &   Y \\
29 & 05231438+0643531   &    1RXS J052315.0+064412                    &   15.86   &    10       &   10.76               &           $3.2\pm5.9$                    &    $-28.4\pm5.9$           &    8      &    113             & \dots                      &   \dots       &     $21.0\pm0.6^{\star}$    &   $+$2.4                                      &   M4      &   $-$3.3                      &   $<$50                       &    Y,?,?                         &   Y    \\
30 & 05313290+0556597   &    1RXS J053132.6+055639                    &   15.90   &    7         &   10.53               &           $-0.8\pm5.7$                  &    $-37.4\pm5.7$           &    8      &    85              & \dots                      &   \dots        &            $24.7\pm1.1$       &   $+$5.9                                     &   M4.5    &   $-$6.4                       &   $<$80                       &    Y,?,?                         &   Y   \\
31 & 05363692+1300369   &                                                               &   16.10   &    11       &   10.39               &           $4.6\pm4.4$                    &    $-36.3\pm4.6$            &    2      &    96             & \dots                      &   \dots        &    $18.0\pm6.6^{\star}$      &   $+$1.0                                     &   M4.5    &   $-$17                        & 350--500                    &   Y,?,Y                         &   Y  \\
32 & 05264073+0712255   &                                                               &   16.32   &    7         &   10.76               &           $1.9\pm4.1$                    &    $-32.8\pm4.4$           &    2        &    99             & \dots                      &   \dots       &           $18.8\pm0.5$        &   $+$0.4                                      &   M4.5    &   $-$4.6                       &   600                           &   Y,?,Y                          &   Y  \\
33 & 05315786--0303367   &                                                              &   13.85   &    7         &   8.54                &           $4.2\pm6.6$                     &    $-34.8\pm4.9$           &    2       &    71             & \dots                      &   \dots        &           $23.1\pm1.0$        &   $+$2.3                                      &   M5$^{\ddag}$       &   $-$9.0                      &   $<$150                     &    Y,?,?                          &   Y \\
34 & 05320596--0301159   &                                                              &   15.61   &    7         &   9.70                &           $9.2\pm4.9$                    &    $-27.4\pm4.9$           &    2        &    85            & \dots                      &   \dots        &            $23.6\pm2.7^{\star}$        &    $+$2.8                                     &   M5       &   $-$11.0                     &   650                          &    Y,?,Y                         &    Y \\
35 & 05174962+0958221   &                                                               &   16.63   &    7         &   10.84               &          $22.3\pm5.9$                   &   $-33.3\pm5.9$           &    8        &    89             & \dots                      &   \dots        &         $18.1\pm1.6$          &   $+$0.6                                      &   M5       &   $-$6.9                       &   530                          &    Y,?,Y                         &   Y  \\
36 & 05235565+1101027   &      1RXS J052355.2+110110                 &   16.65   &    11        &   10.78               &           $7.4\pm6.3$                    &    $-30.3\pm6.3$          &    8        &    110           & \dots                      &   \dots        &           $18.1\pm1.2$        &   $+$0.7                                      &   M5      &   $-$5.8                       &   680                          &   Y,?,Y                          &   Y   \\
37 & 05350092+1125423   &                                                               &   16.84   &    11        &   11.02              &           $7.1\pm6.6$                    &    $-43.0\pm6.7$           &    2       &    79             & \dots                      &   \dots        &     $17.8\pm4.0^{\star}$     &    $+$0.4                                     &   M5       &   $-$6.7$^{\S}$                       &   700                          &   Y,?,Y                         &   Y\\
38 & 05243009+0640349   &                                                               &   16.97   &    10        &   11.13               &           $8.8\pm6.0$                    &    $-36.8\pm6.0$         &    8        &    84             & \dots                      &   \dots        &           $16.1\pm1.0$        &   $-$2.4                                       &   M5      &   $-$14.0                     &   750                           &   Y,?,Y                         &   Y  \\
39 & 05270634+0650377   &                                                               &   17.60  &    11         &   11.64               &           $10.1\pm5.9$                  &    $-35.8\pm5.9$         &    8        &    85             & \dots                      &   \dots        &           $20.8\pm1.0$        &    $+$2.3                                      &   M5      &   $-$8.5                       &   650                           &   Y,?,Y                         &   Y  \\
40 & 05373000+1329344   &                                                               &   16.51  &    10         &    10.78              &            $3.2\pm6.4$                   &    $-32.1\pm6.4$          &    8       &    110           & \dots                      &   \dots        &            $18.5\pm1.2$       &   $+$1.6                                       &   M5.5   &   $-$13.2                     &   630                           &    Y,?,Y                        &    Y   \\
41 & \dots                           &   WISE J052857.68+090104.4                &   \dots  & \dots       &   14.97$^{\dagger}$ &           $-11.0\pm10.0$           &   $-39.0\pm12.0$         &   12      &   93              & \dots                      &   \dots        &   \dots                              &   \dots                                          & L1         & \dots                            &   \dots                        &   \dots                       &   Y  \\
\hline
\end{tabular}
\begin{flushleft}
  Column descriptions for the online version of the Table:
  Column 1 gives the 32~Ori (THOR) member number assigned in this study.
  Columns 2 and 3 list the 2MASS object identifier and the other commonly adopted name.
  Columns 4 and 5 provide the $V$-band magnitude and the source reference.
  Column 6 gives the 2MASS Point Source Catalog (PSC; \protect\citealp{Cutri03})
  $K_{\rm{s}}$-band magnitude, unless otherwise stated.
  Columns 7, 8 and 9 list the proper motion in the right ascension, declination
  used in the selection of candidate members and the source reference.
  Columns 10, 11 and 12 provide the best-fit kinematic distance as calculated in
  Section~\ref{initial_input_catalogue_and_search_criteria}, available literature
  trigonometric parallax distances and the source reference.
  Columns 13 and 14 give the measured radial velocity and radial velocity
  residual.
  Columns 15 and 16 list the spectroscopic spectral type and best-fit $E(B-V)$ reddening.
  Columns 17 and 18 provide the measured H$\alpha$ and Li equivalent widths.
  Columns 19 and 20 give the combined radial velocity, H$\alpha$ emission,
  Li absorption membership diagnostics and the final membership assignment.
  Column 21 lists additional object-specific comments.
  \\
References for $V$-band photometry, proper motions
and trigonometric parallax distances:
(1) \cite{Mermilliod06}; (2) \cite{Zacharias13}; (3) \cite{vanLeeuwen07};
(4) \cite{Hauck98}; (5) \emph{Tycho}-2 $V_{\rm{T}}$ photometry converted to Johnson
$V$ following \cite*{Mamajek06}; (6) Weighted mean of revised \emph{Hipparcos}
reduction \citep{vanLeeuwen07} and \emph{Gaia} DR1 \citep{Gaia16} parallaxes;
(7) \cite{Henden16}; (8) \cite{Zacharias15}; (9) \emph{Gaia} DR1;
(10) \cite{Zacharias05}; (11) \cite{Dolan02}; (12) \cite{Burgasser16}.
Note that all parallax measurements from \emph{Gaia} DR1 include the additional
$\pm0.3$\,mas systematic uncertainty as discussed in \cite{Gaia16}.
\\
$^{\star}$ Mean velocity and standard deviation from multiple epochs (see Table~\ref{tab:rvs} for individual measurements).\\
$^{\S}$ Spectrum also showed strong He\,\textsc{i} 5876 and/or 6678\,\AA\ emission. \\
$^{\ddag}$ 5\,arcsec visual binary; WiFeS spectrum is unresolved.\\
$^{\dagger}$ $K_{\rm{s}}$-band magnitude from the 2MASS Point Source Reject Table \citep[see also \citealp{Burgasser16}]{Cutri03}.
\end{flushleft}
\label{tab:candidate_spectra}
\end{table}
\end{landscape}

\begin{table}
\caption[]{Individual WiFeS radial velocities of candidates observed at more than one epoch.
Stars with a THOR number (Column 1) are confirmed
members of the 32 Ori group. The final column denotes whether the star is a confirmed or suspected spectroscopic binary. Three additional resolved SB systems are listed with their component velocities in Table~\ref{tab:sb2}.\label{tab:rvs}}
\begin{flushleft}
\begin{tabular}{r l c c l}
\hline
THOR & Name   &   Epoch   &  RV & SB?\\
\# & &  (UT) & (\kms)\\
\hline
5 & HD 35499 & 2015 Oct 25 & 	$23.2\pm0.8$ & N\\
& & 2016 Oct 13 &	$18.0 \pm	0.6$\\[2mm]
7 & HD 35695 & 2015 Oct 25 &	$20.3\pm0.5$ & N\\
& & 2016 Oct 13 & 	$19.7\pm	1.3$\\[2mm]
12 & 2M0523+0651 & 2015 Oct 23 &	$15.0\pm	2.2$ & SB2?\\
& & 2016 Oct 13 &	$24.2\pm	1.4$\\
& & 2017 Jan 07 & $15.6\pm	1.9$\\[2mm]
14A & 2M0535+1354 & 2016 Jan 23 & $21.5\pm	0.7$ & SB1?\\
& & 2017 Jan 07 & $18.1\pm	0.9$\\
& & 2017 Jan 07 & $14.7\pm	1.2$\\[2mm]
15 & 2M0533+1400 & 2015 Oct 19 &	$23.1\pm	0.6$ & SB1?\\
& & 2015 Oct 20 &	$18.5\pm	0.7$\\
& & 2017 Jan 07 & $16.1\pm	1.2$\\[2mm]
17A & 2M0527+1446 & 2015 Oct 20 & $14.4\pm	0.8$ & N\\
& & 2016 Jan 26 & $17.0\pm	1.0$\\
& & 2016 Jan 30 & $16.0\pm	1.0$\\[2mm]
18 & SCR 0522$-$0606 & 2016 Oct 14 & $23.7\pm	1.7$ & SB2\\
& & 2017 Jan 07 & $27.4\pm	1.7$\\[2mm]
19 & 2M0525+0625 & 2015 Oct 23 & $19.2\pm	0.7$ & N\\
& & 2015 Oct 25 & $20.0\pm	0.6$\\[2mm]
20 & 2M0519+1038 & 2015 Oct 19 &	$18.4\pm	1.1$ & N\\
& & 2015 Oct 19 & $16.9\pm	0.9$\\[2mm]
23 & 2M0544+0502 & 2016 Jan 23 &	$21.8\pm	1.8$ & N\\
& & 2016 Oct 14 &	$26.9\pm	1.6$\\[2mm]
29 & 2M0523+0643 & 2016 Jan 25 &	$20.4\pm	1.0$ & N\\
& & 2016 Jan 29 &	$21.5\pm	0.9$\\[2mm]
31 & 2M0536+1300 & 2016 Jan 25 &	$16.3\pm	1.5$ & SB2\\
& & 2016 Oct 13 & $28.4\pm	2.1$\\
& & 2016 Oct 14 &	$17.4\pm2.4$\\
& & 2017 Jan 07 &	$10.0\pm	3.0$\\[2mm]
34 & 2M0532$-$0301	&2016 Jan 29 & $26.2\pm	1.3$ & N\\
& & 2017 Jan 08 & 	$20.9\pm 2.6$\\[2mm]
37 & 2M0535+1125  &  2016 Feb 18 & $19.6\pm	1.9$ & N?\\
& & 2016 Feb 21&	$10.9\pm	1.4$\\
& & 2016 Oct 14 &	$21.0\pm	1.7$\\
& & 2017 Jan 08 & $	19.5\pm	0.9$\\[2mm]
\dots & BD+08 900A & 2015 Sep 28 & $19.3\pm	1.1$ & N\\
& & 2015 Sep 28 &	$20.6\pm	1.0$\\
& & 2016 Oct 13 & $18.9\pm0.7$\\
& & 2017 Jan 07 & $22.3\pm4.2^{\dagger}$\\
& & 2017 Jan 07 & $19.3\pm4.5^{\dagger}$\\[2mm]
\dots& BD+08 900B & 2015 Sep 28 & $20.8\pm	1.3$ & SB2?\\
& & 2015 Sep 28 &	$17.9\pm	0.9$\\
& & 2016 Oct 13 & $20.7\pm0.7$\\
& & 2017 Jan 07 & $21.3\pm 5.9^{\dagger}$\\
& & 2017 Jan 07 & $20.9\pm 6.9^{\dagger}$\\[2mm]
\dots& HD 243086 & 2016 Jan 24 &	$23.7\pm	1.7$ & N\\
& & 2016 Feb 21 &	$21.9\pm	1.0$\\[2mm]
\dots& 2M0556+0803	& 2016 Feb 21 &$28.3\pm	1.3$ & SB2\\
& & 2017 Jan 09 & $23.9\pm	2.7$\\
\hline
\end{tabular}\\
$^{\dagger}$ Cross-correlated against M-dwarf standards as no F-type standards were observed during run.
\end{flushleft}
\end{table}

\begin{table}
\caption[]{Individual radial velocities for the three systems which could be resolved into two components. HD 37825 is a clear SB3 system (see Fig.\,\ref{fig:sb3}) with a weaker third component not resolved in the 2017 Jan CCF.  2M0552$-$0044 shows a single-peaked CCF but has a variable, double-peaked H$\alpha$ emission line from which a velocity offset was estimated and added to the primary velocity.}\label{tab:sb2}
\begin{flushleft}
\begin{tabular}{l c c c}
\hline
Name   &   Epoch   &  RV$_{1}$ & RV$_{2}$\\
 &  (UT) & (\kms) & (\kms)\\
\hline
HD 36002 (SB2) & 2017 Jan 11 & $+69.8\pm3.3^{\dagger}$ & $-32.8\pm3.0^{\dagger}$\\
& 2017 Jan 11 & $+58.2\pm2.0^{\dagger}$ & $-27.0\pm3.0^{\dagger}$\\[2mm]
HD 37825 (SB3) & 2016 Jan 24 &	$25.3^{\ddag}$ & \dots \\
&  2016 Feb 21 &	$+28.1\pm0.8$ & \dots \\
& 2016 Oct 13 & $-42.5\pm2.5$ & $+95.3\pm1.9$\\
& 2017 Jan 07 & $-42.1\pm3.0^{\dagger}$ & $+101.9\pm1.8^{\dagger}$\\[2mm]
2M0552$-$0044 & 2015 Oct 23 &	$+65.5\pm	0.7$ & $-99.5\pm5.4$\\
(SB1+H$\alpha$) & 2016 Oct 14 &	$-39.3\pm	1.5$ & $+160.7\pm5.6$\\
& 2017 Jan 07 & $+18.4\pm	1.6$ & \dots \\
& 2017 Jan 07 & $-34.0\pm	1.0$ & $+136.0\pm5.5$\\
& 2017 Jan 08 & $-26.4\pm	0.8$ & $+163.6\pm5.4$\\
& 2017 Jan 08 & $-16.6\pm	1.0$ & $+133.4\pm5.5$\\
& 2017 Jan 09 & $-22.9\pm	0.7$ & $+157.1\pm5.4$\\
\hline
\end{tabular}\\
$^{\dagger}$ Cross-correlated against M-dwarf standards as no F-type standards were observed during run.\\
$^{\ddag}$ Single F-type standard observed during run.
\end{flushleft}
\end{table}

\section{The low-mass population of the 32 Ori group}
\label{32_ori_low_mass_population}

\subsection{Newly identified members}
\label{newly_identified_members}

Whilst any one of the diagnostics discussed in
Section~\ref{spectroscopic_observations} could be used to assign
membership of a candidate to the 32 Ori group, this increases the risk
of erroneously including a member, which when assessed using an
alternative diagnostic, is an obvious non-member.  The most
discriminating diagnostic to differentiate between young and older
stars is the presence of Li, however this is only valid over a small
range of spectral types (see e.g. \citealp{Jeffries06b}). In addition,
stellar activity as traced by H$\alpha$ emission is not solely
restricted to the youngest stars, with older K- and M-type field stars
also exhibiting elevated levels of activity (see
e.g. \citealp{West11}) and so whilst the presence of H$\alpha$ is
necessary, it alone does not provide a reliable membership
diagnostic. Finally, although one would expect stars with the same
space motion as the 32 Ori group to be likely members, within a given
sample there will always be a small fraction of older field stars
which are co-moving purely by coincidence. In contrast, radial
velocities of genuine short-period binary members based on
single-epoch spectra will, in the majority of cases, suggest that
these are not co-moving.

\begin{figure}
\centering
\includegraphics[width=\columnwidth]{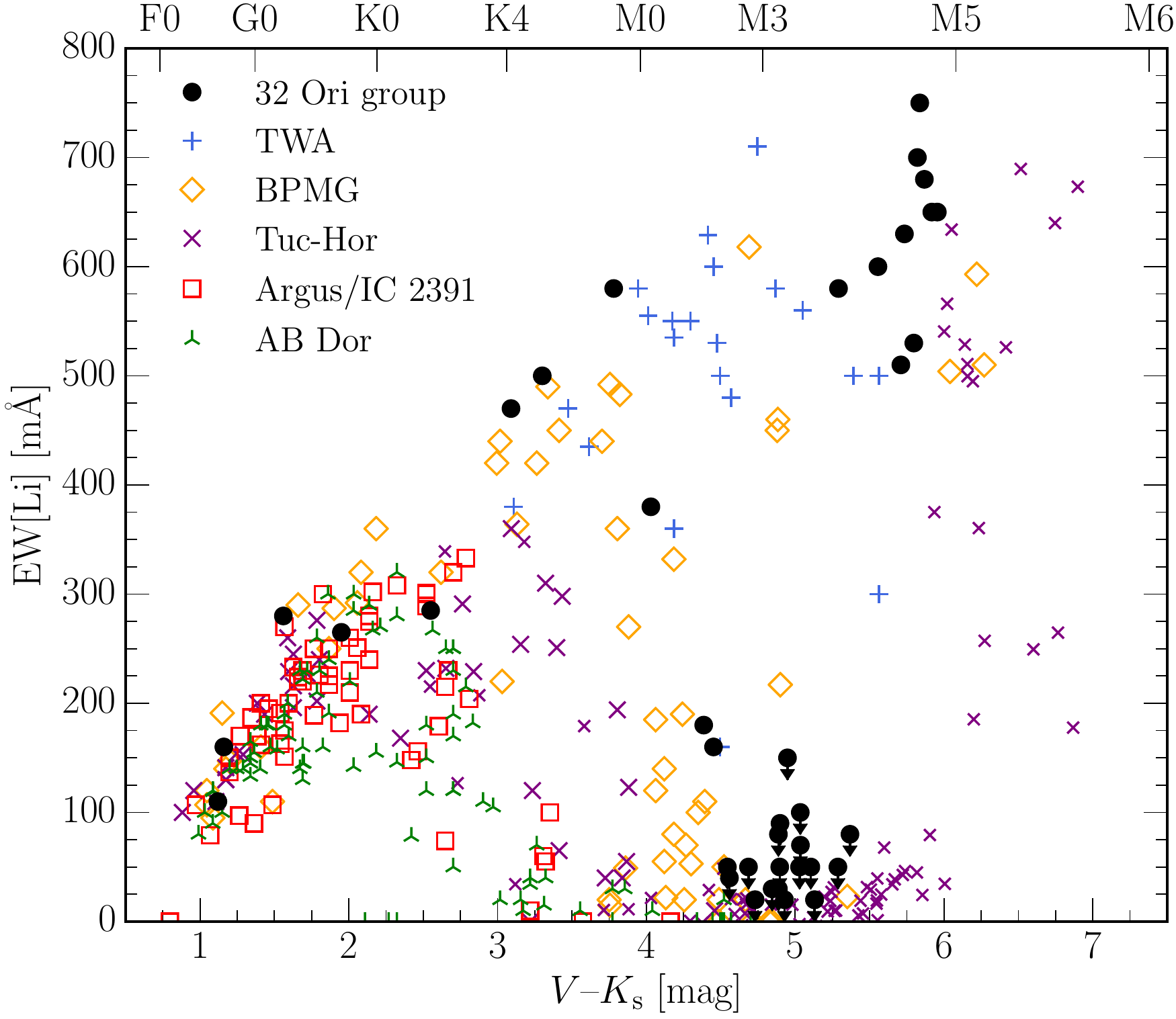}
\caption[]{EW[Li] of newly identified and previously known members
  of the 32~Ori group as
  a function of $V-K_{\rm{s}}$ colour, compared to other nearby young
  moving group members compiled from the studies of \cite{daSilva09},
  \cite{Kraus14}, \cite{Malo14b}, and \cite{Binks16}. Arrows denote
  upper limits for which the measured EW[Li] was zero. The depletion
  pattern closely resembles that of the $\sim$25~Myr-old $\beta$ Pictoris
  moving group (BPMG) and is bracketed by those of the TW Hydrae
  association (TWA; $\sim$10~Myr) and Tucana-Horologium
  ($\sim$45~Myr). The 32~Ori group Li depletion boundary is visible at
  $V-K_{\rm{s}}\approx5.5$\,mag (see Section~\ref{ldb}).}
\label{fig:ew_li}
\end{figure}

\begin{figure}
\centering
\includegraphics[width=\columnwidth]{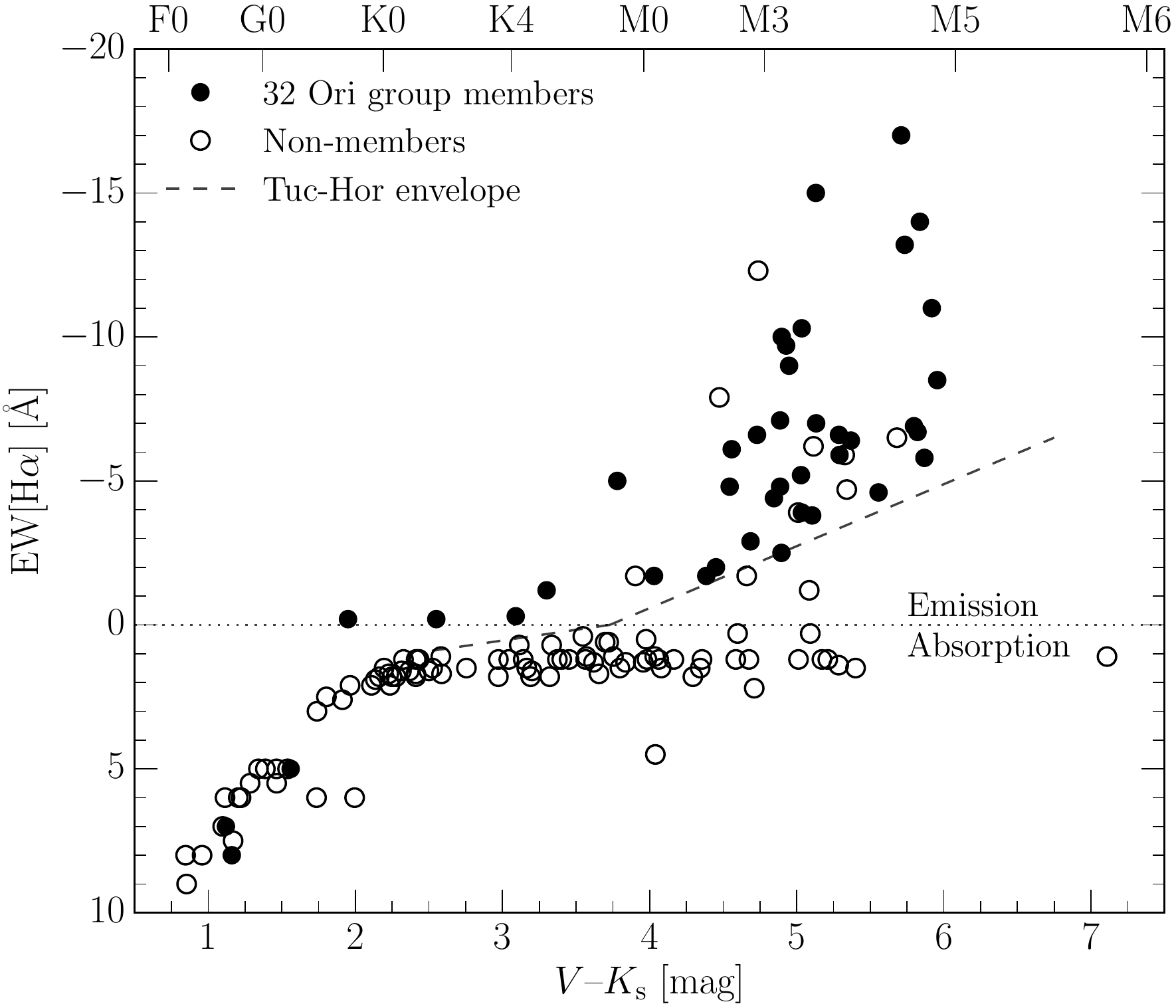}
\caption[]{EW[H$\alpha$] of all observed candidates and known members
  of the 32~Ori group as a function of $V-K_{\rm{s}}$ colour.  At
  late-G spectral types and beyond there is a clear distinction
  between confirmed 32~Ori group members and spectroscopic
  non-members, whereas at earlier spectral types both show similar
  levels of H$\alpha$ absorption. The dashed line denotes the lower
  envelope of H$\alpha$ emission from members of the $\sim$45\,Myr-old
  Tucana-Horologium association \citep{Kraus14}.}
\label{fig:ew_ha}
\end{figure}

To minimise the number of potential interlopers in our final
membership, we combine all three spectroscopic diagnostics (Li,
H$\alpha$, RV) in conjunction with the proper motion match
($\Delta_{\rm PM}$) and kinematic distance determined in
Section~\ref{initial_input_catalogue_and_search_criteria}. We adopt a
threshold of $\mathrm{EW[Li]} \geq 100\,\rm{m\AA}$ across the entire
spectral type range of our sample as an indicator of youth, whereas
those stars with measured $\mathrm{EW[Li]}<100\,\rm{m\AA}$ for
spectral types $\leq$\,K5 and $\geq$\,M5 are deemed to be older field
interlopers. Although there is a range of observed H$\alpha$ emission
levels in both young coeval populations and older field stars, young
stars do exhibit a lower envelope of H$\alpha$ emission (see
e.g. \citealp{Kraus14} in the slightly older Tucana-Horologium
association; hereafter Tuc-Hor) and so we classify any star with
emission below this envelope as an older field star. To illustrate
these criteria, the EW[Li] and EW[H$\alpha$] values of observed
candidates and members of the 32~Ori group are plotted as a function
of $V-K_{\rm{s}}$ colour in Figs.~\ref{fig:ew_li} and \ref{fig:ew_ha}.

Fig.~\ref{fig:ew_li} shows the Li depletion pattern of the 32~Ori
group compared to other nearby young moving groups. Whilst the overall
trend for all such groups is very similar for G-type stars and
earlier, as one moves into the K- and M-type regime the older
groups clearly show evidence of significant Li depletion at earlier
spectral types compared to younger groups. The Li depletion pattern of
the 32~Ori group closely mirrors that of the BPMG and is bracketed by
those of the TW~Hydrae association (TWA) and Tuc-Hor (ages $\sim$10
and 45\,Myr, respectively). In TWA there are a significant number of
Li-rich early-M dwarfs which are clearly absent in the 32~Ori
group, for which the Li depletion pattern appears to turn over at a
spectral type of $\sim$M0. Likewise, there are several Li-rich mid to
late K-type stars in the 32~Ori group which are not observed in the older
Tuc-Hor (turn over at $\sim$K4). This already provides a relative age
ranking suggesting that the 32~Ori group is older than TWA, younger
than Tuc-Hor and approximately coeval with the BPMG (age
$\sim$25~Myr).

\begin{figure}
\centering
\includegraphics[width=\columnwidth]{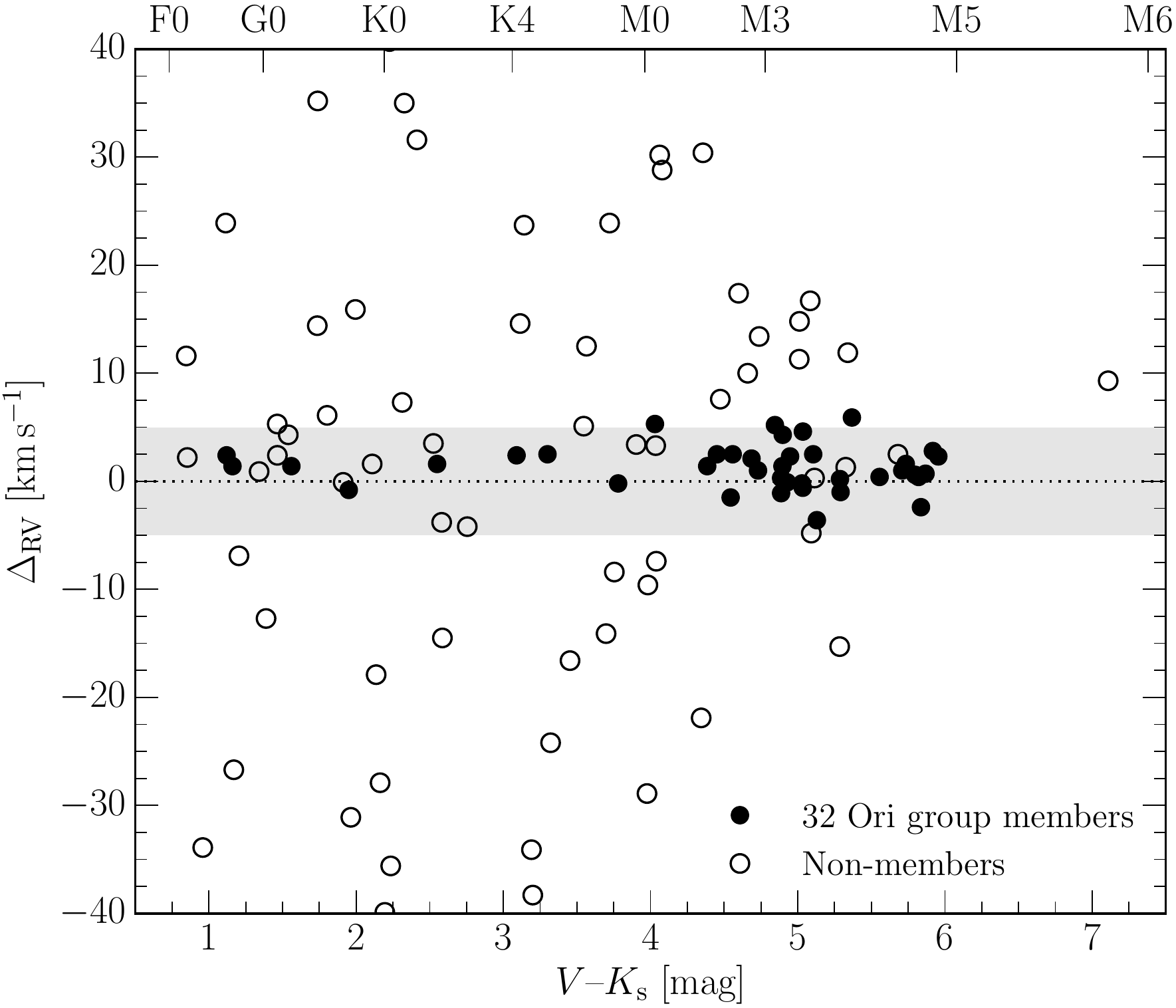}
\caption[]{Radial velocity residual
  ($\Delta_{\mathrm{RV}}=\mathrm{RV}-\mathrm{RV_{expt}}$) of all
  observed candidates and known members of the 32~Ori group as a function of
  $V-K_{\rm{s}}$ colour. The shaded region represents our
  $|\Delta_{\rm{RV}}| \leq 5\,\rm{km\,s^{-1}}$ membership
  criterion. Although three stars appear to lie just outside this
  region, after considering their velocity uncertainties they can be
  considered members.}
\label{fig:delta_rv}
\end{figure}

Given that young moving groups typically have very small intrinsic
velocity dispersions ($\sigma_{\rm 1D} \lesssim 1.5$~\kms; see
\citealp{Mamajek16}), we would expect the observed radial velocity of
a genuine 32~Ori group member to be similar to the line-of-sight projection
of the group space velocity at that position (see
Section~\ref{initial_input_catalogue_and_search_criteria}). We
therefore retained only those candidates for which
$|\Delta_{\mathrm{RV}}|=|\mathrm{RV}-\mathrm{RV_{expt}}| \leq 5$~\kms,
after allowing for the uncertainty on the WiFeS velocity (see
\citealp{Binks16}). When more than one velocity measurement was made
we adopted the unweighted average and standard deviation (also see
discussion in Section~\ref{sec:binaries}). Fig.~\ref{fig:delta_rv}
shows the difference in radial velocity $\Delta_{\mathrm{RV}}$ for all
observed candidates and known members of the 32~Ori group as a
function of $V-K_{\rm{s}}$ colour. Only three stars have WiFeS velocities which
place them just outside the $\Delta_{\rm{RV}}$ limit (all have
$|\Delta_{\rm{RV}}| < 6$\,\kms; see
Table~\ref{tab:candidate_spectra}). After considering their velocity
uncertainties, however, they can all be considered genuine
members. One is the known member and SB2 system V1874 Ori
($\Delta_{\rm RV}=5.3$~\kms) which has a high-resolution systemic
velocity of $18.4\pm0.3$~\kms\ (see Table~\ref{tab:bonafide_members}), giving $\Delta_{\rm
  RV}=1.3$~\kms. The remaining two stars are both new members, one of which
(2M0522+0502) is a possible fast rotator with a broad
CCF.

Combining the Li absorption, H$\alpha$ emission and radial velocity
criteria we identify 28 new members of the 32~Ori group. These new
members, all with spectral types between M1.5 and M5.5, are listed in
Table~\ref{tab:candidate_spectra} along with the 90 non-members and six
interesting systems requiring further study (see
Section~\ref{notes_possible_members}). The penultimate set of columns
in Table~\ref{tab:candidate_spectra} list the results of the
membership tests described above, with the final membership decision
including the proper motion match $\Delta_\textrm{PM}$ and (kinematic) distance.
For convenience, each member is identified by both its 2MASS designation and a shorter 32~Ori group membership number prefixed by the letters THOR,
similar to the TWA (TW Hydrae association) and RECX ($\eta$ Chamaeleontis cluster) nomenclatures commonly used in the literature for other young groups.

\subsection{Recently proposed members from the literature}
\label{sec:burgasser}

\citet{Burgasser16} recently proposed the free-floating low surface
gravity L1 brown dwarf WISE\,J052857.68+090104.4 as the first
substellar member of the 32 Ori group. It is only $3\degr$ from 32~Ori
itself and its estimated distance, proper motion, radial velocity and
spectral characteristics are all consistent with group membership. At
an age of $\sim$25\,Myr, its 1880\,K effective temperature implies a
mass ($M = 14^{+4}_{-3}\,\rm{M_{Jup}}$ ) which straddles the brown
dwarf/planetary-mass boundary.

\citet{Riedel16} also proposed the Li-poor M2.5 star SCR\,0522$-$0606
(2MASS\,J05224069$-$0606238) as a potential 32~Ori group member on the
basis of its proper motion, spatial position and low surface
gravity. Their SALT/RSS radial velocity of $-1.5\pm5.0$\,\kms,
however, is approximately 4$\sigma$ from the $\sim$21\,\kms\ expected
of a genuine member at that position. To test for spectroscopic
binarity we obtained WiFeS/$R7000$ spectra on 2016 October 14 and 2017 January 7,
finding a mean radial velocity of $25.6\pm1.9$~\kms\ and a spectral type
of M3. The spectra showed broad, unresolved CCFs
($\textrm{FWHM}=3.6$~px) and double-peaked He\,\textsc{i}
$5876\,\rm{\AA}$ emission lines, both suggestive of binarity. The star
is outside our $10\degr$ survey radius, but based on its UCAC4 proper
motion, we calculate $\Delta_{\rm PM}=7.9$~\masyr\ at $d_{\rm
  kin}=88$\,pc. At this distance SCR\,0522$-$0606 has
$M_{V}=9.55\,\rm{mag}$ and an elevated CMD position consistent with an
equal-mass binary. Given its low gravity, distance, radial velocity
and reasonable proper motion, we confirm membership of
SCR\,0522$-$0606 in the 32~Ori group.

Including these two objects (as well as HD\,36002, see
Section~\ref{2m0528}), there are currently 46 known 32~Ori group
members. Note that given the limited areal coverage and depth of the
current study, the true stellar census of the group is almost
certainly incomplete (especially at lower masses) and this number should be treated as a lower
limit.

\subsection{Notable systems}

\subsubsection{2MASS\,J05284050+0113333 (THOR 4B)}
\label{2m0528}

\begin{figure*}
\centering
\includegraphics[width=\textwidth]{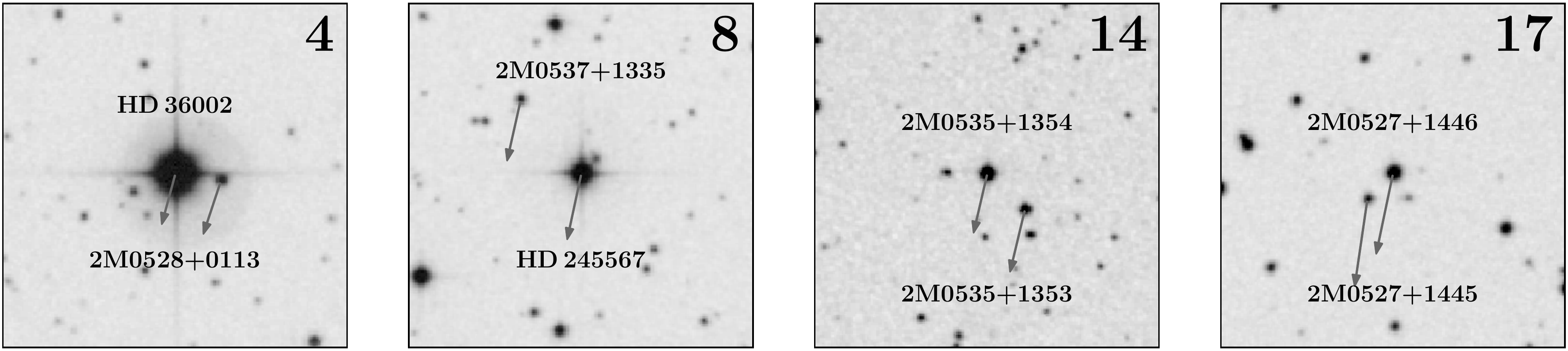}
\caption[]{Red Digitized Sky Survey images of the four wide
  pairs confirmed as members of the 32~Ori group during this
  study (the large number in the upper right corner of each image corresponds
  to the THOR number assigned as part of this study; see Table~\ref{tab:candidate_spectra}).
  Each image is $3\times3$\,arcmin and oriented north up, east left. Arrows show the UCAC4 proper motions
  projected over a period of 1000\,yr. Each pair clearly shares a
  common proper motion, in addition to a common distance and radial
  velocity. From left to right, their physical separations are
  approximately 2600, 5500, 3000 and 1600\,au.}
\label{fig:close_pairs}
\end{figure*}

2M0528+0113 (M3.5, $d_{\rm kin}=99$\,pc), is only 24~arcsec from and
co-moving with the early-type star HD\,36002 (see
Fig.~\ref{fig:close_pairs}), which was in our final candidate list
($d_{\rm kin}=105$\,pc, $\Delta_\textrm{PM}=0.2$\,\masyr) but not initially
observed. It was, however, observed by both
\emph{Hipparcos} ($d=103_{-8}^{+9}$\,pc; \citealp{vanLeeuwen07}) and \emph{Gaia}
($112_{-7}^{+8}$\,pc; including the additional $\pm0.3$\,mas systematic
uncertainty on the parallax as described in \citealp{Gaia16}). Its \emph{Gaia} DR1 proper
motion is $<$\,2\,\masyr\ from that expected of a 32~Ori group member at the weighted mean distance of 109$_{-5}^{+6}$\,pc and 2M0528+0113 is $<$\,3\,\masyr\ discrepant, well within its 5.8\,\masyr\ URAT1 uncertainties. We obtained two WiFeS R7000 spectra of HD\,36002 on 2017 January 11 and find it to be an SB2 system (see Table\,\ref{tab:sb2}) with an estimated spectral type of A7. Given
their close separation, congruent distances and proper motions, we
propose HD\,36002 and 2M0528+0113 are a genuine co-moving pair
separated by $\sim$2600\,au and thus HD\,36002 is also a member of the
32~Ori group (THOR 4A; see Table~\ref{tab:candidate_spectra}). 

\subsubsection{2MASS J05372061+1335310 (THOR 8B)}

This M3 star is 50\,arcsec north-east of the previously known G5
member HD\,245567 (THOR 8A; see Fig.~\ref{fig:close_pairs}).  At a kinematic
distance of 113\,pc, 2M0537+1335 has a space motion only
1.5\,\kms\ from the mean 32~Ori group velocity.  HD\,245567 has a
kinematic distance of 104\,pc, which agrees with that of 2M0537+1335 (physical
separation of $\sim$5500\,au) to within the uncertainties propagated from
their respective proper motions. Neither star has a \emph{Hipparcos}
or \emph{Gaia} DR1 parallax. A wider possible co-moving companion, the M5.5
2MASS J05373000+1329344, is 6\,arcmin from HD\,245567 at a distance of
110\,pc, corresponding to a separation of $\sim$40\,kau.

\subsubsection{2MASS J05351761+1354180 (THOR 14A)}

This new Li-rich M1.5 member is only 27\,arcsec from
2MASS J05351625+1353594 (THOR 14B, M3.5; see Fig.~\ref{fig:close_pairs}). The
\emph{ROSAT} source 1RXS J053516.6+135404 is more likely associated
with the lower mass component. Despite observed radial velocities
3--5\,\kms\ larger than expected (noting that 2M0535+1354 is a suspected SB1; see Table\,\ref{tab:rvs}), both stars are excellent
proper motion matches to the group ($\Delta_{\rm
  PM}<2$\,\masyr) at distances of 116 and 110\,pc, respectively, and
are unequivocal members of 32~Ori with a separation of
$\sim$3000\,au.

\subsubsection{2MASS J05274313+1446121 (=DIL7; THOR 17A)} 

2M0527+1446 (spectral type M3) is associated with the \emph{ROSAT}
source 1RXS J052743.4+144609 and was catalogued as an H$\alpha$
emission line object (DIL7) in the north-western outskirts of the
$\lambda$~Orionis star-forming region by \citet*{Duerr82}. It forms a
19\,arcsec wide pair with the M5 member 2MASS J05274404+1445584
(THOR~17B; see Fig.~\ref{fig:close_pairs}). Both stars are excellent kinematic
matches to the mean 32~Ori group space motion ($\Delta_{UVW}<2$\,\kms)
at inferred distances of 87 and 82\,pc, respectively.
At such distances their angular separation corresponds to
$\sim$1600\,au. The pair are unlikely to be $\lambda$~Orionis members,
given both the much larger distance to the association and its
significantly younger age ($\sim$400\,pc and $\sim$10\,Myr,
respectively; see \citealp{Bell13}) and thus we assign them as members
of the 32~Ori group. Interestingly, both this pair and THOR 14AB have
components either side of the 32~Ori group lithium depletion zone but
in the opposite sense; THOR 17A (M3) and THOR 14B (M3.5) lie in
the Li-poor region, while THOR 17B (M5; $\textrm{EW[Li]}=580$\,m\AA)
and THOR 14A (M1.5; 180\,m\AA) are Li-rich. 
Component identifications and Washington Double Star
(WDS) catalogue parameters for these four new wide systems are listed in Table~\ref{tab:wds}.

\begin{table*}
  \caption[]{Wide binaries discovered in this study. Column 1 lists the THOR number for each component as assigned in this study,
    whereas Column 3 provides the Washington Double Star (WDS) catalogue designation assigned based on the position of the primary.}
\begin{tabular}{r l c c c c c}
\hline
THOR & Component               & WDS                         & Epoch (Ref.)    & PA     & Sep.      & $V$\\
\#   & designation             & designation                      & (yr)            & (\degr)  & (arcsec) & (mag)\\
\hline
4A   & HD~36002                & \multirow{2}{*}{05287+0113} & 2000.08 (2MASS) & 261.46 & 24.01 &  7.46\\
4B   & 2MASS~J05284050+0113333 &                             & 2015.00 (\emph{Gaia})  & 261.16 & 24.01 & 14.74\\
\hline
8A$^{\dagger}$   & HD~245567               & \multirow{2}{*}{05373+1335} & 1998.74 (2MASS) & 39.46  & 49.79 &  9.54\\
8B$^{\ddag}$   & 2MASS~J05372061+1335310 &                             & 2015.00 (\emph{Gaia})  & 39.47  & 49.82 & 14.96\\
\hline
14A  & 2MASS~J05351761+1354180 & \multirow{2}{*}{05353+1354} & 1998.74 (2MASS) & 226.71 & 27.12 & 13.53\\
14B  & 2MASS~J05351625+1353594 &                             & 2015.00 (\emph{Gaia})  & 226.07 & 27.27 & 14.95\\
\hline
17A  & 2MASS~J05274313+1446121 & \multirow{2}{*}{05277+1446} & 1998.73 (2MASS) & 135.95 & 18.98 & 14.13\\
17B  & 2MASS~J05274404+1445584 &                             & 2015.00 (\emph{Gaia})  & 135.05 & 18.95 & 16.26\\
\hline
\end{tabular}\\
\begin{flushleft}
$^{\dagger}$ HD~245567 is itself a 0.3\,arcsec close binary (see \citealp{Metchev09}) with the WDS catalogue labelling the components Aa and Ab.\\
$^{\ddag}$ The WDS catalogue lists four additional components to HD~245567 (labelled B--E) with separations of between $\sim$\,3 and
11\,arcsec, however astrometry indicates that these companions are unphysical (see \citealp{Metchev09}). We therefore simply adopt
the label B for 2M0537+1335.
\end{flushleft}
\label{tab:wds}
\end{table*}

\subsubsection{2MASS\,J05363692+1300369 (THOR 31)}

\begin{figure}
\centering
\includegraphics[width=\columnwidth]{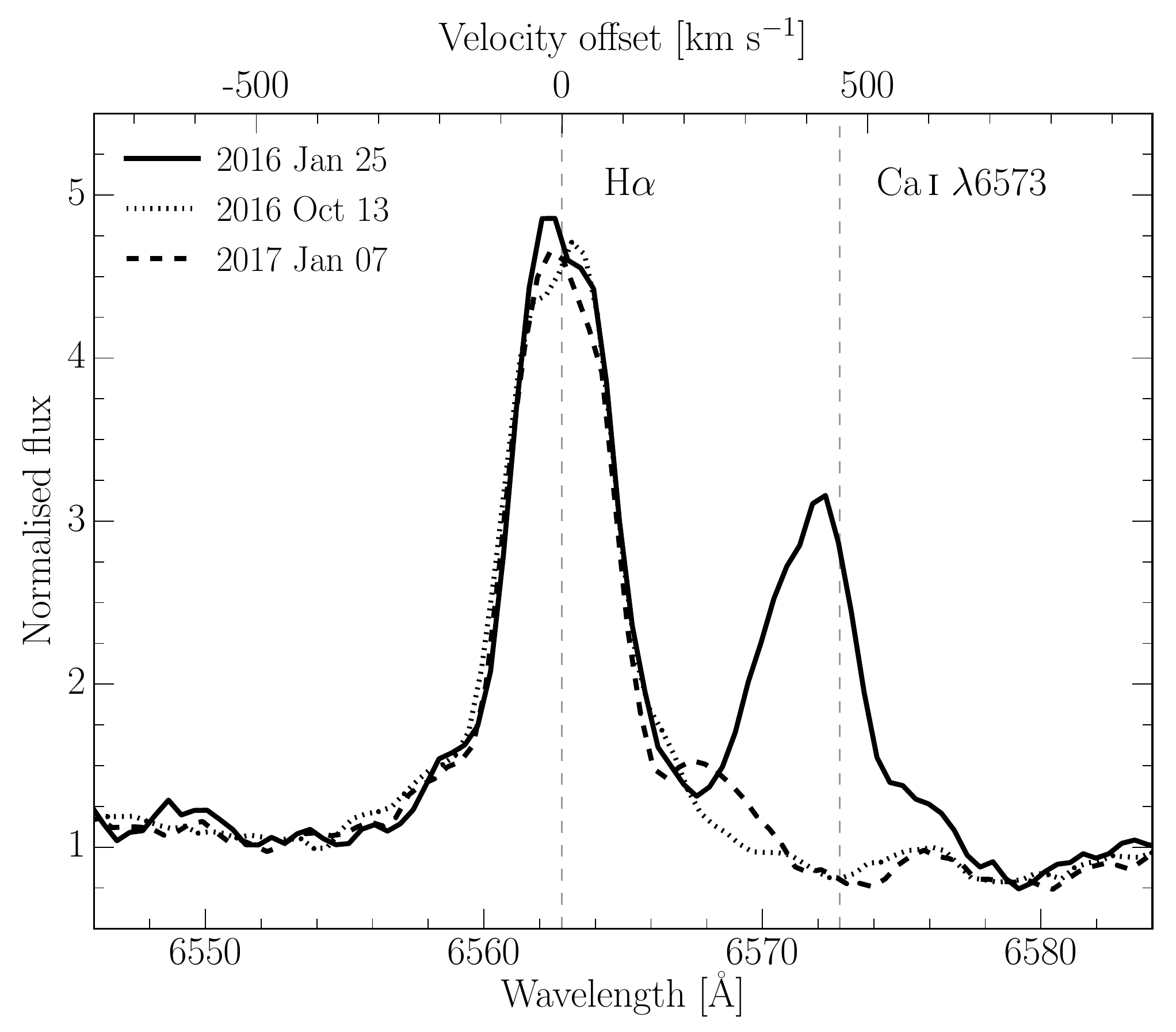}
\caption[]{Multi-epoch H$\alpha$-region spectra of 2MASS
  J05363692+1300369 (THOR 31) showing the strong emission feature $\sim$\,400\,\kms\ redward of H$\alpha$. All spectra have been
  shifted to the heliocentric rest frame, with the rest wavelengths of
  H$\alpha$ and Ca\,\textsc{i} at 6573~\AA\ given by the dashed lines. Note the
  double-peaked (SB2) structure of the H$\alpha$ emission.}
\label{fig:cai}
\end{figure}

The 2016 January 25 spectrum of this Li-rich SB2 member (combined
spectral type M4.5) showed a strong ($\textrm{EW}\approx-5$\,\AA)
emission line $\sim$\,400~\kms\ redward of H$\alpha$, near the rest wavelength of 6573\,$\rm{\AA}$ Ca\,\textsc{i} (see Fig.~\ref{fig:cai}). Inspection of the raw image shows that the emission is associated with a single source and not the sky background or
extended nebular emission. The feature 
is clearly variable as only Ca\,\textsc{i} absorption is seen in two 2016 October observations a night apart, but a weaker emission `bump' at +200\,\kms\ is visible in the 2017 January 7 spectrum. At all times the shape and strength of H$\alpha$ remained approximately constant.  The emission features cannot be red-shifted H$\alpha$ from the companion, whose velocity offset is only $\sim$\,50\,\kms\ based on the broad but unresolved CCF and double-peaked H$\alpha$ line.

Given the lack of excess H$\alpha$, other Ca emission lines (e.g. Ca\,\textsc{i} 6717~\AA) or any emission lines in the January spectrum typically associated with flare activity (especially He\,\textsc{i} 5876 and 6678~\AA), we do not believe this line is associated with flare-driven Ca\,\textsc{i} 6573~\AA. Instead, we propose it is red-shifted H$\alpha$ emission arising from an eruptive prominence or coronal mass ejection (CME) moving away from the system at close to its escape velocity (c.f. $\sim$\,350~\kms\ for $M=0.1\,\rm{M_{\odot}}$ and $R=0.3\,\rm{R_{\odot}}$, appropriate for a 25\,Myr star of $T_{\rm eff}=3000$\,K; \citealp{Baraffe15}).  Balmer line emission from CMEs has been detected in a handful of active M dwarfs (e.g. \citealp*{Houdebine90}; \citealp{Guenther97,Fuhrmeister04}) but is typically  associated with contemporaneous flare activity and a \emph{blue-shifted} asymmetry in the line profile.  That we observed red-shifted emission with a projected velocity of 400~\kms\ outside of an obvious flare event suggests the ejected material was long-lived and large enough to not be completely occulted by the stellar disc. The role of the binary companion in this scenario also remains unclear. High-cadence monitoring of 2M0536+1300 would be useful to firmly establish the nature of the January 25 emission and the frequency of such events. 

\subsection{Potential members requiring further study}
\label{notes_possible_members}

Below we present notes on six systems which we have assigned a
membership of `?' in Table~\ref{tab:candidate_spectra} (provided
in full as Supporting Information with the online version of the paper). Improved spectroscopic, astrometric or photometric data is required to
unequivocally assign membership of these stars to the 32 Ori group.

\subsubsection{2MASS J05525572$-$0044266}
\label{2massj0552-0044}

\begin{table}
\caption[]{Preliminary orbital and physical parameters of 2MASS J05525572$-$0044266, derived from a least squares Keplerian orbit fit to the primary and secondary velocities from Table~\ref{tab:sb2}. Uncertainties were propagated from the covariance matrix of the fit. Similar orbital parameters are obtained fitting only the primary velocities.}
\label{tab:eb}
\begin{tabular}{l l}
\hline
Period, $P$ & 0.8589884~d (fixed)\\
Eccentricity, $e$ & $0.10\pm0.11$\\
Primary velocity semi-amplitude, $K_{1}$ & $51.6\pm3.0$~\kms \\
Secondary velocity semi-amplitude, $K_{2}$ & $137.6\pm7.1$~\kms \\
Systemic velocity, $\gamma$ & $20.9\pm2.3$~\kms\\
Epoch of periastron passage (MJD), $\tau$ & $57319.228\pm 0.064$\\
Orientation of periastron, $\omega$ & $211\pm33^{\circ}$ \\
Mass ratio, $q=M_{2}/M_{1}$ & $0.375\pm0.029$\\
Total mass, $(M_{1}+M_{2})\sin^3 i$ & $0.602\pm0.074\,\rm{M_{\odot}}$\\
Semi-major axis, $a\sin i$ & $0.01494\pm0.00061$ au\\
& $3.21\pm0.13$ R$_{\odot}$\\
Assuming inclination $i=90^{\circ}$:\\[1mm]
Primary mass, $M_{1}$ & $0.438\pm0.058\,\rm{M_{\odot}}$\\
Secondary mass, $M_{2}$ & $0.164\pm0.019\,\rm{M_{\odot}}$\\
\hline
\end{tabular}
\end{table}

\begin{figure}
\centering
\includegraphics[width=\columnwidth]{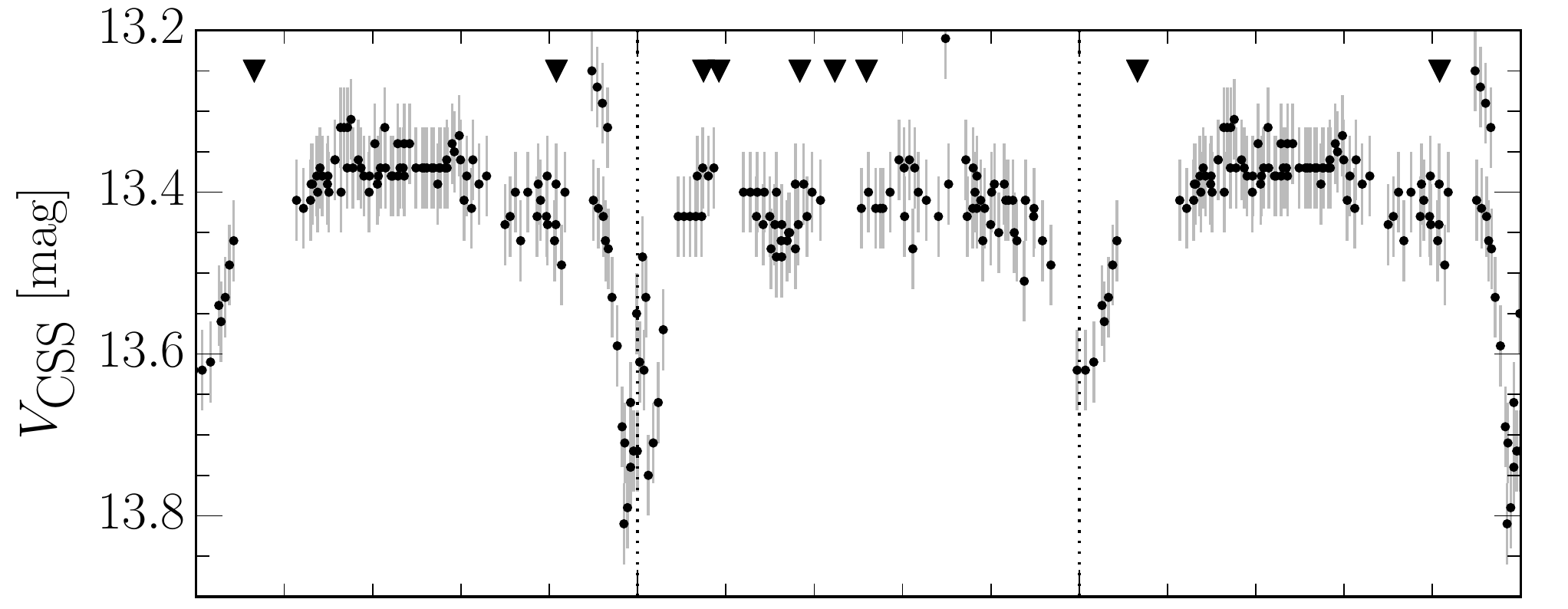}\\
\includegraphics[width=\columnwidth]{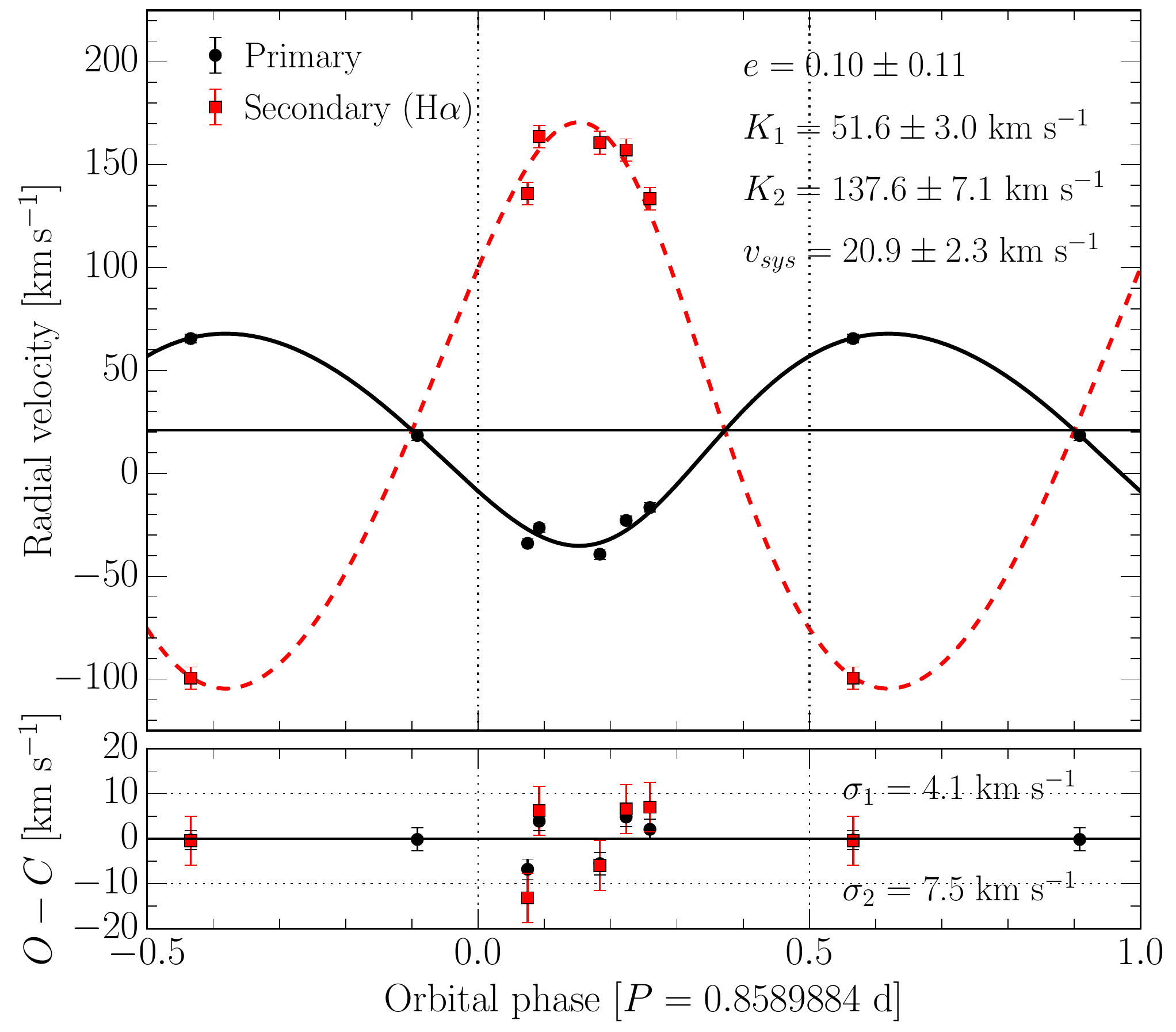}
\caption[]{\emph{Top:} Catalina Sky Survey $V_{\rm CSS}$-band light curve of the eclipsing binary 2MASS J05525572$-$0044266.
The data have been phased with a period of 0.8589884~d \protect\citep{Drake14} and offset such that the primary eclipse occurs at $\phi=0$. Triangles denote the
phases at which the WiFeS spectra were taken. \emph{Middle:} Keplerian radial velocity curves  (see Table~\ref{tab:eb}) fitted to the WiFeS data and phased with the light curve. The derived systemic velocity (20.9\,\kms) is shown by the horizontal line. The large offset from this velocity at zero phase immediately implies a non-zero eccentricity. \emph{Bottom:}  Velocity residuals from the best fit. Note that both the secondary velocities and their residuals are correlated with the primary velocities, as expected (see text).}
\label{fig:eb}
\end{figure}

2M0552$-$0044 was classified as a detached, Algol-type eclipsing
binary with a period of 0.86\,d by the Catalina Sky Survey
\citep[CSS; ][]{Drake14}, and as such makes it a rare example of an eclipsing
M dwarf (unresolved spectral type M3).
Its phased $V_{\rm CSS}$ light curve is plotted in the top panel of Fig.\,\ref{fig:eb}. We confirm binarity from two radial 
velocities of +65.5 and $-$39.3\,\kms\ ($\Delta t=357$\,d, see
Table~\ref{tab:sb2}) and a variable, double-peaked H$\alpha$ emission
line.  We obtained five more WiFeS observations of the star during 2017 January 7--9 for the purposes of establishing a preliminary orbital solution. Although the secondary component was not visible in the CCF, we estimated its velocity at each epoch by measuring the velocity offset between H$\alpha$ peaks and adding this to the primary velocity derived from cross-correlation. One observation did not show a double H$\alpha$ line, implying its radial velocity (18.4\,\kms) is close to systemic. Using these velocities we fit Keplerian orbits of period 0.8589884\,d \citep{Drake14} using standard least squares methods and derive the solution listed in Table~\ref{tab:eb} and the middle panel of Fig.\,\ref{fig:eb}. The system has a total mass of $(M_{1}+M_{2})\sin^3 i=0.6$~M$_{\odot}$ and separation $a\sin i=3.2~R_{\odot}$. Assuming $i=90^{\circ}$, this corresponds to component masses of approximately 0.44 and 0.16~M$_{\odot}$. Given the agreement between the fitted systemic velocity of $20.9\pm2.3$~\kms\ and the 20.6~\kms\ expected of a 32~Ori group member at that position, we consider 2M0552$-$0044 a highly likely member pending further velocity measurements and improved photometry.

The CSS light curve contains 192 measurements over 7.3 years and is not well sampled around the eclipses. Rotationally modulated photometric variation due to star-spots appearing and disappearing during that time also likely contributes to scatter in the regions outside the eclipses. This may affect the period determination and makes deriving temperature ratios and radii problematic. Moreover,  the APASS $V$-band magnitude ($15.15\pm0.06\,\rm{mag}$) differs
significantly from the CSS photometry.  The
accuracy of these data is limited by the transformation to $V_{\rm CSS}$
from the unfiltered survey photometry, and is only appropriate for the
G-dwarf calibrators used by the survey. From 445 Landolt standard
stars, \citet{Drake13} derived $V = V_{\rm CSS} + 0.31(B-V)^{2} +
0.04$, which for $(B-V)_{\rm APASS}=1.51$ yields
$V=14.2$~mag. After re-deriving this relation we find it is
well-defined for standards as red as 2M0552$-$0044, so the remaining discrepancy is likely due to binarity affecting the transformation. We therefore 
 adopt the APASS photometry, which provides a CMD position
appropriate for a 25~Myr-old M3 star at $\sim$90\,pc.

We are obtaining high-cadence photometry and further radial velocity observations of the system and will present a full reanalysis and characterisation, including final 32~Ori group membership, in an upcoming work. Until then we note that 2M0552$-$0044 is Li-poor, satisfies both selection criteria ($\Delta_{\rm{PM}}=2.5$\,\masyr, $d_{\rm{kin}}=92$\,pc), and is likely associated with the X-ray source 1RXS\,J055257.7$-$004424.

\subsubsection{BD+08\,900AB (=HD 34081AB)}

A 4\,arcsec A7+F2 near-equal brightness binary,
BD+08\,900AB is resolved in UCAC4 and we obtained spectra for both
components during good seeing (1.5\,arcsec FWHM) WiFeS observations in
2015 September and 2017 January. BD+08\,900B may be an SB2 while the primary is a
possible fast rotator (see Section~\ref{sec:binaries}). Both stars have radial velocities which are
consistent with 32~Ori group membership at $\lesssim$\,2\,\kms\ and
BD+08\,900B is an excellent proper motion match ($\Delta_{\rm
  PM}=1.5$\,\masyr) at a kinematic distance of 82\,pc. BD+08\,900A's
UCAC4 proper motion is significantly smaller and a poor match at any
distance. The combined system was observed with \emph{Hipparcos}
($d=120_{-16}^{+22}$\,pc) and \emph{Gaia} ($97^{+6}_{-5}$\,pc). Adopting
the astrometry of the latter, BD+08\,900AB is only 3\,\kms\ from the
32~Ori group mean space motion and the
$\textrm{EW[Li]}=130$\,m\AA\ we measure for BD+08\,900B is similar to
young F-type stars in Fig.~\ref{fig:ew_li}.  Until
the binary nature of BD+08\,900B is confirmed we refrain from assigning the system to
the 32 Ori group, but note it is a strong kinematic candidate.

\subsubsection{HD\,37825}

HD\,37825 is a new SB3 system (see Fig.~\ref{fig:sb3}) which we
observed at four epochs (see Table~\ref{tab:rvs}), one of which (2016
February 21) exhibited narrow, single lines and a radial velocity of
28\,\kms, which must be near-systemic. This is in reasonable agreement with the $\sim$20\,\kms\ expected of a
32~Ori group member at that position and there is also a moderately good
proper motion match to the group ($\Delta_{\rm PM}=6.8$\,\masyr) at a
kinematic distance of 84\,pc. The 2016 October and 2017 January spectra showed clearly resolved double lines with a weaker tertiary component. However, only two components are visible in the CCF, separated by $\sim$\,140~\kms. The system does not have a
\emph{Hipparcos} or \emph{Gaia} DR1 parallax or proper motion. The
$\textrm{EW[Li]}=130$\,m\AA\ we measured in the single-lined spectrum
is typical of a young F5 star (see Fig.\,\ref{fig:ew_li}). Higher resolution spectroscopy of the system and a velocity curve would be useful in confirming its membership.

\subsubsection{2MASS J05053333+0044034}
\label{2massj0505+0044}

2M0505+0044 satisfies both UCAC4 selection criteria
($\Delta_{\rm{PM}}=5.0$\,\masyr, $d_{\rm{kin}}=90$\,pc) and also shows
depleted lithium, consistent with its M2.5 spectral type. The star's
radial velocity, however, is rather discrepant
($\Delta_{\rm{RV}}=13.4$\,\kms). Coupled with a broad H$\alpha$ line
($\Delta v\approx 300$\,\kms\ at 10 per cent of peak flux;
$\textrm{EW[H$\alpha$]}=-12.3$\,\AA) and strong He\,\textsc{i} 5876 and
$6678\,\rm{\AA}$, Na\,\textsc{d}, and X-ray emission
(1RXS~J050533.4+004421), this could be indicative of spectroscopic
binarity. The star's CCF is not broadened, suggesting it is not a fast
rotator or SB2. Choosing to ignore the
$|\Delta_{\rm{RV}}|<5$\,\kms\ velocity threshold would have
implications for other candidates and so we retain 2M0505+0044 as a
possible member requiring further velocity measurements.

\subsubsection{2MASS J05561307+0803034}
\label{2massj0556+0803}

2M0556+0803 easily satisfies both selection criteria
($\Delta_{\rm{PM}}=0.5\,\rm{mas\,yr^{-1}}$ and
$d_{\rm{kin}}=97\,\rm{pc}$), whereas its mean radial velocity is
somewhat discrepant ($\Delta_{\rm{RV}}=7.6\,\rm{km\,s^{-1}}$). On its
own, this would suggest non-member status, however, the star's CCF is
rather broad ($\textrm{FWHM}=3.8$\,px) and a double-peaked
He\,\textsc{i} $5876\,\rm{\AA}$ emission line suggests SB2
binarity. Furthermore, 2M0556+0803 lies in a region of the CMD
surrounded by other Li-depleted group members (to within the
uncertainties on its photographic $V$-band magnitude), consistent with
an M4 spectral type.  We therefore retain 2M0556+0803 as a possible
member to be re-examined when further spectroscopic observations
become available.

\subsubsection{2MASS J05572121+0502158}
\label{2massj0557+0502}

2M0557+0502 satisfies the proper motion selection criterion
($\Delta_{\rm{PM}}=1.3$\,\masyr) and at a kinematic distance of
115\,pc was one of a small number of distance outliers in our sample
observed with WiFeS. The star's radial velocity, however, is strongly
inconsistent with membership ($\Delta_{\rm{RV}}=30.4$\,\kms). The
presence of a strong Li line ($\rm{EW[Li]}=290$\,m\AA) suggests
2M0557+0502 is young, but we also find its rising spectrum is
significantly reddened [$E(B-V)\approx0.6$\,mag], with an underlying
spectral type of K2.
This level of reddening is more than an order of magnitude greater than the
group mean of $E(B-V)=0.03$\,mag (see
Section~\ref{reddening}). Assuming 
$(V-K_{\rm{s}})_{\circ}\approx2.3\,\rm{mag}$ appropriate for an early K-dwarf and a
de-reddened $M_{V}\approx7.2$\,mag at 115~pc, 2M0557+0502 lies
1--2\,mag below the 32~Ori group CMD sequence and is unlikely to be a
member, especially if it is an SB1. Given the strong reddening and
large radial velocity it could be a Li-rich background giant but we
seek a better spectroscopic characterisation before finalising its
evolutionary and membership status.

\subsection{Comparison to previous non-spectroscopic surveys}
\label{comparison_to_previous_surveys}

Since its initial discovery by \cite{Mamajek07}, there have been no
surveys dedicated \emph{specifically} to discovering new members of
the 32~Ori group, with new members being individually added on the
basis of common attributes like proper motion and radial velocity
(\citealp{Mace09,Franciosini11}). Given that in this work we aim to
characterise the stellar population of the 32~Ori group, it is
important that we place our new members in context by re-examining
previous non-spectroscopic memberships  from large surveys
of Galactic open clusters and associations.

As part of their global catalogue of Milky Way clusters,
\cite{Kharchenko13} recovered the 32~Ori group, arguing that it
comprises 40 members within a radius of 2.2\degr, lies at a distance
of 95\,pc and has an age of $32\,\rm{Myr}$. They found a mean proper
motion and radial velocity of ($\mu_{\alpha}\cos\delta, \mu_{\delta})
\simeq (10.0, -32.2)\,\rm{mas\,yr^{-1}}$ ($\sigma_{\mu} \simeq
\pm0.8\,\rm{mas\,yr^{-1}}$) and $13.1\,\rm{km\,s^{-1}}$,
respectively. \citeauthor{Kharchenko13} assigned membership
probabilities using a combination of three metrics; one kinematic
(based on the proper motion of a given object with respect to the
cluster mean) and two photometric (based on positions in the
$K_{\rm{s}}$, $J-H$ and $J-K_{\rm{s}}$ CMDs). Stars which
satisfied all three metrics with a probability greater than 61 per
cent were classified as `most probable' members. Cross-matching their
40 most probable members with the 45 stellar group members of the
current study, we find only three objects in common: HR~1807, HD~35714
and HD~36338. In other words, the \citeauthor{Kharchenko13} selection
criteria omit 32~Ori as a member of the 32~Ori group! Furthermore,
cross-matching their members against potential kinematic members from
UCAC4 and URAT1 (see
Section~\ref{initial_input_catalogue_and_search_criteria}), we find 13
and 40 objects in common, respectively. Aside from HR~1807, HD~35714
and HD~36338, none of the other candidates have CMD positions or
$\Delta_{\rm PM}$ values consistent with membership in the 32 Ori
group.

The main reason the \cite{Kharchenko13} membership for late-type
objects is particularly unreliable is due to their reliance on
near-infrared 2MASS CMDs. At low effective temperatures ($T_{\rm{eff}}
\lesssim$\,4000\,K), isochrones in both the $K_{\rm{s}}$, $J-H$
and $J-K_{\rm{s}}$ CMDs become vertical and essentially degenerate
with age. This removes any meaningful photometric distance information
and means that the study was reliant solely on deeper but less
accurate PPMXL proper motions for membership. We note that within
their cluster radius of $2.2\degr$ there are 15 new members from
Table~\ref{tab:candidate_spectra} which could have provided
matches. Of these, 13 failed the kinematic test and two failed the
photometric tests. Clearly these tests are too restrictive and the
PPMXL proper motions not accurate enough to rely on for membership
determinations.

The 32~Ori group is also listed (as Mamajek 3) in the catalogue of
optically-visible open clusters by \citet[v.2.5;
  2005]{Dias02}. Recently, \cite{Dias14} used UCAC4 astrometry to
provide membership probabilities for individual stars by fitting the
observed proper motion distribution in a region surrounding each
cluster with two elliptical bivariate populations.  Based on an
apparent diameter of 250\,arcmin, \cite{Dias14} identified over
$2.8\times10^4$ UCAC4 counterparts, of which $2.3\times10^4$ were
assigned to the 32~Ori group, making it the twelfth most populous
cluster in the entire catalogue. From these members they estimated a
mean group proper motion of ($\mu_{\alpha}\cos\delta, \mu_{\delta}) =
(0.55, -2.75)\,\rm{mas\,yr^{-1}}$, which is not only inconsistent with
our proper motion (see Table~\ref{tab:group_mean}), but also the
proper motion listed in the original \cite{Dias02}
catalogue. Cross-matching our 45 stellar members with the sources
identified by \citeauthor{Dias14}, we find only 15 objects in common
(11 of which are previously known members, see
Table~\ref{tab:bonafide_members}; 28 lie outside the adopted diameter
of the group). None has a membership probability greater than 69 per
cent. As with the \cite{Kharchenko13} study, 32~Ori itself appears to
be a non-member of its own group, making it hard to place much
credibility in the \citeauthor{Dias14} results.  Their membership
probabilities are based solely on the proper motion of an object
relative to the mean, which alone is not sufficient to unambiguously
demonstrate membership.

\section{Properties of the 32 Ori group}
\label{properties_32_ori_group}

Below we use the previously known and new members of the 32~Ori group
to investigate its global properties; namely interstellar reddening,
age, circumstellar disc frequency and spatial structure.

\subsection{Reddening and extinction}
\label{reddening}

Before attempting to determine an age for the 32~Ori group from its
CMD, the effects of interstellar reddening must first be accounted
for. The 32~Ori group lies at a distance of $\sim$\,90\,pc, and whilst
reddenings for stars at this distance are typically low
[$E(B-V)\lesssim0.05\,\rm{mag}$], the complex shape of the Local
Bubble and surrounding clouds is such that the reddening is not
necessarily negligible \citep{Reis11}.

In Table~\ref{tab:red} we list $E(B-V)$ estimates for the
bright B-, A- and F-type members of the 32~Ori group using the best
available spectral types and photometry. For the two B-type members
(unresolved 32~Ori and HR~1807) we adopt $UBV$ photometry 
from \cite{Mermilliod06} and combine this with the
Q-method \citep{Pecaut13} to derive negligible reddenings of
$E(B-V)=0.01$ and 0.00\,mag, respectively. Using resolved $BV$
photometry from \emph{Tycho}-2, and adopting spectral types of B5 and B7
for 32~Ori A and B from \cite{Edwards76}, and A7 and A7.5 for
HD~36002 and HD~36823, we estimate $E(B-V)=0.01$, 0.06, 0.03 and
0.02\,mag, respectively. For the five other members, we fit the
$BVJHK_{\rm{s}}$ photometry \citep[and for HR\,1807, $U$-band
  from][]{Johnson66} to the intrinsic dwarf colour locus of
\citet{Pecaut13}, calculating $\chi^2$ fits for spectral energy distributions (SEDs) spaced in
$T_{\rm{eff}}$ by 10\,K and in steps of $E(B-V)=0.01\,\rm{mag}$. The
quoted $E(B-V)$ values reflect the distribution of those fits for
which the $\chi^2$ probability $q>5$ per cent over the range of
$T_{\rm{eff}}$ reflecting a $\pm$1 subtype uncertainty in spectral
type (where B9.5V to A0V was counted as a full step in subtype). The
uncertainties in the $E(B-V)$ estimates from the SED fitting are
typically $\pm$0.02\,mag.

Based on the values in Table~\ref{tab:red} we adopt a median reddening
toward the group of $E(B-V)=0.03\pm0.01$\,mag, which translates to a
$V$-band extinction of $A_{V}=0.10\pm0.03$\,mag and a
$K_{\rm{s}}$-band extinction of $A_{K_{\rm{s}}}=0.011\pm0.004$\,mag
\citep[following][]{Bilir08}.  The range of values [$E(B-V) \lesssim$
  0.06 mag] are suggestive of small amounts of patchy extinction
across the core of the group. Note that the $K_s$-band extinction is
smaller than the typical 2MASS photometric errors.

\begin{table}
\caption[]{Reddening estimates for the known B-, A- and F-type members of the 32~Ori group. \label{tab:red}}
\begin{tabular}{r l c l}
\hline
THOR & Name      & $E(B-V)$ & Method\\
      \#    & (mag)    &       \\
\hline
1AB & 32 Ori AB & 0.01  & Q-method ($UBV$)\\
1A & 32 Ori A  & 0.01  & $B-V$, SpT\\
1B & 32 Ori B  & 0.06  & $B-V$, SpT\\
2 & HR 1807  & 0.00  & Q-method ($UBV$)\\
2 & HR 1807  & $0.01\pm0.01$ & SED ($UBVJHK_{\rm{s}}$, SpT) \\
3A & HD 35714  & $0.05\pm0.02$ & SED ($BVJHK_{\rm{s}}$, SpT)\\
3B & HD 36823  &  0.02  &  $B-V$, SpT\\
4A & HD 36002  &  0.03  &  $B-V$, SpT\\
5 & HD 35499  & $0.04\pm0.02$ & SED ($BVJHK_{\rm{s}}$, SpT)\\
6 & HD 36338  & $0.03\pm0.02$ & SED ($BVJHK_{\rm{s}}$, SpT)\\
7 & HD 35695  & $0.06\pm0.02$ & SED ($BVJHK_{\rm{s}}$, SpT)\\
\hline
& {\bf Median} & {\boldmath$0.03\pm0.01$} & \\
\hline
\end{tabular}
\end{table}

\subsection{Isochronal age}
\label{iso}

\begin{figure*}
\centering \includegraphics[width=\textwidth]{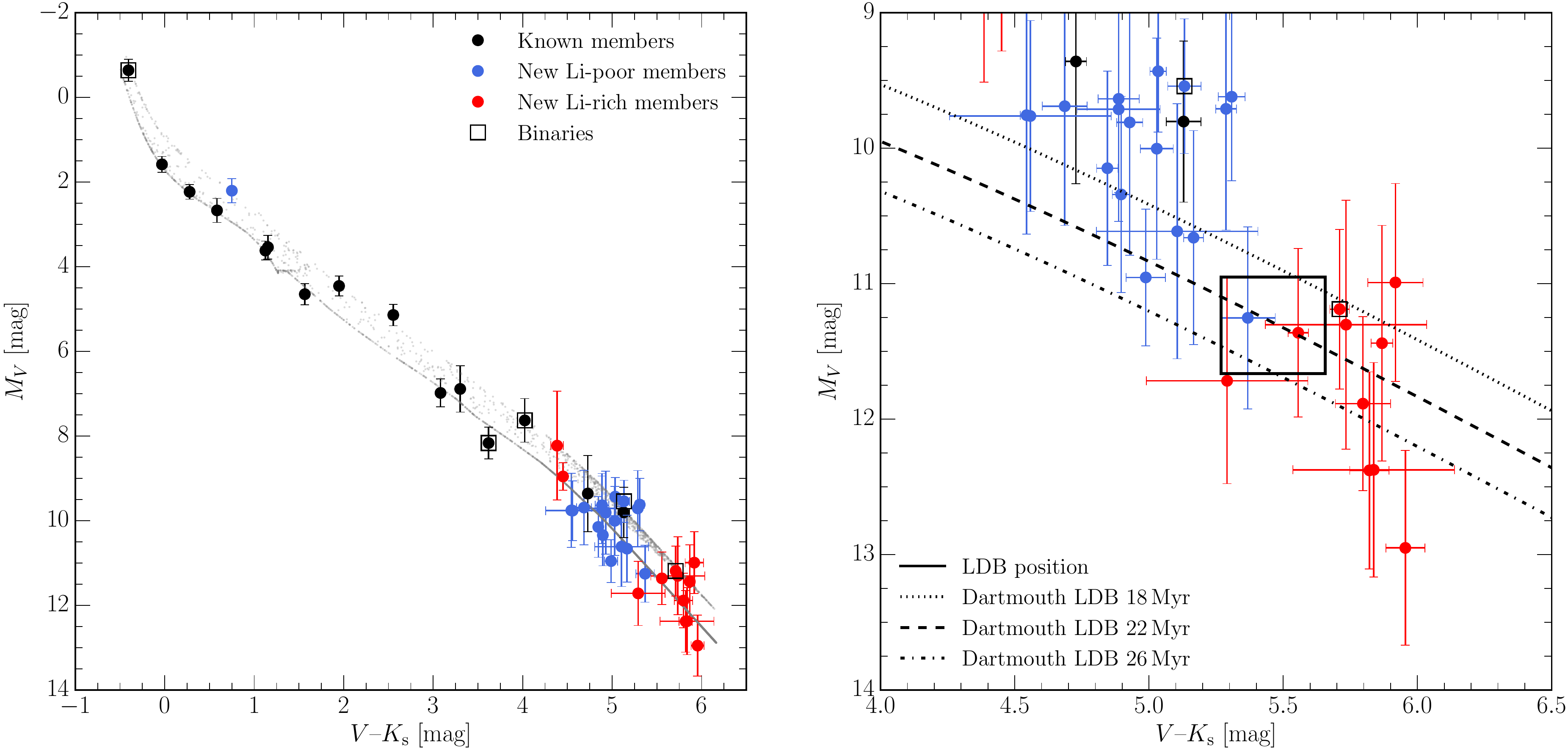}
\caption[]{\emph{Left:} Example best-fit $M_{V}, V-K_{\rm{s}}$ CMD of
  the 32~Ori group with the Dartmouth models overlaid. The grey points
  represent individual stars in our two-dimensional probability
  distribution (see text) from which we derive a best-fit age of
  23\,Myr. The lithium depletion boundary (LDB) is clearly identified
  at $V-K_{\rm{s}} \simeq 5.5\,\rm{mag}$ (spectral type M4.5).
  \emph{Right:} The position of the 32~Ori group LDB as
  defined in this study (see text). Overlaid are several lines
  corresponding to loci of constant luminosity at which Li is depleted
  at the 99 per cent level as predicted by the Dartmouth evolutionary
  models.  The best-fit LDB age of 22\,Myr in this panel is in
  excellent agreement with the isochronal age based on the same
  stellar evolutionary models.}
\label{fig:cmd_ldb}
\end{figure*}

Fig.~\ref{fig:cmd_ldb} shows the $M_{V}$, $V-K_{\rm{s}}$ CMD of
the 45 stellar members of the 32~Ori group. We preferentially use
trigonometric parallaxes to transform the apparent $V$-band magnitudes
listed in Table~\ref{tab:candidate_spectra} to $M_{V}$, however in the
majority of cases these are not available and so we adopt the
kinematic distances from our selection process (see
Section~\ref{initial_input_catalogue_and_search_criteria}).  Note that
for parallaxes from the recent \emph{Gaia} DR1 we include the
additional systematic uncertainty of $\pm0.3$\,mas as discussed in
\cite{Gaia16}. Our adoption of the best-fit kinematic distances is in
contrast to \cite{Bell15} in which we simply adopted the mean group
distance for the 11 stars listed in Table~\ref{tab:bonafide_members}
without parallaxes (barring HD~36823 which was not included in that
study), and has resulted in a less tight colour-magnitude sequence in
Fig.~\ref{fig:cmd_ldb}. Uncertainties on $M_{V}$ incorporate the
uncertainties in the proper motions used to calculate the kinematic
distances (typically several mas$\,\rm{yr^{-1}}$) and result in
distance uncertainties in the range of 8--30 per cent.

\begin{table*}
\caption[]{Isochronal and LDB ages for the 32~Ori group. \label{tab:iso_ldb_ages}}
\begin{tabular}{l c c c}
\hline
Method      &   Model   &   Age (Myr)   &   Mean age (Myr)\\
\hline
\multirow{4}{*}{Isochrone}   &   Dartmouth   &   $23^{+7}_{-3}$   &   \\
   &   BHAC15       &   $28^{+3}_{-4}$   &   $25\pm5$ ($1\sigma$)\\
   &   Pisa   &   $20^{+4}_{-1}$   &   [$\pm4$ (statistical), $\pm3$ (systematic)]\\
   &   PARSEC      &   $29\pm3$   &   \\
\hline
\multirow{4}{*}{LDB}   &   Dartmouth   &   $22^{+4}_{-3}$   &   \\
   &   BHAC15      &   $22^{+4}_{-3}$   &   $23\pm4$ ($1\sigma$)\\
   &   Pisa              &   $21\pm3$   &   [$\pm4$ (statistical), $\pm1$ (systematic)]\\
   &   Dartmouth ($\langle Bf\rangle=2.5\,\rm{kG}$)   &   $26\pm4$   &   \\
\hline
\multirow{2}{*}{{\bf Final adopted age}} & \multicolumn{3}{|c|}{{\boldmath$24\pm4$ ($1\sigma$)}}\\
   &   \multicolumn{3}{|c|}{{\boldmath[$\pm4$} {\bf(statistical),} \boldmath{$\pm2$} \bf{(systematic)]}}\\
\hline
\end{tabular}
\end{table*}

We derive an isochronal age for the 32~Ori group using the same method
as that described in previous papers (see
e.g. \citealp{Bell14,Bell15}), namely by fitting two-dimensional
probability distributions to the CMD using the $\tau^{2}$ fitting
statistic of \cite{Naylor06} and \cite{Naylor09}. In brief, the
probability distributions are created using stellar evolutionary
models and not only include binarity but also incorporate an empirical
colour-$T_{\rm{eff}}$ relation and bolometric corrections from
observations of low-mass Pleiades members, as well as theoretical
corrections for the dependence on surface gravity. For our isochronal
age analysis, we have adopted the following interior models: Dartmouth
\citep{Dotter08}, PARSEC \citep{Bressan12}, BHAC15 \citep{Baraffe15},
and Pisa \citep*{Tognelli15}.  Note that the Pisa models are based on
the same calculations as described in \cite{Tognelli15}, however they
cover a mass range of $0.08<M/\rm{M_{\odot}}<9$.  As per
Section~\ref{reddening}, we have included the effects of interstellar
reddening, adopting a group mean
$E(B-V)=0.03\,\rm{mag}$. Table~\ref{tab:iso_ldb_ages} lists the
individual isochronal ages for the 32~Ori group in addition to our
mean isochronal age of $25\pm5\,\rm{Myr}$ ($\pm4\,\rm{Myr}$
statistical, $\pm3\,\rm{Myr}$ systematic).

\subsection{Lithium depletion boundary age}
\label{ldb}

The lithium depletion boundary (LDB) is defined as the sharp
transition between stars which have contracted sufficiently that their
core temperatures have reached the critical value to burn Li and thus
exhibit depleted levels of Li in their photospheres, and those stars
at slightly lower masses which have not yet reached this critical
temperature and show undepleted levels. Over the past few years the
LDB has been extolled as the least model-dependent, absolute
age-dating technique for coeval populations with ages of between
$\sim$20 and 200\,Myr (see e.g. \citealp{Soderblom14}).

As shown in Fig.\,\ref{fig:cmd_ldb}, the LDB in the 32~Ori group is
defined by the Li-poor M4.5 member 2MASS J05313290+0556597 (THOR 30) and the
Li-rich ($\textrm{EW[Li]}=600$~m\AA) M4.5 member 2MASS
J05264073+0712255 (THOR 32). We note that neither of these stars appear to be an
unresolved binary which could significantly affect the determination
of the LDB luminosity. To derive an LDB age for the group we adopt the
method described in \cite{Binks14,Binks16}, which involves defining a
region in the CMD at which the LDB is located and then fitting curves
of constant luminosity corresponding to 99 per cent Li depletion.
These curves are created using stellar evolutionary models and after
transforming $L_{\rm{bol}}$ and $T_{\rm{eff}}$ to absolute $V$-band
magnitude and $V-K_{\rm{s}}$ colour using the pre-MS bolometric
correction and colour relations of \cite{Pecaut13}. Our threshold for
Li-poor stars (EW[Li]\,$< 100\,\rm{m\AA}$) ensures that, in comparison
to the Li-rich stars which exhibit undepleted levels of Li consistent
with measurements of young stars at birth (see
e.g. \citealp{Palla07}), the difference in depletion between the
Li-poor and Li-rich stars is greater than a factor of 100, and so
calculating the LDB age in such a manner is entirely justified (see
also \citealp{Jeffries05,Tognelli15}).

To define the LDB region we simply adopt the central position between
the two stars and the separation in both colour and magnitude,
yielding $V-K_{\rm{s}}=5.462\pm0.094$ and
$M_{V}=11.308\pm0.055$\,mag. These values are consistent with those
reported by \cite{Binks14} for the BPMG. The LDB age is then
calculated by fitting the Li depletion curves to this point and the
uncertainty on the age calculated from the size of the region in both
colour and magnitude. Note that we have also included an additional
uncertainty of 0.1\,mag in colour and 0.3\,mag in $M_{V}$ to reflect
likely uncertainties in both the photometric calibration (especially
the inhomogeneous $V$-band photometry) and kinematic
distances. Table~\ref{tab:iso_ldb_ages} lists the individual LDB ages
derived from the following sets of evolutionary models: Dartmouth
(including a new prescription for magnetic fields, $\langle
Bf\rangle=2.5\,\rm{kG}$; \citealp{Feiden13,Feiden14}), BHAC15 and
Pisa, which together yield a mean LDB age of $23\pm4\,\rm{Myr}$
($\pm4\,\rm{Myr}$ statistical, $\pm1\,\rm{Myr}$ systematic). The
individual LDB ages from the non-magnetic Dartmouth, BHAC15 and Pisa
models all agree to within 1\,Myr, whereas the magnetic Dartmouth
models imply an older, yet consistent, age for the group. These
findings are similar to those reported by \cite{Malo14b} and
\cite{Binks16} for the BPMG.

\subsection{Final adopted age}
\label{final_adopted_age}

The age analyses above clearly demonstrate that the isochronal and LDB
ages for the 32~Ori group are in agreement. Combining these two age
determinations, we calculate a final adopted age of $24\pm4\,\rm{Myr}$
($\pm4\,\rm{Myr}$ statistical, $\pm2\,\rm{Myr}$ systematic).  We note
that this age is essentially identical to the $23\pm3\,\rm{Myr}$ age
for the BPMG derived by \cite{Mamajek14}.  The uncertainty in our
final adopted age is driven by the statistical uncertainty which stems
from i) the reasonably large uncertainties on the kinematic distances
and ii) the inhomogeneous $V$-band photometry collated from the
available literature. Both of these points will be directly addressed
by \emph{Gaia} data releases in the coming years, by providing
parallaxes and well-calibrated, homogeneous $G$-band photometry for
all members of the group. This will naturally lead to a tighter
sequence in the CMD compared to that shown in Fig.~\ref{fig:cmd_ldb}
and will allow more statistically-robust age determinations.

\subsection{Circumstellar disc frequency}
\label{disc_frequency}

Given that the 32~Ori group appears to be essentially coeval with the
BPMG, which is known to harbour several optically-thin debris discs
(e.g. $\beta$ Pic itself, 51~Eri and AU~Mic), it is conceivable that
such discs may also be present in the 32~Ori group. These could
present ideal targets for direct imaging in an attempt to discover gas
giant planets and further constrain potential planetary formation
mechanisms.  As discussed in the Introduction, the \emph{Spitzer} IRAC and MIPS
survey described by \cite{Shvonski16} demonstrates that 4/14
members (not including HD~36823) exhibit $24\,\rm{\mu m}$ excess
emission, and hence these stars are likely debris disc candidates. In
this Section we search for further evidence of circumstellar material
around the combined 32~Ori group census.

\begin{figure*}
\centering
\includegraphics[width=\textwidth]{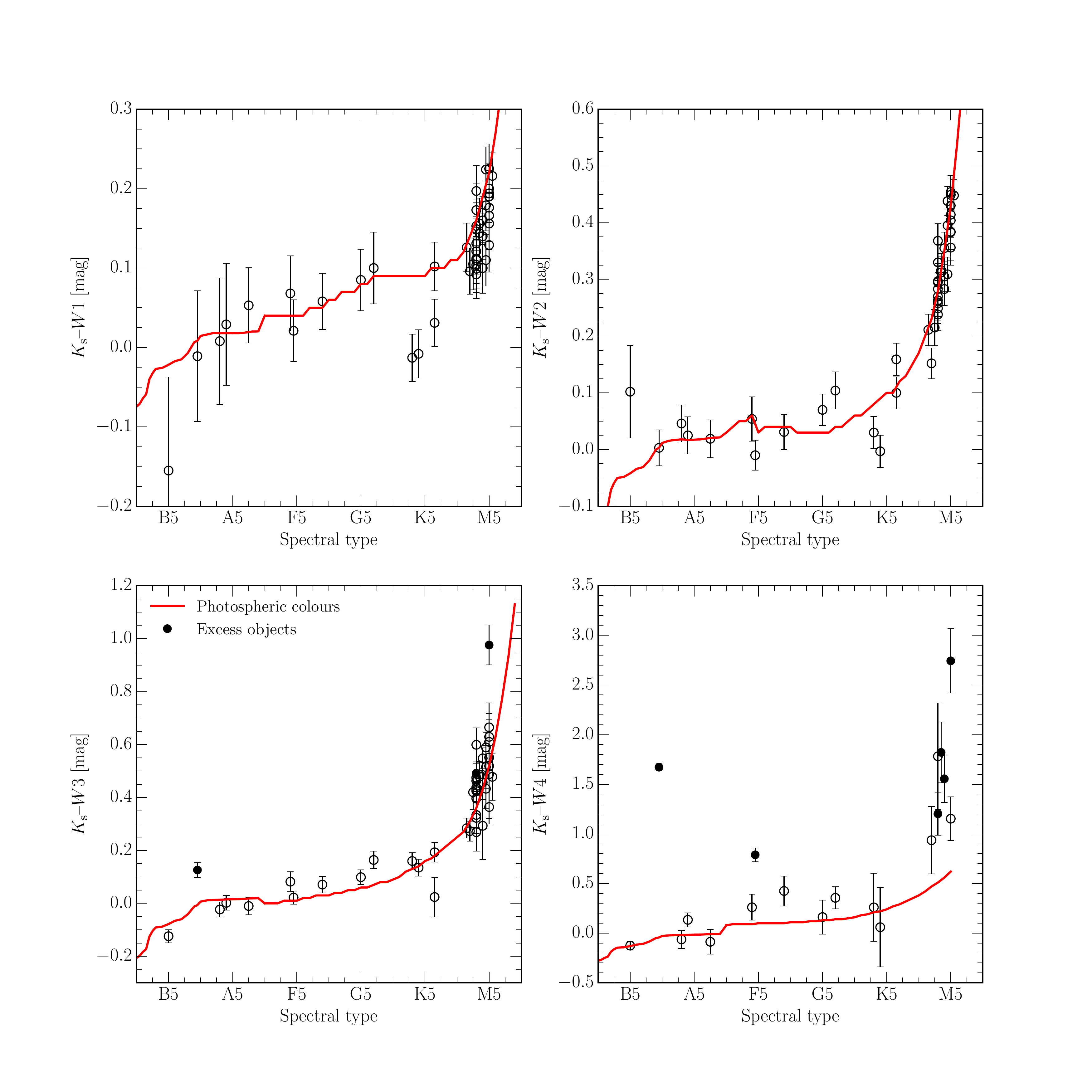}
\caption[]{$K_{\rm{s}}-W1$, $K_{\rm{s}}-W2$, $K_{\rm{s}}-W3$ and
  $K_{\rm{s}}-W4$ infrared colours as a function of spectral type for
  the 45 stellar members of the 32~Ori group (only 19 of which have
  detections in the $W4$ bandpass). The solid line in each panel
  represents the expected intrinsic photospheric colours. For spectral types
  later than F0 these are taken from the pre-MS relations provided by \cite{Pecaut13},
  whereas for earlier spectral types these are based on the synthetic colours
  from the \textsc{atlas9 odfnew} atmospheric models of \cite{Castelli04}.
  Objects deemed to have an infrared excess arising
  from circumstellar material must lie 3$\sigma$ above the
  photospheric colours in a given bandpass as well as every other
  bandpass at longer wavelengths. Six stars are identified as exhibiting
  excess $W4$ emission, three of which also demonstrate $W3$ excesses.}
\label{fig:excess}
\end{figure*}

To identify potential circumstellar discs we cross-matched our
membership against the All\emph{WISE} catalogue \citep{Cutri14}. All
45 stars have counterparts in the $W1$ $(3.4\,\rm{\mu m})$, $W2$
$(4.5\,\rm{\mu m})$ and $W3$ $(12\,\rm{\mu m})$ bands, however only 19
were detected in the $W4$ $(22\,\rm{\mu m})$
band. Fig.~\ref{fig:excess} plots the $K_{\rm{s}}-W1$,
$K_{\rm{s}}-W2$, $K_{\rm{s}}-W3$ and $K_{\rm{s}}-W4$ infrared colours
as a function of spectral type for our 32~Ori group members. To
determine whether a given star exhibits an infrared excess we use the
pre-MS colour sequence of \cite{Pecaut13}, which essentially defines
photospheric colours for stars with ages of $\simeq$\,5--40\,Myr. Due to a
combination of saturation effects and a paucity of early-type stars within
the Local Bubble, the \citeauthor{Pecaut13} photospheric colours only cover
spectral types of F0 and later in the \emph{WISE} bandpasses. For earlier
spectral types we adopted the \textsc{atlas9} \textsc{odfnew} synthetic
colour indices as calculated by \citeauthor{Pecaut13}, which are based on
the atmospheric models of \cite{Castelli04} and for which we assumed
log$\,g=4.5$\,dex, and interpolated within these for the $T_{\rm{eff}}$
corresponding to the specific spectral type as prescribed by the dwarf
spectral type-$T_{\rm{eff}}$ relation of \citeauthor{Pecaut13}. We
define an excess in a given \emph{WISE} band as any object which lies
greater than 3$\sigma$ above the \citeauthor{Pecaut13} relation, where
$\sigma$ is the photometric uncertainty on the observed colour. We
stipulate, however, that for a source to be labelled an excess object
it must also exhibit excesses in each of the longer wavelength
\emph{WISE} bandpasses.

\begin{figure*}
\centering
\includegraphics[width=\textwidth]{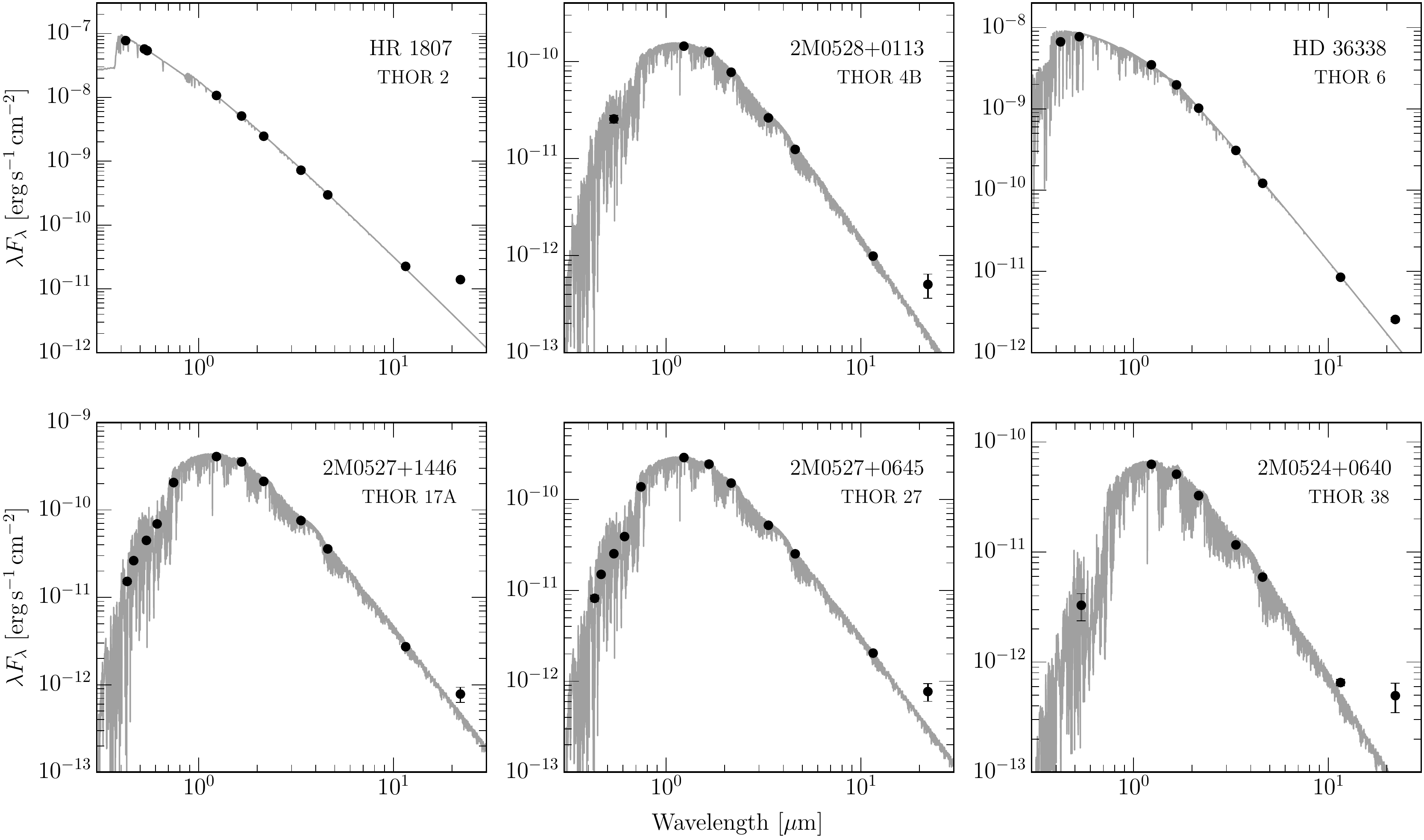}
\caption[]{Spectral energy distributions of six members of the
  32~Ori group for which we have identified excess $W4$ emission at
  22\,$\mu$m. Note that the uncertainties on the individual points are
  typically smaller than the symbols used. In all cases except
  HR~1807, for which we used the \textsc{atlas9} \textsc{odfnew}
  models, the BT-Settl CIFIST models have been adopted to compare
  against the photometric data.}
\label{fig:sed}
\end{figure*}

The terminology of circumstellar discs can vary depending on which
criteria one adopts, thus making like-for-like comparisons between
different regions problematic. We therefore adopt the observational
criteria described in \cite{Luhman12} which has been used in several
recent studies of nearby young moving groups and associations (see
e.g. \citealp{Kraus14,Pecaut16}). Following \citeauthor{Luhman12}, we
can eliminate any full, transitional or evolved discs which all
require an excess of $E(K_{\rm{s}}-W4) > 3.2\,\rm{mag}$, yielding a
disc fraction of $<$\,3.9 per cent (68 per cent confidence level) for
such discs. Debris discs are classified as objects with excesses of
$E(K_{\rm{s}}-W4) < 3.2\,\rm{mag}$, which comfortably covers all of
the excess objects identified in the bottom-right panel of
Fig.~\ref{fig:excess}. Of the 19 objects with $W4$ detections, we
identify six -- HR~1807 (THOR 2), 2MASS J05284050+0113333
(THOR 4B), HD~36338 (THOR 6), 2MASS J05274313+1446121
(THOR 17A), 2MASS J05274855+0645459 (THOR 27) and
2MASS J05243009+0640349 (THOR 38) --
as exhibiting excess emission and derive a debris disc
fraction of $32^{+12}_{-8}$ per cent based on binomial statistics (see \citealp{Cameron11}).
Of these six stars, THOR 2, 17A and 38 also display varying degrees of excess
$W3$ emission.
The mid-infrared All\emph{WISE} photometry and excesses for these debris
disc candidates are listed in Table~\ref{tab:excesses}. In Fig.~\ref{fig:sed} we show
the SEDs for these objects, created
from a combination of optical, near- and mid-IR photometric data
including \emph{Tycho}-2 $(BV_{\rm{T}})$, APASS DR9 $BVgri$, 2MASS
$JHK_{\rm{s}}$ and All\emph{WISE} $W1$--$W4$. Using the Virtual Observatory SED
Analyser (VOSA; \citealp{Bayo08}) we fitted the observed SEDs with the
solar-metallicity BT-Settl CIFIST atmospheric models of
\cite*{Allard11}, computed adopting the revised solar abundances of
\cite{Caffau11}. Note that due to the upper limit of 7000\,K on the
CIFIST models, in the case of HR~1807 we instead used the
\textsc{atlas9} \textsc{odfnew} models of \cite{Castelli04}.

\begin{landscape}
\begin{table}
\caption[]{Mid-infrared photometry and excesses for 32 Ori group debris disc candidates. \label{tab:excesses}}
\begin{tabular}{r l c c c c c c c c}
\hline
THOR   &    WISE J                          &   $W1$                       &   $W2$                    &   $W3$                      &   $W4$                     &   $E(K_{\rm{s}}-W1)$   &   $E(K_{\rm{s}}-W2)$   &   $E(K_{\rm{s}}-W3)$   &   $E(K_{\rm{s}}-W4)$\\ 
\#            &   designation                     &   (mag)                       &   (mag)                    &   (mag)                      &   (mag)                     &   (mag)                         &   (mag)                          &   (mag)                         &              (mag)\\
\hline
2 &   052638.83+065206.9     &   $6.449\pm0.079$    &   $6.435\pm0.022$  &   $6.312\pm0.016$   &   $4.766\pm0.031$   &   $-0.019\pm0.082$   &   $0.000\pm0.032$   &   $0.132\pm0.028$   &   $1.715\pm0.039$\\
4B    &   052840.51+011333.1   &   $10.047\pm0.023$   &   $9.879\pm0.020$   &   $9.705\pm0.042$   &   $8.370\pm0.304$   &   $-0.031\pm0.031$   &   $-0.003\pm0.029$   &   $0.091\pm0.047$   &   $1.286\pm0.305$\\
6          &   053115.70+053946.1      &   $7.374\pm0.035$   &   $7.405\pm0.020$   &   $7.373\pm0.018$   &   $6.606\pm0.067$   &   $-0.019\pm0.039$   &   $-0.055\pm0.026$   &   $0.012\pm0.025$   &   $0.694\pm0.069$\\
17A    &   052743.14+144611.7   &   $8.899\pm0.023$   &   $8.728\pm0.021$   &   $8.604\pm0.028$   &   $7.894\pm0.216$   &   $0.037\pm0.032$   &   $0.088\pm0.030$   &   $0.132\pm0.036$   &   $0.692\pm0.217$\\
27    &   052748.55+064545.6   &   $9.306\pm0.023$   &   $9.111\pm0.020$   &   $8.918\pm0.034$   &   $7.911\pm0.238$   &   $-0.030\pm0.030$   &   $0.005\pm0.028$   &   $0.118\pm0.039$   &   $0.995\pm0.239$\\
38    &   052430.10+064034.6   &   $10.933\pm0.023$   &   $10.684\pm0.020$   &   $10.157\pm0.072$   &   $8.390\pm0.324$   &   $0.020\pm0.032$   &   $0.019\pm0.030$   &   $0.456\pm0.075$   &   $2.123\pm0.325$\\
\hline
\end{tabular}
\end{table}
\end{landscape}

The obvious dataset against which to compare our census of
circumstellar discs is the recently published \emph{Spitzer} survey
by \cite{Shvonski16}. Our \emph{WISE} analysis corroborates that
of \citeauthor{Shvonski16} with neither study finding evidence for the
presence of full or warm dusty discs based on a lack of excess detections
in all four IRAC bands as well as the \emph{WISE} $W1$ and $W2$ bands. Of the four debris disc
candidates identified by \citeauthor{Shvonski16}, we find that both HR~1807 (THOR 2)
and HD~36338 (THOR 6) are $22\,\rm{\mu m}$ excess sources, however we do not
detect any excess emission from either HD~35499 (THOR 5) or TYC-112-1486-1 (THOR 10),
despite \citeauthor{Shvonski16} determining both exhibit $24\,\rm{\mu m}$
emission at $>4\sigma$ above typical photospheric values.
The cause of this discrepancy is unclear as neither
object is flagged as variable in the All\emph{WISE} catalogue (but see \citealp{Melis12})
and visual inspection of the co-added images suggest nothing unusual (e.g. nearby
sources, non-uniform background). In the case of TYC 112-1486-1 this could be
a sensitivity issue as the signal-to-noise ratio of the $W4$-band measurement is
only $\sim$3. The remaining four debris disc candidates
identified in this study are all new members and thus were not included in
the \citeauthor{Shvonski16} study.

Of the six objects identified as exhibiting $W4$ excesses,
the four M dwarf debris disc candidates are of particular note. Studies
have demonstrated that as stellar mass decreases, so does the fraction of
associated debris discs (see e.g. \citealp{Lestrade09}). To date,
only three M dwarf debris discs have been confirmed via scattered
light observations, all with ages similar to or younger than the 32~Ori
group (AU~Mic, TWA~7 and TWA~25; see
\citealp{Choquet16}). If the new debris disc candidates presented in this study
(as well as those in the BPMG; see \citealp{Binks16b}\footnote{Recently, \cite{Silverberg16}
claimed to have identified an M dwarf debris disc candidate
(WISE J080822.18-644357.3) in the 45\,Myr-old Carina
association. Their membership is based solely on the SPM4 proper motion,
which when combined with the BANYAN II Bayesian membership tool \citep{Gagne14}, provides a membership probability of
93.9 per cent. Given this reliance on a single observable, it is therefore perturbing to note that the PPMXL
$\mu_{\delta}$ value differs from that provided by SPM4 by $\simeq$\,19\,$\rm{mas\,yr^{-1}}$.
Prior to confirming membership, we await an improved proper motion and spectroscopic observations of the star,
most importantly a radial velocity in agreement with the $\sim$\,21\,\kms\ expected of a Carina member at that position.
})
can be confirmed by ancillary
mid- and far-IR observations, then these would not only represent some
of the oldest and lowest-mass stars with such discs, but they would also
provide ideal targets
for follow-up direct imaging which may help to enhance our understanding
of the processes governing the production and dynamics of dust and
planetesimals in such systems during an epoch which is believed to be
important for terrestrial planet formation.

\begin{figure}
\centering
\includegraphics[width=\columnwidth]{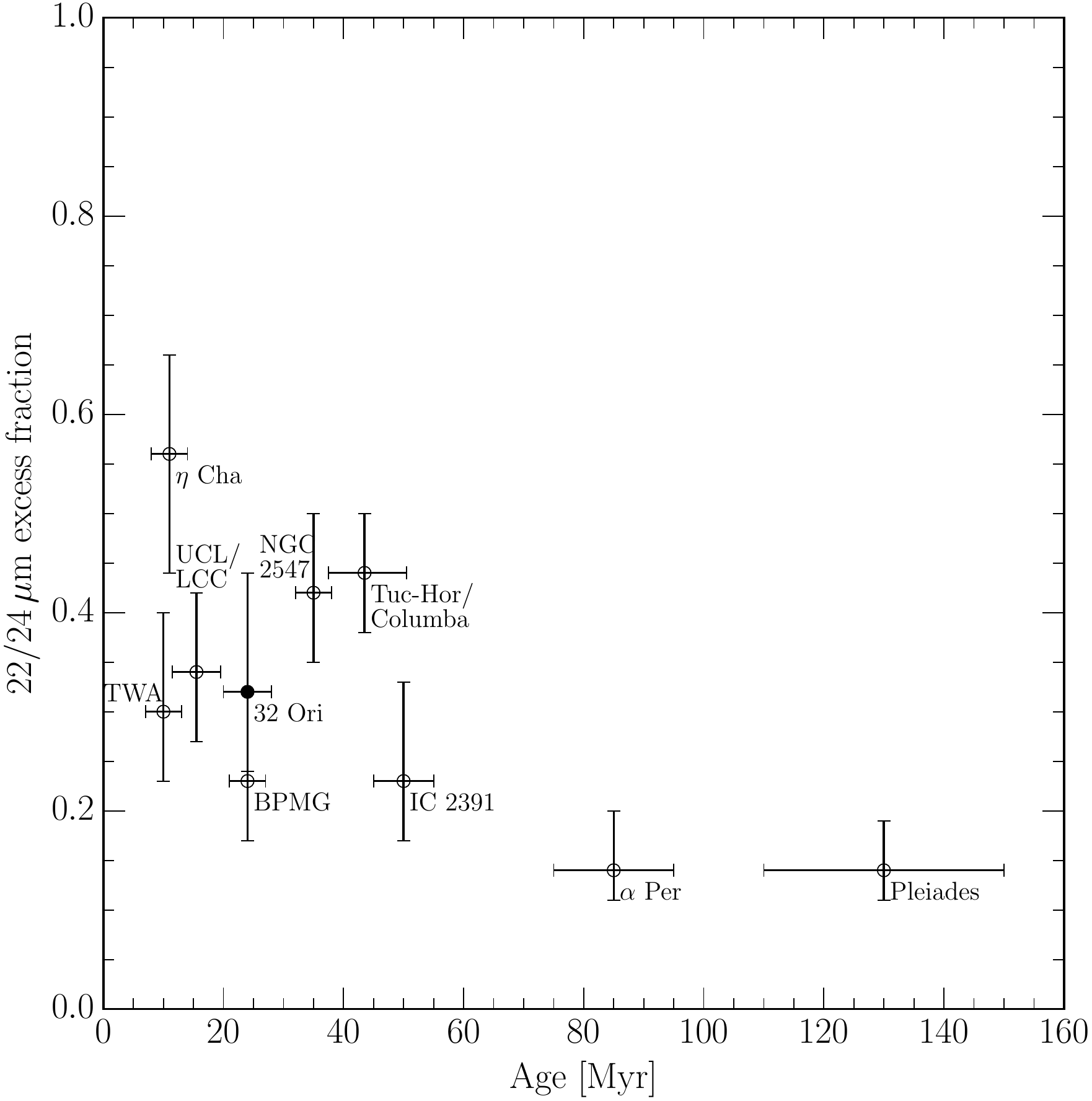}
\caption[]{Debris disc fraction (as estimated from excess emission
at either 22 or $24\,\rm{\mu m}$) as a function of age for young nearby
groups/associations and clusters. The disc fractions for all other regions
have been taken from \cite{Zuckerman12}.}
\label{fig:disc_fraction}
\end{figure}

Fig.~\ref{fig:disc_fraction} shows the debris disc fraction of the
32 Ori group compared to other nearby young groups/associations
and clusters (based on excess emission at either 22 or $24\,\rm{\mu m}$).
The disc fractions for all other regions have been taken from the
compilation of \cite{Zuckerman12} and the uncertainties calculated as
above using binomial statistics. The ages and uncertainties have been compiled
from \cite*{Barrado04b} for IC~2391, $\alpha$~Per and the Pleiades,
\cite{Jeffries05} for NGC~2547, \cite{Bell15} for TWA, $\eta$~Cha, the BPMG and
Tuc-Hor/Columba, and \cite{Pecaut16} for UCL/LCC.

\subsection{Spatial structure of the 32~Ori group}
\label{spatial_structure}

\begin{figure*}
\centering
\includegraphics[width=\textwidth]{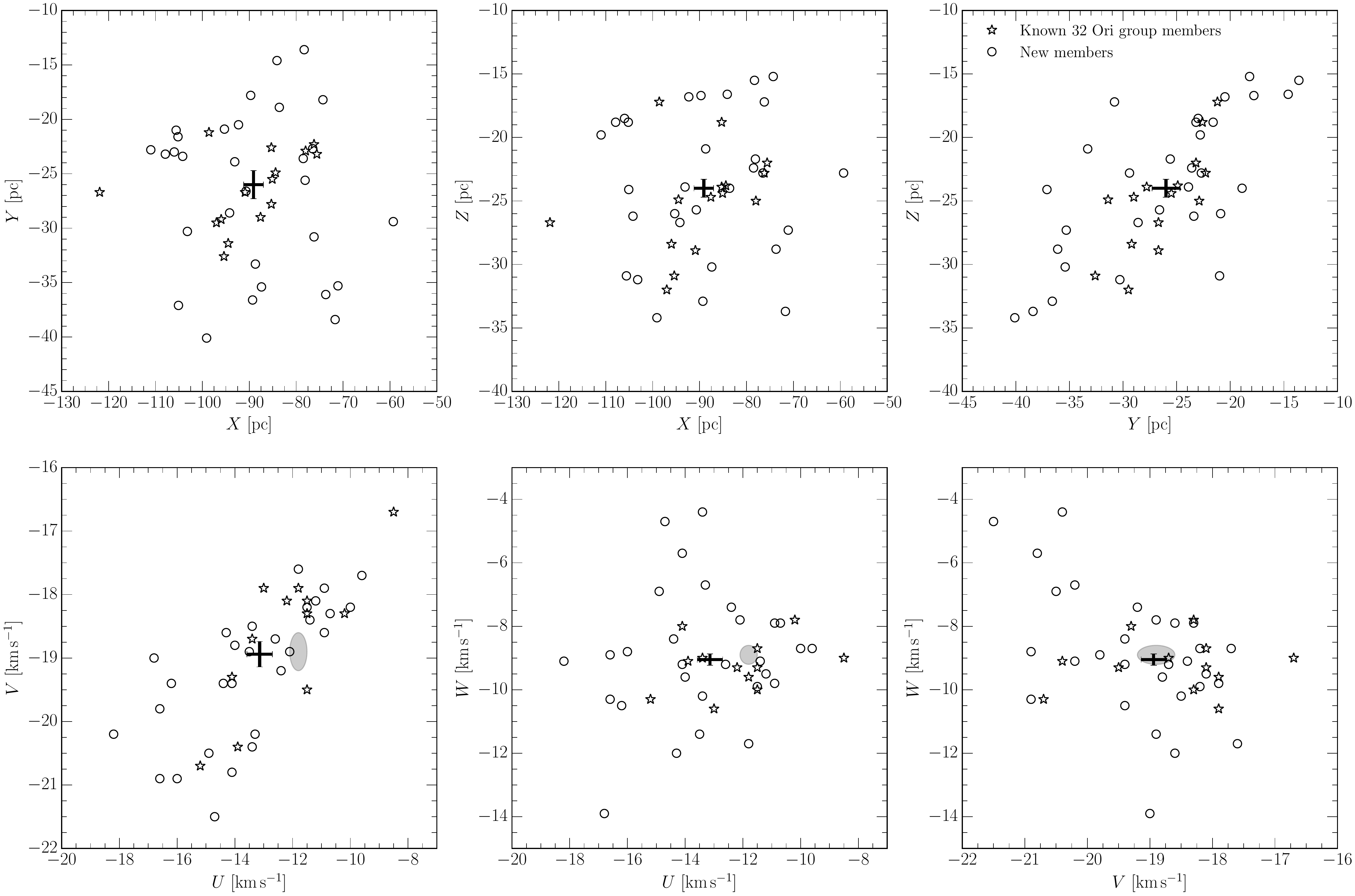}
\caption[]{$XYZ$ spatial (top row) and $UVW$ velocity (bottom row)
  distribution for members of the 32~Ori group. Note that the known
  members HR~1807, HD~35714, HD~36823 and the new member HD~36002 are
  not included in the velocity plots because of unreliable or unavailable literature radial velocities.  The black crosses
  represent the mean 32~Ori group space position and velocity of all members with the
  error bars corresponding to the standard error of the mean. The
  shaded region denotes the mean group $UVW$ velocity as listed in Table~\ref{tab:group_mean}.}
\label{fig:xyzuvw}
\end{figure*}

Fig.~\ref{fig:xyzuvw} shows the $XYZ$ spatial and $UVW$ velocity
distributions for the 45 stellar members of the 32~Ori group. Note
that due to unreliable literature radial velocities for both HR~1807
and HD~35714, and missing radial velocities for HD~36823 and
HD~36002 (SB2), these stars are not
shown in the $UVW$ panels of Fig.~\ref{fig:xyzuvw}. From these distributions
we estimate a mean group position of $(X,Y,Z) = (-89.1\pm2.1, -26.0\pm1.3, -24.0\pm0.7$)\,pc
and velocity of $(U,V,W) = (-13.1\pm0.4, -18.9\pm0.2, -9.0\pm0.2)\,\rm{km\,s^{-1}}$,
where the uncertainties on each represent the standard error of the mean.
Due to the use of kinematic distances for the majority of the stars
shown in Fig.~\ref{fig:xyzuvw}, we would advise that, until additional
\emph{Gaia} parallaxes become available, the mean group $UVW$
velocity listed in Table~\ref{tab:group_mean} be adopted for future
searches for additional members.

Examining the $XYZ$ distribution we see that the geometry of the
32~Ori group is broadly ellipsoidal and elongated toward the Galactic
centre, with ($\Delta_{X}, \Delta_{Y}, \Delta_{Z}$)\,$\sim$\,$(60, 25,
20)$\,pc. This is quite different from more filamentary/sheet-like
geometries like those observed in TWA and Tuc-Hor (see
\citealp*{Weinberger13}; \citealp{Kraus14}). Based on our derived age
for the group, and assuming that all members formed within a region
$\ll$1\,pc, the current dispersion in the $XYZ$ plane suggests a
one-dimensional internal velocity dispersion on the order of
1--$2\,\rm{km\,s^{-1}}$, which is consistent with those of other young
moving groups and associations in the Local Bubble \citep{Mamajek16}.
Furthermore, as argued by \cite{Kraus14} for Tuc-Hor, the age of the
32 Ori group is much less than one Galactic orbital period and so we
would not expect the tidal field of the Milky Way to have
significantly influenced the current geometry of the group. Hence, the
ellipsoidal shape more likely reflects the initial star formation
conditions in which the group formed and could be more indicative of
an originally compact cluster which has since become gravitationally
unbound and is slowly dispersing into the Galactic field, as proposed
for the $\epsilon$~Cha group \citep*{Murphy13}.

\section{Conclusions}
\label{conclusions}

We have undertaken the first large-scale systematic stellar census of
the nearby but poorly-studied 32~Orionis group. The main
results from this study are as follows:

\begin{enumerate}
  
\item Based on spectroscopic follow-up of candidate members selected
  from UCAC4 and URAT1, we have identified a total of 30 new members
  of the 32~Ori group; 29 M dwarf members and one A-type star which
  forms a co-moving common proper motion pair with one of the new members.
    Members were confirmed by combining Li
  absorption, H$\alpha$ emission and radial velocity information with
  kinematic distances and proper motions. This study has increased the
  number of known group members by a factor of three, bringing the
  total number of 32 Ori group members to 46.

\item We have unambiguously identified the Li depletion boundary (LDB)
  of the 32~Ori group. Using stellar evolutionary models we derive
  ages from both isochronal fitting ($25\pm5\,\rm{Myr}$) and LDB
  analyses ($23\pm4\,\rm{Myr}$), which we combine to calculate a final
  adopted age for the 32~Ori group of $24\pm4\,\rm{Myr}$
  ($\pm4\,\rm{Myr}$ statistical, $\pm2\,\rm{Myr}$ systematic). This
  age implies that the 32~Ori group is coeval with the somewhat closer
  $\beta$~Pictoris moving group.

\item We have searched for the presence of circumstellar discs around
  the 45 stellar members of the 32~Ori group using the
  All\emph{WISE} catalogue. As with other groups and associations of
  similar age, we find no evidence for prevailing warm, dusty discs,
  however we have identified several possible debris discs based on
  excess emission in the \emph{WISE} $W4$-band at 22~$\mu$m. From our
  limited sample of stars with $W4$ detections, we estimate a debris
  disc fraction of $32^{+12}_{-8}$ per cent for the group.
  
\end{enumerate}

\section*{Acknowledgements}
CPMB acknowledges support from the Swiss National Science Foundation
(SNSF).
SJM acknowledges support from a University of New South Wales Vice
Chancellor's Fellowship.
EEM acknowledges support from the University of Rochester School of
Arts and Sciences, NSF grant AST-1313029, and the NASA NExSS program.
Part of this research was carried out at the Jet Propulsion
Laboratory, California Institute of Technology, under a contract with
the National Aeronautics and Space Administration.
The authors would like to thank the ANU TAC for the generous
allocation of telescope time which made this study possible as well as
Alex Binks for discussions regarding LDB ages, and John Stauffer and 
Sarah Schmidt for discussions regarding the emission feature seen in 2M0536+1300.
The authors would also like to thank the referee Ben Zuckerman for comments
which have improved the paper.
This research has made extensive use of the VizieR and SIMBAD services
provided by CDS as well as the Tool for OPerations on Catalogues And
Tables (TOPCAT) software package \citep{Taylor05}.
This publication makes use of VOSA, developed under the Spanish
Virtual Observatory project supported from the Spanish MICINN through
grant AyA2011-24052.
This work has made use of data from the European Space Agency (ESA)
mission {\it Gaia} (\url{http://www.cosmos.esa.int/gaia}), processed
by the {\it Gaia} Data Processing and Analysis Consortium (DPAC,
\url{http://www.cosmos.esa.int/web/gaia/dpac/consortium}).
Funding for the DPAC has been provided by national institutions, in
particular the institutions participating in the {\it Gaia}
Multilateral Agreement.

\bibliographystyle{mn3e}
\bibliography{references}

\begin{thebibliography}{120}
\providecommand{\natexlab}[1]{#1}

\bibitem[{{Abt} \& {Morrell}(1995)}]{Abt95}
{Abt} H.~A., {Morrell} N.~I., 1995, \apjs, 99, 135

\bibitem[{{Alcal{\'a}} et~al.(2000){Alcal{\'a}}, {Covino}, {Torres}, {Sterzik},
  {Pfeiffer} \& {Neuh{\"a}user}}]{Alcala00}
{Alcal{\'a}} J.~M., {Covino} E., {Torres} G., {Sterzik} M.~F., {Pfeiffer}
  M.~J., {Neuh{\"a}user} R., 2000, \aap, 353, 186

\bibitem[{{Alcal{\'a}} et~al.(1996)}]{Alcala96}
{Alcal{\'a}} J.~M. et~al., 1996, \aaps, 119, 7

\bibitem[{{Allard} et~al.(2011){Allard}, {Homeier} \& {Freytag}}]{Allard11}
{Allard} F., {Homeier} D., {Freytag} B., 2011, in C.~{Johns-Krull}, M.K.
  {Browning}, A.A. {West}, eds, 16th Cambridge Workshop on Cool Stars, Stellar
  Systems, and the Sun. Astronomical Society of the Pacific Conference Series,
  Vol. 448, p.~91

\bibitem[{{Baraffe} et~al.(2015){Baraffe}, {Homeier}, {Allard} \&
  {Chabrier}}]{Baraffe15}
{Baraffe} I., {Homeier} D., {Allard} F., {Chabrier} G., 2015, \aap, 577, 42

\bibitem[{{Barbier-Brossat} \& {Figon}(2000)}]{Barbier-Brossat00}
{Barbier-Brossat} M., {Figon} P., 2000, \aaps, 142, 217

\bibitem[{{Barrado y Navascu{\'e}s} et~al.(2004){Barrado y Navascu{\'e}s},
  {Stauffer} \& {Jayawardhana}}]{Barrado04b}
{Barrado y Navascu{\'e}s} D., {Stauffer} J.~R., {Jayawardhana} R., 2004, \apj,
  614, 386

\bibitem[{{Bayo} et~al.(2008){Bayo}, {Rodrigo}, {Barrado Y Navascu{\'e}s},
  {Solano}, {Guti{\'e}rrez}, {Morales-Calder{\'o}n} \& {Allard}}]{Bayo08}
{Bayo} A., {Rodrigo} C., {Barrado Y Navascu{\'e}s} D., {Solano} E.,
  {Guti{\'e}rrez} R., {Morales-Calder{\'o}n} M., {Allard} F., 2008, \aap, 492,
  277

\bibitem[{{Bell} et~al.(2013){Bell}, {Naylor}, {Mayne}, {Jeffries} \&
  {Littlefair}}]{Bell13}
{Bell} C.~P.~M., {Naylor} T., {Mayne} N.~J., {Jeffries} R.~D., {Littlefair}
  S.~P., 2013, \mnras, 434, 806

\bibitem[{{Bell} et~al.(2014){Bell}, {Rees}, {Naylor}, {Mayne}, {Jeffries},
  {Mamajek} \& {Rowe}}]{Bell14}
{Bell} C.~P.~M., {Rees} J.~M., {Naylor} T., {Mayne} N.~J., {Jeffries} R.~D.,
  {Mamajek} E.~E., {Rowe} J., 2014, \mnras, 445, 3496

\bibitem[{{Bell} et~al.(2015){Bell}, {Mamajek} \& {Naylor}}]{Bell15}
{Bell} C.~P.~M., {Mamajek} E.~E., {Naylor} T., 2015, \mnras, 454, 593

\bibitem[{{Bessell}(1999)}]{Bessell99}
{Bessell} M.~S., 1999, \pasp, 111, 1426

\bibitem[{{Bhatt} \& {Cami}(2015)}]{Bhatt15}
{Bhatt} N.~H., {Cami} J., 2015, \apjs, 216, 22

\bibitem[{{Bilir} et~al.(2008){Bilir}, {Ak}, {Karaali}, {Cabrera-Lavers},
  {Chonis} \& {Gaskell}}]{Bilir08}
{Bilir} S., {Ak} S., {Karaali} S., {Cabrera-Lavers} A., {Chonis} T.~S.,
  {Gaskell} C.~M., 2008, \mnras, 384, 1178

\bibitem[{{Binks} \& {Jeffries}(2014)}]{Binks14}
{Binks} A.~S., {Jeffries} R.~D., 2014, \mnras, 438, L11

\bibitem[{{Binks} \& {Jeffries}(2016{\natexlab{a}})}]{Binks16b}
{Binks} A.~S., {Jeffries} R.~D., 2016{\natexlab{a}}, ArXiv e-prints

\bibitem[{{Binks} \& {Jeffries}(2016{\natexlab{b}})}]{Binks16}
{Binks} A.~S., {Jeffries} R.~D., 2016{\natexlab{b}}, \mnras, 455, 3345

\bibitem[{{Bobylev} et~al.(2006){Bobylev}, {Goncharov} \&
  {Bajkova}}]{Bobylev06}
{Bobylev} V.~V., {Goncharov} G.~A., {Bajkova} A.~T., 2006, Astronomy Reports,
  50, 733

\bibitem[{{Bochanski} et~al.(2007){Bochanski}, {West}, {Hawley} \&
  {Covey}}]{Bochanski07}
{Bochanski} J.~J., {West} A.~A., {Hawley} S.~L., {Covey} K.~R., 2007, \aj, 133,
  531

\bibitem[{{Bouy} \& {Alves}(2015)}]{Bouy15}
{Bouy} H., {Alves} J., 2015, \aap, 584, A26

\bibitem[{{Bressan} et~al.(2012){Bressan}, {Marigo}, {Girardi}, {Salasnich},
  {Dal Cero}, {Rubele} \& {Nanni}}]{Bressan12}
{Bressan} A., {Marigo} P., {Girardi} L., {Salasnich} B., {Dal Cero} C.,
  {Rubele} S., {Nanni} A., 2012, \mnras, 427, 127

\bibitem[{{Burgasser} et~al.(2016)}]{Burgasser16}
{Burgasser} A.~J. et~al., 2016, \apj, 820, 32

\bibitem[{{Caffau} et~al.(2011){Caffau}, {Ludwig}, {Steffen}, {Freytag} \&
  {Bonifacio}}]{Caffau11}
{Caffau} E., {Ludwig} H.~G., {Steffen} M., {Freytag} B., {Bonifacio} P., 2011,
  \solphys, 268, 255

\bibitem[{{Cameron}(2011)}]{Cameron11}
{Cameron} E., 2011, \pasa, 28, 128

\bibitem[{{Canup}(2004)}]{Canup04}
{Canup} R.~M., 2004, \araa, 42, 441

\bibitem[{{Cardelli} et~al.(1989){Cardelli}, {Clayton} \&
  {Mathis}}]{Cardelli89}
{Cardelli} J.~A., {Clayton} G.~C., {Mathis} J.~S., 1989, \apj, 345, 245

\bibitem[{{Castelli} \& {Kurucz}(2004)}]{Castelli04}
{Castelli} F., {Kurucz} R.~L., 2004, in N.E. {Piskunov}, W.W. {Weiss}, D.F.
  {Gray}, eds, Modelling of Stellar Atmospheres. IAU Symposium, Vol. 210, p.~20

\bibitem[{{Choquet} et~al.(2016)}]{Choquet16}
{Choquet} {\'E}. et~al., 2016, \apjl, 817, L2

\bibitem[{{Cutri} \& {et al.}(2014)}]{Cutri14}
{Cutri} R.~M., {et al.}, 2014, VizieR Online Data Catalog, 2328

\bibitem[{{Cutri} et~al.(2003){Cutri}, {Skrutskie}, {van Dyk} \& {et
  al.}}]{Cutri03}
{Cutri} R.~M., {Skrutskie} M.~F., {van Dyk} S., {et al.}, 2003, {2MASS Point
  Source Catalogue. Available at \url{http://www.ipac.caltech.edu/2mass/}}

\bibitem[{{da Silva} et~al.(2009){da Silva}, {Torres}, {de La Reza}, {Quast},
  {Melo} \& {Sterzik}}]{daSilva09}
{da Silva} L., {Torres} C.~A.~O., {de La Reza} R., {Quast} G.~R., {Melo}
  C.~H.~F., {Sterzik} M.~F., 2009, \aap, 508, 833

\bibitem[{{de la Reza} et~al.(1989){de la Reza}, {Torres}, {Quast}, {Castilho}
  \& {Vieira}}]{delaReza89}
{de la Reza} R., {Torres} C.~A.~O., {Quast} G., {Castilho} B.~V., {Vieira}
  G.~L., 1989, \apjl, 343, L61

\bibitem[{{Dias} et~al.(2002){Dias}, {Alessi}, {Moitinho} \&
  {L{\'e}pine}}]{Dias02}
{Dias} W.~S., {Alessi} B.~S., {Moitinho} A., {L{\'e}pine} J.~R.~D., 2002, \aap,
  389, 871

\bibitem[{{Dias} et~al.(2014){Dias}, {Monteiro}, {Caetano}, {L{\'e}pine},
  {Assafin} \& {Oliveira}}]{Dias14}
{Dias} W.~S., {Monteiro} H., {Caetano} T.~C., {L{\'e}pine} J.~R.~D., {Assafin}
  M., {Oliveira} A.~F., 2014, \aap, 564, 79

\bibitem[{{Dolan} \& {Mathieu}(2002)}]{Dolan02}
{Dolan} C.~J., {Mathieu} R.~D., 2002, \aj, 123, 387

\bibitem[{{Dopita} et~al.(2007){Dopita}, {Hart}, {McGregor}, {Oates}, {Bloxham}
  \& {Jones}}]{Dopita07}
{Dopita} M., {Hart} J., {McGregor} P., {Oates} P., {Bloxham} G., {Jones} D.,
  2007, \apss, 310, 255

\bibitem[{{Dotter} et~al.(2008){Dotter}, {Chaboyer}, {Jevremovi{\'c}},
  {Kostov}, {Baron} \& {Ferguson}}]{Dotter08}
{Dotter} A., {Chaboyer} B., {Jevremovi{\'c}} D., {Kostov} V., {Baron} E.,
  {Ferguson} J.~W., 2008, \apjs, 178, 89

\bibitem[{{Drake} et~al.(2013)}]{Drake13}
{Drake} A.~J. et~al., 2013, \apj, 763, 32

\bibitem[{{Drake} et~al.(2014)}]{Drake14}
{Drake} A.~J. et~al., 2014, \apjs, 213, 9

\bibitem[{{Ducourant} et~al.(2005){Ducourant}, {Teixeira}, {P{\'e}ri{\'e}},
  {Lecampion}, {Guibert} \& {Sartori}}]{Ducourant05}
{Ducourant} C., {Teixeira} R., {P{\'e}ri{\'e}} J.~P., {Lecampion} J.~F.,
  {Guibert} J., {Sartori} M.~J., 2005, \aap, 438, 769

\bibitem[{{Duerr} et~al.(1982){Duerr}, {Imhoff} \& {Lada}}]{Duerr82}
{Duerr} R., {Imhoff} C.~L., {Lada} C.~J., 1982, \apj, 261, 135

\bibitem[{{Edwards}(1976)}]{Edwards76}
{Edwards} T.~W., 1976, \aj, 81, 245

\bibitem[{{Elliott} et~al.(2014){Elliott}, {Bayo}, {Melo}, {Torres}, {Sterzik}
  \& {Quast}}]{Elliott14}
{Elliott} P., {Bayo} A., {Melo} C.~H.~F., {Torres} C.~A.~O., {Sterzik} M.,
  {Quast} G.~R., 2014, \aap, 568, 26

\bibitem[{{Feiden} \& {Chaboyer}(2013)}]{Feiden13}
{Feiden} G.~A., {Chaboyer} B., 2013, \apj, 779, 183

\bibitem[{{Feiden} \& {Chaboyer}(2014)}]{Feiden14}
{Feiden} G.~A., {Chaboyer} B., 2014, \apj, 789, 53

\bibitem[{{Franciosini} \& {Sacco}(2011)}]{Franciosini11}
{Franciosini} E., {Sacco} G.~G., 2011, \aap, 530, 150

\bibitem[{{Friedemann}(1992)}]{Friedemann92}
{Friedemann} C., 1992, Bulletin d'Information du Centre de Donnees Stellaires,
  40, 31

\bibitem[{{Fuhrmeister} \& {Schmitt}(2004)}]{Fuhrmeister04}
{Fuhrmeister} B., {Schmitt} J.~H.~M.~M., 2004, \aap, 420, 1079

\bibitem[{{Gagn{\'e}} et~al.(2014){Gagn{\'e}}, {Lafreni{\`e}re}, {Doyon},
  {Malo} \& {Artigau}}]{Gagne14}
{Gagn{\'e}} J., {Lafreni{\`e}re} D., {Doyon} R., {Malo} L., {Artigau} {\'E}.,
  2014, \apj, 783, 121

\bibitem[{{Gaia Collaboration} et~al.(2016)}]{Gaia16}
{Gaia Collaboration} et~al., 2016, \aap, 595, 2

\bibitem[{{Gizis} et~al.(2002){Gizis}, {Reid} \& {Hawley}}]{Gizis02}
{Gizis} J.~E., {Reid} I.~N., {Hawley} S.~L., 2002, \aj, 123, 3356

\bibitem[{{Gontcharov}(2006)}]{Gontcharov06}
{Gontcharov} G.~A., 2006, Astronomy Letters, 32, 759

\bibitem[{{Gregorio-Hetem} et~al.(1992){Gregorio-Hetem}, {Lepine}, {Quast},
  {Torres} \& {de La Reza}}]{Gregorio-Hetem92}
{Gregorio-Hetem} J., {Lepine} J.~R.~D., {Quast} G.~R., {Torres} C.~A.~O., {de
  La Reza} R., 1992, \aj, 103, 549

\bibitem[{{Guenther} \& {Emerson}(1997)}]{Guenther97}
{Guenther} E.~W., {Emerson} J.~P., 1997, \aap, 321, 803

\bibitem[{{Hauck} \& {Mermilliod}(1998)}]{Hauck98}
{Hauck} B., {Mermilliod} M., 1998, \aaps, 129, 431

\bibitem[{{Henden} et~al.(2012){Henden}, {Levine}, {Terrell}, {Smith} \&
  {Welch}}]{Henden12}
{Henden} A.~A., {Levine} S.~E., {Terrell} D., {Smith} T.~C., {Welch} D., 2012,
  Journal of the American Association of Variable Star Observers, 40, 430

\bibitem[{{Henden} et~al.(2016){Henden}, {Templeton}, {Terrell}, {Smith},
  {Levine} \& {Welch}}]{Henden16}
{Henden} A.~A., {Templeton} M., {Terrell} D., {Smith} T.~C., {Levine} S.,
  {Welch} D., 2016, VizieR Online Data Catalog: AAVSO Photometric All Sky
  Survey (APASS) DR9, 2336

\bibitem[{{Hohle} et~al.(2010){Hohle}, {Neuh{\"a}user} \& {Schutz}}]{Hohle10}
{Hohle} M.~M., {Neuh{\"a}user} R., {Schutz} B.~F., 2010, Astronomische
  Nachrichten, 331, 349

\bibitem[{{Houdebine} et~al.(1990){Houdebine}, {Foing} \&
  {Rodono}}]{Houdebine90}
{Houdebine} E.~R., {Foing} B.~H., {Rodono} M., 1990, \aap, 238, 249

\bibitem[{{Janson} et~al.(2011){Janson}, {Bonavita}, {Klahr}, {Lafreni{\`e}re},
  {Jayawardhana} \& {Zinnecker}}]{Janson11}
{Janson} M., {Bonavita} M., {Klahr} H., {Lafreni{\`e}re} D., {Jayawardhana} R.,
  {Zinnecker} H., 2011, \apj, 736, 89

\bibitem[{{Jeffries}(2006)}]{Jeffries06b}
{Jeffries} R.~D., 2006, in {Randich, S.~\& Pasquini, L.}, ed., Chemical
  Abundances and Mixing in Stars in the Milky Way and its Satellites. p. 163

\bibitem[{{Jeffries} \& {Oliveira}(2005)}]{Jeffries05}
{Jeffries} R.~D., {Oliveira} J.~M., 2005, \mnras, 358, 13

\bibitem[{{Jeffries} et~al.(2013){Jeffries}, {Naylor}, {Mayne}, {Bell} \&
  {Littlefair}}]{Jeffries13}
{Jeffries} R.~D., {Naylor} T., {Mayne} N.~J., {Bell} C.~P.~M., {Littlefair}
  S.~P., 2013, \mnras, 434, 2438

\bibitem[{{Johnson}(1966)}]{Johnson66}
{Johnson} H.~L., 1966, \araa, 4, 193

\bibitem[{{Kharchenko} et~al.(2013){Kharchenko}, {Piskunov}, {Schilbach},
  {R{\"o}ser} \& {Scholz}}]{Kharchenko13}
{Kharchenko} N.~V., {Piskunov} A.~E., {Schilbach} E., {R{\"o}ser} S., {Scholz}
  R.~D., 2013, \aap, 558, 53

\bibitem[{{Kraus} et~al.(2014){Kraus}, {Shkolnik}, {Allers} \& {Liu}}]{Kraus14}
{Kraus} A.~L., {Shkolnik} E.~L., {Allers} K.~N., {Liu} M.~C., 2014, \aj, 147,
  146

\bibitem[{{Lagrange} et~al.(2010)}]{Lagrange10}
{Lagrange} A.~M. et~al., 2010, Science, 329, 57

\bibitem[{{Lestrade} et~al.(2009){Lestrade}, {Wyatt}, {Bertoldi}, {Menten} \&
  {Labaigt}}]{Lestrade09}
{Lestrade} J.~F., {Wyatt} M.~C., {Bertoldi} F., {Menten} K.~M., {Labaigt} G.,
  2009, \aap, 506, 1455

\bibitem[{{Luhman} \& {Mamajek}(2012)}]{Luhman12}
{Luhman} K.~L., {Mamajek} E.~E., 2012, \apj, 758, 31

\bibitem[{{Mace} et~al.(2009){Mace}, {Prato}, {Wasserman}, {Schaefer}, {Franz}
  \& {Simon}}]{Mace09}
{Mace} G.~N., {Prato} L., {Wasserman} L.~H., {Schaefer} G.~H., {Franz} O.~G.,
  {Simon} M., 2009, \aj, 137, 3487

\bibitem[{{Malo} et~al.(2013){Malo}, {Doyon}, {Lafreni{\`e}re}, {Artigau},
  {Gagn{\'e}}, {Baron} \& {Riedel}}]{Malo13}
{Malo} L., {Doyon} R., {Lafreni{\`e}re} D., {Artigau} {\'E}., {Gagn{\'e}} J.,
  {Baron} F., {Riedel} A., 2013, \apj, 762, 88

\bibitem[{{Malo} et~al.(2014)}]{Malo14b}
{Malo} L., {Doyon} R., {Feiden} G.~A., {Albert} L., {Lafreni{\`e}re} D.,
  {Artigau} {\'E}., {Gagn{\'e}} J., {Riedel} A., 2014, \apj, 792, 37

\bibitem[{{Mamajek}(2007)}]{Mamajek07}
{Mamajek} E.~E., 2007, in B.G. {Elmegreen}, J.~{Palous}, eds, Triggered Star
  Formation in a Turbulent ISM. IAU Symposium, Vol. 237, p. 442

\bibitem[{{Mamajek}(2016)}]{Mamajek16}
{Mamajek} E.~E., 2016, in J.H. {Kastner}, B.~{Stelzer}, S.A. {Metchev}, eds,
  Young Stars and Planets Near the Sun. IAU Symposium, Vol. 314, p.~21

\bibitem[{{Mamajek} \& {Bell}(2014)}]{Mamajek14}
{Mamajek} E.~E., {Bell} C.~P.~M., 2014, \mnras, 445, 2169

\bibitem[{{Mamajek} et~al.(2006){Mamajek}, {Meyer} \& {Liebert}}]{Mamajek06}
{Mamajek} E.~E., {Meyer} M.~R., {Liebert} J., 2006, \aj, 131, 2360

\bibitem[{{Melis} et~al.(2012){Melis}, {Zuckerman}, {Rhee}, {Song}, {Murphy} \&
  {Bessell}}]{Melis12}
{Melis} C., {Zuckerman} B., {Rhee} J.~H., {Song} I., {Murphy} S.~J., {Bessell}
  M.~S., 2012, \nat, 487, 74

\bibitem[{{Melis} et~al.(2014){Melis}, {Reid}, {Mioduszewski}, {Stauffer} \&
  {Bower}}]{Melis14}
{Melis} C., {Reid} M.~J., {Mioduszewski} A.~J., {Stauffer} J.~R., {Bower}
  G.~C., 2014, Science, 345, 1029

\bibitem[{{Mentuch} et~al.(2008){Mentuch}, {Brandeker}, {van Kerkwijk},
  {Jayawardhana} \& {Hauschildt}}]{Mentuch08}
{Mentuch} E., {Brandeker} A., {van Kerkwijk} M.~H., {Jayawardhana} R.,
  {Hauschildt} P.~H., 2008, \apj, 689, 1127

\bibitem[{{Mermilliod}(2006)}]{Mermilliod06}
{Mermilliod} J.~C., 2006, VizieR Online Data Catalog: Homogeneous Means in the
  $UBV$ System, 2168

\bibitem[{{Metchev} \& {Hillenbrand}(2009)}]{Metchev09}
{Metchev} S.~A., {Hillenbrand} L.~A., 2009, \apjs, 181, 62

\bibitem[{{Murphy} \& {Lawson}(2015)}]{Murphy15}
{Murphy} S.~J., {Lawson} W.~A., 2015, \mnras, 447, 1267

\bibitem[{{Murphy} et~al.(2013){Murphy}, {Lawson} \& {Bessell}}]{Murphy13}
{Murphy} S.~J., {Lawson} W.~A., {Bessell} M.~S., 2013, \mnras, 435, 1325

\bibitem[{{Naylor}(2009)}]{Naylor09}
{Naylor} T., 2009, \mnras, 399, 432

\bibitem[{{Naylor} \& {Jeffries}(2006)}]{Naylor06}
{Naylor} T., {Jeffries} R.~D., 2006, \mnras, 373, 1251

\bibitem[{{Nidever} et~al.(2002){Nidever}, {Marcy}, {Butler}, {Fischer} \&
  {Vogt}}]{Nidever02}
{Nidever} D.~L., {Marcy} G.~W., {Butler} R.~P., {Fischer} D.~A., {Vogt} S.~S.,
  2002, \apjs, 141, 503

\bibitem[{{Palla} et~al.(2007){Palla}, {Randich}, {Pavlenko}, {Flaccomio} \&
  {Pallavicini}}]{Palla07}
{Palla} F., {Randich} S., {Pavlenko} Y.~V., {Flaccomio} E., {Pallavicini} R.,
  2007, \apjl, 659, L41

\bibitem[{{Pecaut} \& {Mamajek}(2013)}]{Pecaut13}
{Pecaut} M.~J., {Mamajek} E.~E., 2013, \apjs, 208, 9

\bibitem[{{Pecaut} \& {Mamajek}(2016)}]{Pecaut16}
{Pecaut} M.~J., {Mamajek} E.~E., 2016, \mnras, 461, 794

\bibitem[{{Pickles}(1998)}]{Pickles98}
{Pickles} A.~J., 1998, \pasp, 110, 863

\bibitem[{{Reis} et~al.(2011){Reis}, {Corradi}, {de Avillez} \&
  {Santos}}]{Reis11}
{Reis} W., {Corradi} W., {de Avillez} M.~A., {Santos} F.~P., 2011, \apj, 734, 8

\bibitem[{{Riedel} et~al.(2016){Riedel}, {Alam}, {Rice}, {Cruz} \&
  {Henry}}]{Riedel16}
{Riedel} A.~R., {Alam} M.~K., {Rice} E.~L., {Cruz} K.~L., {Henry} T.~J., 2016,
  ArXiv e-prints

\bibitem[{{R{\"o}ser} et~al.(2010){R{\"o}ser}, {Demleitner} \&
  {Schilbach}}]{Roeser10}
{R{\"o}ser} S., {Demleitner} M., {Schilbach} E., 2010, \aj, 139, 2440

\bibitem[{{Rucinski} \& {Krautter}(1983)}]{Rucinski83}
{Rucinski} S.~M., {Krautter} J., 1983, \aap, 121, 217

\bibitem[{{Shaya} \& {Olling}(2011)}]{Shaya11}
{Shaya} E.~J., {Olling} R.~P., 2011, \apjs, 192, 2

\bibitem[{{Shvonski} et~al.(2010){Shvonski}, {Mamajek}, {Meyer} \&
  {Kim}}]{Shvonski10}
{Shvonski} A.~J., {Mamajek} E.~E., {Meyer} M.~R., {Kim} J.~S., 2010, in
  American Astronomical Society Meeting Abstracts \#215. Bulletin of the
  American Astronomical Society, Vol.~42, p. \#428.22

\bibitem[{{Shvonski} et~al.(2016){Shvonski}, {Mamajek}, {Kim}, {Meyer} \&
  {Pecaut}}]{Shvonski16}
{Shvonski} A.~J., {Mamajek} E.~E., {Kim} J.~S., {Meyer} M.~R., {Pecaut} M.~J.,
  2016, ArXiv e-prints: 1612.06924

\bibitem[{{Silverberg} et~al.(2016)}]{Silverberg16}
{Silverberg} S.~M. et~al., 2016, \apjl, 830, L28

\bibitem[{{Soderblom} et~al.(1993){Soderblom}, {Jones}, {Balachandran},
  {Stauffer}, {Duncan}, {Fedele} \& {Hudon}}]{Soderblom93}
{Soderblom} D.~R., {Jones} B.~F., {Balachandran} S., {Stauffer} J.~R., {Duncan}
  D.~K., {Fedele} S.~B., {Hudon} J.~D., 1993, \aj, 106, 1059

\bibitem[{{Soderblom} et~al.(2014){Soderblom}, {Hillenbrand}, {Jeffries},
  {Mamajek} \& {Naylor}}]{Soderblom14}
{Soderblom} D.~R., {Hillenbrand} L.~A., {Jeffries} R.~D., {Mamajek} E.~E.,
  {Naylor} T., 2014, Protostars and Planets VI, 219

\bibitem[{{Stauffer} et~al.(2007){Stauffer}, {Hartmann}, {Fazio} \&
  et~al.}]{Stauffer07}
{Stauffer} J.~R., {Hartmann} L.~W., {Fazio} G.~G., et~al., 2007, \apjs, 172,
  663

\bibitem[{{Stauffer} et~al.(1998)}]{Stauffer98a}
{Stauffer} J.~R., {Schild} R., {Barrado y Navascues} D., {Backman} D.~E.,
  {Angelova} A.~M., {Kirkpatrick} J.~D., {Hambly} N., {Vanzi} L., 1998, \apj,
  504, 805

\bibitem[{{Taylor}(2005)}]{Taylor05}
{Taylor} M.~B., 2005, in P.~{Shopbell}, M.~{Britton}, R.~{Ebert}, eds,
  Astronomical Data Analysis Software and Systems XIV. Astronomical Society of
  the Pacific Conference Series, Vol. 347, p.~29

\bibitem[{{Tognelli} et~al.(2015){Tognelli}, {Prada Moroni} \&
  {Degl'Innocenti}}]{Tognelli15}
{Tognelli} E., {Prada Moroni} P.~G., {Degl'Innocenti} S., 2015, \mnras, 449,
  3741

\bibitem[{{Tonry} \& {Davis}(1979)}]{Tonry79}
{Tonry} J., {Davis} M., 1979, \aj, 84, 1511

\bibitem[{{Torres} et~al.(2000){Torres}, {da Silva}, {Quast}, {de la Reza} \&
  {Jilinski}}]{Torres00}
{Torres} C.~A.~O., {da Silva} L., {Quast} G.~R., {de la Reza} R., {Jilinski}
  E., 2000, \aj, 120, 1410

\bibitem[{{Torres} et~al.(2008){Torres}, {Quast}, {Melo} \&
  {Sterzik}}]{Torres08}
{Torres} C.~A.~O., {Quast} G.~R., {Melo} C.~H.~F., {Sterzik} M.~F., 2008,
  {Handbook of Star Forming Regions: Volume II, The Southern Sky}. p. 757

\bibitem[{{van Leeuwen}(2007)}]{vanLeeuwen07}
{van Leeuwen} F., 2007, \aap, 474, 653

\bibitem[{{Voges} et~al.(1999)}]{Voges99}
{Voges} W. et~al., 1999, \aap, 349, 389

\bibitem[{{Webb} et~al.(1999){Webb}, {Zuckerman}, {Platais}, {Patience},
  {White}, {Schwartz} \& {McCarthy}}]{Webb99}
{Webb} R.~A., {Zuckerman} B., {Platais} I., {Patience} J., {White} R.~J.,
  {Schwartz} M.~J., {McCarthy} C., 1999, \apjl, 512, L63

\bibitem[{{Weinberger} et~al.(2013){Weinberger}, {Anglada-Escud{\'e}} \&
  {Boss}}]{Weinberger13}
{Weinberger} A.~J., {Anglada-Escud{\'e}} G., {Boss} A.~P., 2013, \apj, 762, 118

\bibitem[{{West} et~al.(2011)}]{West11}
{West} A.~A. et~al., 2011, \aj, 141, 97

\bibitem[{{White} et~al.(2007){White}, {Gabor} \& {Hillenbrand}}]{White07}
{White} R.~J., {Gabor} J.~M., {Hillenbrand} L.~A., 2007, \aj, 133, 2524

\bibitem[{{Zacharias} et~al.(2005){Zacharias}, {Monet}, {Levine}, {Urban},
  {Gaume} \& {Wycoff}}]{Zacharias05}
{Zacharias} N., {Monet} D.~G., {Levine} S.~E., {Urban} S.~E., {Gaume} R.,
  {Wycoff} G.~L., 2005, VizieR Online Data Catalog: NOMAD Catalog, 1297

\bibitem[{{Zacharias} et~al.(2013){Zacharias}, {Finch}, {Girard}, {Henden},
  {Bartlett}, {Monet} \& {Zacharias}}]{Zacharias13}
{Zacharias} N., {Finch} C.~T., {Girard} T.~M., {Henden} A., {Bartlett} J.~L.,
  {Monet} D.~G., {Zacharias} M.~I., 2013, \aj, 145, 44

\bibitem[{{Zacharias} et~al.(2015)}]{Zacharias15}
{Zacharias} N. et~al., 2015, \aj, 150, 101

\bibitem[{{Zorec} et~al.(2009){Zorec}, {Cidale}, {Arias}, {Fr{\'e}mat},
  {Muratore}, {Torres} \& {Martayan}}]{Zorec09}
{Zorec} J., {Cidale} L., {Arias} M.~L., {Fr{\'e}mat} Y., {Muratore} M.~F.,
  {Torres} A.~F., {Martayan} C., 2009, \aap, 501, 297

\bibitem[{{Zuckerman} \& {Song}(2004)}]{Zuckerman04a}
{Zuckerman} B., {Song} I., 2004, \araa, 42, 685

\bibitem[{{Zuckerman} \& {Webb}(2000)}]{Zuckerman00}
{Zuckerman} B., {Webb} R.~A., 2000, \apj, 535, 959

\bibitem[{{Zuckerman} et~al.(2012){Zuckerman}, {Melis}, {Rhee}, {Schneider} \&
  {Song}}]{Zuckerman12}
{Zuckerman} B., {Melis} C., {Rhee} J.~H., {Schneider} A., {Song} I., 2012,
  \apj, 752, 58

\end{thebibliography}

\end{document}